\documentclass[letterpaper,11pt]{article}
\pdfoutput=1
\usepackage{jheppub}
\usepackage[utf8]{inputenc}
\usepackage{mathtools,amssymb,braket,mathrsfs,xspace,xcolor,float,color,siunitx,comment,afterpage,multirow,array,hhline,amsthm,framed,placeins,physics,subfig,graphicx}
\usepackage{tikz}
\usetikzlibrary{positioning,arrows,calc}
\usepackage{afterpage}
\usepackage{pdflscape}
\usepackage{booktabs}
\usepackage{amsmath,amssymb}
\usepackage{tocvsec2}
\usepackage{xspace}
\usepackage[shortlabels]{enumitem}
\usepackage[capitalise]{cleveref}

\newcommand{\sectionAuthor}[2]{
    \section{#1}
     \emph{Author: #2} \vspace{0.4cm}
}
\newcommand{\sectionAuthors}[2]{
    \section{#1}
     \emph{Authors: #2} \vspace{0.4cm}
}

\newcommand{\s}       {\ensuremath{\sqrt{s}}\xspace}

\newcommand{\kt}      {\ensuremath{k_{t}}\xspace}
\newcommand{\ktof}[1] {\ensuremath{k_{t,{#1}}}\xspace}
\newcommand{\pt}      {\ensuremath{p_{t}}\xspace}
\newcommand{\pTjet}   {\ensuremath{p_{t\mathrm{,jet}}}\xspace}

\newcommand{\mjet}    {\ensuremath{m_\mathrm{jet}}\xspace}

\newcommand{\TeV}     {\textrm{Te\kern-.07emV}\xspace}
\DeclareSIUnit\electronvolt{{e\kern-.07em V}}
\newcommand{\lhcb}    {\mbox{LHCb}\xspace}

\newcommand{\Z}       {\ensuremath{{Z}}\xspace}
\newcommand{\W}       {\ensuremath{{W}}\xspace}
\newcommand{\Dzero}   {\ensuremath{{D}^0}\xspace}
\newcommand{\Bpm}     {\ensuremath{{B}^\pm}\xspace}

\newcommand{\jpsi}    {\ensuremath{{J\mskip -3mu/\mskip -2mu\psi}}\xspace}
\newcommand{\mup}     {\ensuremath{\mu^+}\xspace}
\newcommand{\mun}     {\ensuremath{\mu^-}\xspace}

\newcommand{\pip}     {\ensuremath{\pi^{+}}\xspace}

\newcommand{\kam}     {\ensuremath{{K}^{-}}\xspace}
\newcommand{\kapm}    {\ensuremath{{K}^{\pm}}\xspace}

\newcommand{\pythia}{\textsc{Pythia}\xspace}
\newcommand{\pythiav}[1]{\pythia \textsc{#1}}
\newcommand{\sherpa}{\textsc{Sherpa}\xspace}
\newcommand{\comix}{\textsc{Comix}\xspace}
\newcommand{\amegic}{\textsc{Amegic++}\xspace}
\newcommand{\csshower}{\textsc{CSShower}\xspace}
\newcommand{\lhapdf}{\textsc{LHAPDF}\xspace}

\newcommand{\herwig}{\textsc{Herwig}\xspace}
\newcommand{\herwigv}[1]{\herwig \textsc{#1}}
\newcommand{\matchbox}{\textsc{Matchbox}\xspace}

\newcommand{\mcatnlo}{\ensuremath{\mathrm{\textsc{MC@NLO}}}\xspace}

\newcommand{\ThePI}{\textsc{TheP8I}\xspace}
\newcommand{\cut}{\text{cut}}

\newcommand\mgaNLO{\textsc{MadGraph5\_aMC@NLO}\xspace}
\newcommand\fastjet{\textsc{FastJet}\xspace}
\newcommand{\nnlojet}{\textsc{NNLOjet}\xspace}

\newcommand{\rivet}{\textsc{Rivet}\xspace}

\newcommand{\rF}{\mathrm{F}}
\newcommand{\rR}{\mathrm{R}}
\newcommand{\rT}{\mathrm{T}}
\newcommand{\muf}{\mu_\rF}
\newcommand{\mur}{\mu_\rR}

\title{Flavoured jet algorithms: a comparative study}

\author[1]{Arnd Behring,}
\affiliation[1]{Max-Planck-Institut für Physik, Boltzmannstraße 8, 85748 Garching, Germany}
\emailAdd{abehring@mpp.mpg.de}

\author[2]{Simone Caletti,}
\affiliation[2]{Institute for Theoretical Physics, ETH, CH-8093 Zürich, Switzerland}
\emailAdd{scaletti@phys.ethz.ch}

\author[3]{Francesco Giuli,}
\affiliation[3]{Dipartimento di Fisica, Universit\`a degli Studi di Roma Tor Vergata and INFN, Sezione di Roma Tor Vergata, Via della Ricerca Scientifica 1, Rome 00133, Italy}
\emailAdd{francesco.giuli@roma2.infn.it}

\author[4,5]{Rados\l aw Grabarczyk,}
\affiliation[4]{Department of Physics, University of Oxford, Keble Road, Oxford, OX1 3RH, UK}
\emailAdd{radoslaw.grabarczyk@physics.ox.ac.uk}
\affiliation[5]{Rudolf Peierls Centre for Theoretical Physics, Clarendon Laboratory, Parks Road, Oxford OX1 3PU, UK}

\author[6]{Andreas Hinzmann,}
\affiliation[6]{Deutsches Elektronen-Synchrotron DESY, Notkestr. 85, 22607 Hamburg, Germany}
\emailAdd{andreas.hinzmann@desy.de}

\author[7]{Alexander Huss,}
\affiliation[7]{CERN, 1211 Geneva 23, Switzerland}
\emailAdd{alexander.huss@cern.ch}

\author[8]{Joey Huston,}
\affiliation[8]{Department of Physics and Astronomy, Michigan State University, East Lansing, MI 48824, USA}
\emailAdd{huston@msu.edu}

\author[7]{Ezra D. Lesser,}
\emailAdd{ezra.lesser@cern.ch}

\author[9]{Simone Marzani,}
\affiliation[9]{Dipartimento di Fisica, Universit\`a di Genova and INFN, Sezione di Genova, Via Dodecaneso 33, 16146, Italy}
\emailAdd{simone.marzani@ge.infn.it}

\author[10]{Davide Napoletano,}
\affiliation[10]{Universit\`{a} degli Studi di Milano-Bicocca \& INFN, Piazza della Scienza 3, Milano 20126, Italy}
\emailAdd{davide.napoletano@unimib.it}

\author[11]{Rene Poncelet,}
\affiliation[11]{Institute of Nuclear Physics, ul. Radzikowskiego 152, 31--342 Krakow, Poland}
\emailAdd{rene.poncelet@ifj.edu.pl}

\author[7,12]{Daniel Reichelt,}
\affiliation[12]{Institute for Particle Physics Phenomenology, Durham University, Durham DH1 3LE, UK}
\emailAdd{d.reichelt@cern.ch}

\author[6,9]{Alberto Rescia,}
\emailAdd{alberto.rescia@ge.infn.it}

\author[5,14]{Gavin P. Salam,}
\affiliation[14]{All Souls College, Oxford OX1 4AL, UK}
\emailAdd{gavin.salam@physics.ox.ac.uk}

\author[15]{Ludovic Scyboz,}
\affiliation[15]{School of Physics and Astronomy, Monash University, Wellington Rd, Clayton VIC-3800, Australia}
\emailAdd{ludovic.scyboz@monash.edu}

\author[9]{Federico Sforza,}
\emailAdd{federico.sforza@ge.infn.it}

\author[7,16]{Andrzej Si\'odmok,}
\affiliation[16]{Jagiellonian University, ul. prof. Stanisława Łojasiewicza 11, 30-348 Kraków, Poland}
\emailAdd{andrzej.siodmok@uj.edu.pl}

\author[10]{Giovanni Stagnitto,}
\emailAdd{giovanni.stagnitto@unimib.it}

\author[16]{James Whitehead,}
\emailAdd{james.whitehead@uj.edu.pl}

\author[17]{Ruide Xu}
\affiliation[17]{University of Michigan, Ann Arbor, MI 48109, USA}
\emailAdd{ruidexu@umich.edu}

\preprint{CERN-TH-2025-113, IFJPAN-IV-2025-13, MCNET-25-14, \\
  \hspace*{\fill} MPP-2025-118, PUBDB-2025-01862}

\abstract{
The accurate identification of heavy-flavour jets---those which originate from bottom or charm quarks---is crucial for precision studies of the Standard Model and searches for new physics. However, assigning flavour to jets presents significant challenges, primarily due to issues with infrared and collinear (IRC) safety.
This paper aims to address these challenges by evaluating recently-proposed jet algorithms designed to be IRC-safe and applicable in high-precision measurements.
We compare these algorithms across benchmark heavy-flavour production processes and kinematic regimes that are relevant for LHC phenomenology. 
Exploiting both fixed-order calculations in QCD as well as parton shower simulations, we analyse the infrared sensitivity of these new algorithms at different stages of the event evolution and compare to flavour labelling strategies currently adopted by LHC collaborations. 
The results highlight that, while all algorithms lead to more robust flavour assignments compared to current techniques, they vary in performance depending on the observable and energy regime. The study lays groundwork for robust, flavour-aware jet analyses in current and future collider experiments to maximise the physics potential of experimental data by reducing discrepancies between theoretical and experimental methods.
}

\begin{document}

\flushbottom
\maketitle

\clearpage

\section{Introduction}
\label{sec:intro}

Jets, i.e.~sprays of highly energetic particles, play a vital role in the precision program of particle physics~\cite{Salam:2010nqg, Marzani:2024mtt}.
They are a useful concept because they are well-defined across several energy scales that characterise high-energy scattering processes, and allow for comparisons between theoretical calculations and experimental measurements. Fixed-order calculations derive jets from quarks and gluons that arise directly from the hard interaction. These partonic jets, with relatively low multiplicity, then evolve down to lower energy scales, fragmenting into more partons, a process usually referred to as the \emph{parton shower}. Low-energy partons then eventually form particles that can be seen in the detectors. At the experimental end of the spectrum, jets are derived from tracks and/or calorimeter clusters.
Jet definitions which are largely stable under the scale evolution between hard scattering and measurement in a detector capture essential features of the short-distance, high-energy scattering, while being fairly insensitive to soft or collinear emissions. Jet observables can also be defined in ways which are insensitive to details of how quarks and gluons coalesce into hadrons and how those interact with the detector.
Progress on both the experimental and theoretical sides has reduced the uncertainties of measurements and improved the precision of predictions. 
This has led to jet observables becoming indispensable tools for high-precision analyses.

To date, jets at the LHC are most commonly defined using the anti-\kt algorithm~\cite{Cacciari:2008gp}, which sequentially clusters particles to construct geometrically regular-shaped jets. Anti-\kt jets are particularly resilient against nonperturbative effects such as multiparton interactions (MPI) that give rise to the Underlying Event (UE) and, to some extent, pile-up, which is caused by multiple proton-proton interactions per bunch crossing.
While hard interactions can be described with high precision, exploiting perturbative QCD calculations, nonperturbative effects are known with limited precision and partly rely on phenomenological models.
Thus, precision next-to-next-to-leading order (NNLO) calculations in QCD have been successfully compared to experimental data at large jet transverse momentum, with relatively small nonperturbative corrections needed~\cite{ATLAS:2017ble,ATLAS:2022nrp,CMS:2024iaa,CMS:2021yzl,AbdulKhalek:2020jut,ATLAS:2024png,CMS:2024hwr,ATLAS:2023tgo}.

Jets are initiated by energetic particles and it is natural to inspect the extent to which they maintain information about those particles through their evolution down to smaller (infrared) energy scales.
Jet substructure~\cite{Marzani:2019hun} aims to answer this by characterising the radiation pattern of jets initiated by different particles. In this context, the question of meaningfully distinguishing quark-initiated jets from gluonic ones has been the subject of many dedicated studies, see, e.g., Refs.~\cite{Gallicchio:2011xq,FerreiradeLima:2016gcz, Cheng:2017rdo,Caletti:2021ysv,Metodiev:2018ftz,Dreyer:2021hhr,Butter:2022xyj,Dolan:2023abg,Knobbe:2023njd}, including a few that originated at past editions of the Les Houches workshop ``Physics at \TeV Colliders (Phys\TeV)''~\cite{Proceedings:2018jsb,Gras:2017jty,Amoroso:2020lgh}.
Furthermore, while the flavour of the hard-scattered quark evolves through the parton shower and fragmentation processes, it carries valuable information about the short-distance physics that produced it.

At high-energy colliders, such as the LHC, it is interesting to study bottom ($b$) and charm ($c$) quark flavour.%
\footnote{Jets initiated by top quarks are also widely studied but fall into a different category due to the short lifetime of top quarks; however, recent evidence of a $t\bar{t}$ state has been observed~\cite{CMS:2025kzt}. Studies for future colliders also consider the possibility of identifying strange quark jets (see, e.g., Ref.~\cite{FCC:2025lpp}).}
This is experimentally feasible because heavy-flavoured hadrons (henceforth $b$ and $c$ hadrons, respectively) typically decay while passing through the detector, giving rise to recognisable experimental signatures. 

Heavy-flavour jets play a crucial role in many physics analyses. For instance, the Higgs boson decays primarily into two $b$ quarks ($H \rightarrow b \bar{b}$) which each subsequently fragment into jets, and studying this process plays a central role in understanding the Higgs mechanism. Furthermore, any definitive statement about the presence of an intrinsic heavy-flavour component in the proton~\cite{LHCb-PAPER-2021-029, NNPDF_IC} requires precision calculations of perturbative cross sections involving heavy quarks. 
Finally, top quarks predominantly decay into \W bosons and bottom quarks, making the flavour of the $b$ quark an important signal for identifying decays of top quarks.

It is therefore of great interest to be able to assign a quark flavour to a jet.
An intuitive expectation could be that this should correspond to the flavour of the partons present at the beginning of the jet evolution.
However, assigning such a flavour label is highly nontrivial.
Experimental analyses usually assign a flavour label based on the presence of flavoured hadrons in the vicinity of a jet, i.e., information available at the late stage of evolution.
This is theoretically problematic, as flavoured quarks can be introduced to a jet via radiation contributions enhanced by collinear and soft singularities.
This weakens the correlation between jet flavour assigned at the hard scattering (based on partons) and the flavour assigned experimentally based on the presence of flavoured hadrons.
Moreover, defining jet flavour in fixed-order calculations is complicated by the fact that solely counting the flavour of the constituent partons is not an infrared- and collinear-safe (IRC-safe) procedure, resulting in the noncancellations of infrared singularities between real and virtual corrections.
This issue was recognised long ago~\cite{Banfi:2006hf,Banfi:2007gu} and addressed thanks to the introduction of an IRC-safe version of the \kt-algorithm~\cite{Catani:1993hr,Ellis:1993tq}, known as the flavour-\kt algorithm.

With the advent of NNLO calculations for jet processes, together with the precision physics program at the LHC, the issue of assigning flavour labels to jets in an IRC-safe manner had to be reconsidered. Given the ubiquity of the anti-\kt jet algorithm in modern analyses, the need for flavoured jet algorithms that maintain the kinematic features of anti-\kt jets arose.
Over the past few years, four proposals for novel anti-\kt-like flavoured jet algorithms that are IRC-safe up to NNLO~\cite{Caletti:2022hnc} or even to all orders~\cite{Czakon:2022wam, Gauld:2022lem, Caola:2023wpj} have appeared.%
\footnote{Another approach that has been pursued consists of abandoning collinear safety and introducing a ``flavour fragmentation function'' calculable to any order~\cite{Caletti:2022glq, Larkoski:2023upz}; see \cref{subsec:WTA_flav} for details.}
At the 2023 edition of the Les Houches Phys\TeV workshop, these algorithms were compared to each other extensively~\cite{Andersen:2024czj}. 
The exercise reported here stems from discussions started at Les Houches. It aims to review the behaviour of these new algorithms, compare them in benchmark scenarios and improve the understanding of how they can be used in practice.
The hope is that this improved understanding will lead to a better knowledge of heavy-flavour jet final states at the LHC, and will foster follow-up studies, for example at future Les Houches workshops.


\section{Setting the stage}
\label{sec:objectives}
QCD interactions are flavour-blind, i.e., gluons couple to quarks irrespective of their flavour and, consequently, their mass. Furthermore, the highest precision in perturbative calculations is usually achieved by considering massless quarks. This enhances theoretical degeneracy with respect to flavour.
Yet flavour is an indispensable physical signature of experimental final states, easily seen at colliders: hadrons containing $b$ and $c$ valence quarks are distinguished by their heavy rest masses and long lifetimes relative to strongly-decaying hadrons and vector bosons.
This makes them an attractive target for experimental measurements.

Unlike for general jet observables, where parton--hadron duality provides a theoretical motivation for comparing parton-level jets with their experimental counterparts constructed from particle tracks and/or calorimeter clusters, na\"ive flavour assignment according to parton identity is intrinsically IRC unsafe. 
This has historically prevented the comparison of experimental measurements of jet flavour with precision calculations.
Recently, a number of algorithms have been proposed to resolve this, providing IRC-safe methods for assigning flavour ``labels'' to jets clustered at parton level.
These flavoured jet algorithms differ substantially from conventional experimental procedures, which despite being experimentally well-defined and measurable, are formally IRC unsafe for massless partons or develop large mass-dependent logarithms for massive quarks.

In this section we summarise current approaches to assign jet flavour, the theoretical properties which must be satisfied by any candidate algorithm for it to be calculable within perturbative QCD, and outline possible criteria by which available alternatives may be assessed.

\subsection{Approaches to jet labelling and tagging}\label{sec:exp-approach}
\afterpage{%
  \begin{landscape}
    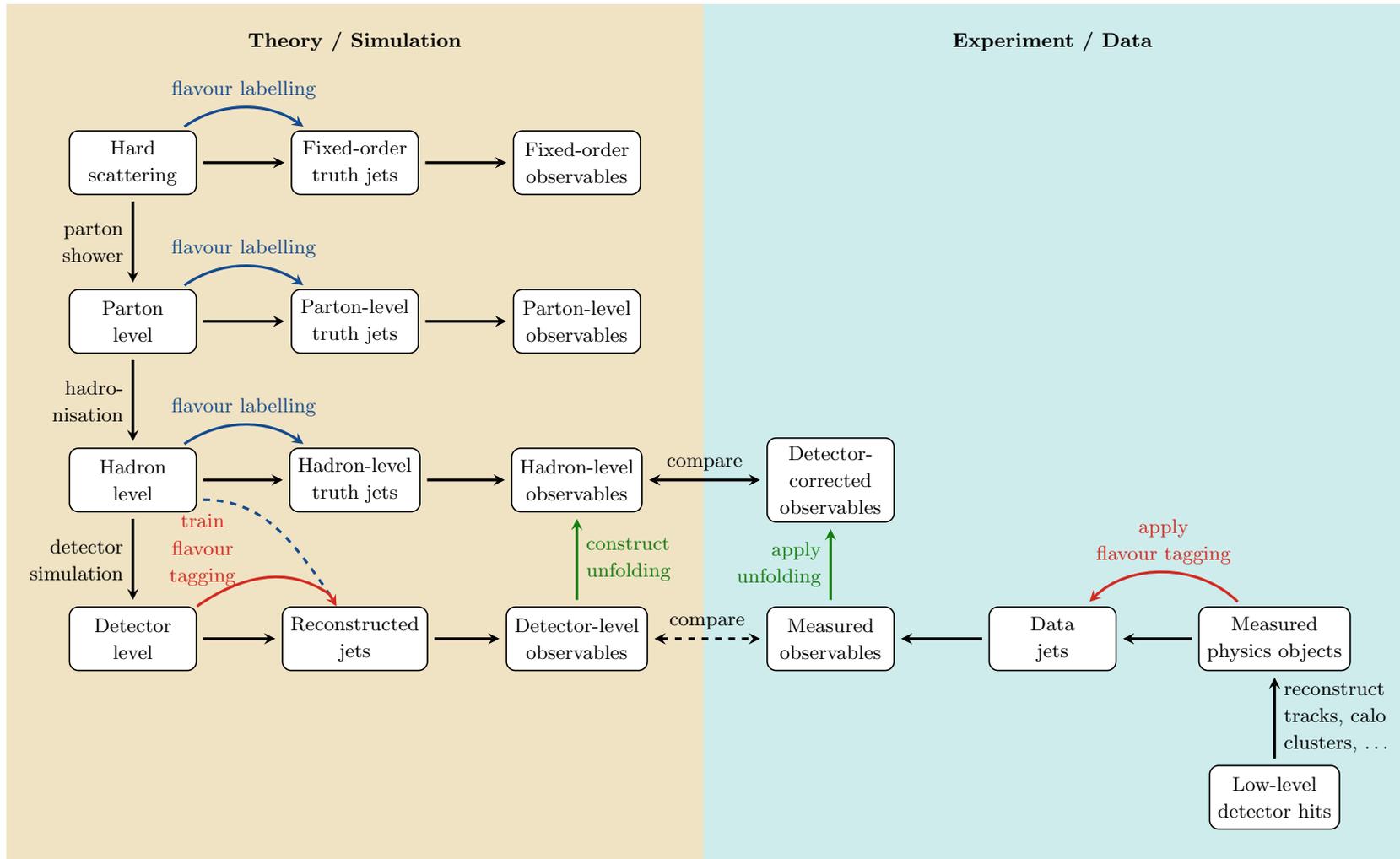
\begin{figure}
      \centering
      \definecolor{bg-exp-cyan}{HTML}{CEECEC}
      \definecolor{bg-th-tan}{HTML}{F0E3C4}
      \definecolor{labelling-blue}{HTML}{144B8F}
      \definecolor{unfolding-green}{HTML}{157D12}
      \definecolor{tagging-red}{HTML}{D42A21}
      \begin{tikzpicture}[
          very thick,
          on grid,
          node distance = 2.5cm and 3.5cm,
          anchor = center,
          align=center,
          font=\footnotesize,
          arr/.style={
            ->,
            >=stealth,
            shorten > = 1mm,
            shorten < = 1mm,
            very thick
          },
          concept box/.style={
            draw,
            semithick,
            fill=white,
            rounded corners,
            align=center,
            minimum height=1cm,
            minimum width=2cm
          }
        ]
        \pgfdeclarelayer{background}
        \pgfdeclarelayer{unfolding}
        \pgfsetlayers{background,unfolding,main}

        \begin{pgfonlayer}{background}
          \fill[bg-th-tan,anchor=north west] (0,0) rectangle (11,-13.5);
          \fill[bg-exp-cyan,anchor=north west] (11,0) rectangle (22,-13.5);
        \end{pgfonlayer}
        \node (th heading) at (5.5,-0.6)  {\textbf{Theory / Simulation}};
        \node (ex heading) at (16.5,-0.6) {\textbf{Experiment / Data}};

        \node[concept box] (th hard ev)     at (2,-2.5)               {Hard \\ scattering};
        \node[concept box] (th parton ev)   [below=of th hard ev]   {Parton \\ level};
        \node[concept box] (th hadron ev)   [below=of th parton ev] {Hadron \\ level};
        \node[concept box] (th detector ev) [below=of th hadron ev] {Detector \\ level};
        \node[concept box] (ex data ev)  [right=18cm of th detector ev] {Measured \\ physics objects};
        \node[concept box] (ex data2 ev) [below=of ex data ev] {Low-level \\ detector hits};
        \draw[arr]
          (th hard ev)
          -- node[left,align=right] {parton \\ shower}
          (th parton ev);
        \draw[arr]
          (th parton ev)
          -- node[left,align=right,text width=4em] {ha\-dro\-ni\-sa\-tion}
          (th hadron ev);
        \draw[arr]
          (th hadron ev)
          -- node[left,align=right] {detector \\ simulation}
          (th detector ev);
        \draw[arr]
          (ex data2 ev)
          -- node[right,align=left] {reconstruct \\ tracks, calo \\ clusters, \dots}
          (ex data ev);

        \node[concept box] (th fo jets)     [right=of th hard ev]     {Fixed-order \\ truth jets};
        \node[concept box] (th parton jets) [right=of th parton ev]   {Parton-level \\ truth jets};
        \node[concept box] (th truth jets)  [right=of th hadron ev]   {Hadron-level \\ truth jets};
        \node[concept box] (th reco jets)   [right=of th detector ev] {Reconstructed \\ jets};
        \node[concept box] (ex data jets)   [left =of ex data ev]     {Data \\ jets};
        \draw[arr] (th hard ev)     -- (th fo jets);
        \draw[arr] (th parton ev)   -- (th parton jets);
        \draw[arr] (th hadron ev)   -- (th truth jets);
        \draw[arr] (th detector ev) -- (th reco jets);
        \draw[arr] (ex data ev)     -- (ex data jets);

        \node[concept box] (th fo obs)       [right=of th fo jets]
          {Fixed-order \\ observables};
        \node[concept box] (th parton obs)   [right=of th parton jets]
          {Parton-level \\ observables};
        \node[concept box] (th hadron obs)   [right=of th truth jets]
          {Hadron-level \\ observables};
        \node[concept box] (th detector obs) [right=of th reco jets]
          {Detector-level \\ observables};
        \node[concept box] (ex detector obs) [left =of ex data jets]
          {Measured \\ observables};
        \node[concept box] (ex hadron obs)   [above=of ex detector obs]
          {Detector-\\corrected \\ observables};
        \draw[arr] (th fo jets)     -- (th fo obs);
        \draw[arr] (th parton jets) -- (th parton obs);
        \draw[arr] (th truth jets)  -- (th hadron obs);
        \draw[arr] (th reco jets)   -- (th detector obs);
        \draw[arr] (ex data jets)   -- (ex detector obs);

        \draw[arr,labelling-blue]
          (th hard ev)
          to[out=35,in=145] node[above] {flavour labelling}
          (th fo jets);
        \draw[arr,labelling-blue]
          (th parton ev)
          to[out=35,in=145] node[above] {flavour labelling}
          (th parton jets);
        \draw[arr,labelling-blue]
          (th hadron ev)
          to[out=35,in=145] node[above] {flavour labelling}
          (th truth jets);

        \coordinate (train b tagging) at ($(th reco jets)+(-0.4,0.7)$);
        \draw[labelling-blue,dashed,very thick]
          ($(th hadron ev.south east)+(0.1,0.2)$)
          to[out=0,in=125]
          (train b tagging);
        \draw[tagging-red,very thick]
          (th detector ev.north east)
          to[out=35,in=145]
          (train b tagging);
        \draw[arr,tagging-red,shorten < = 0mm,shorten > = 0mm]
          (train b tagging)
          --
          (th reco jets);
        \node[text=tagging-red]
          (train b tagging label)
          at ($(train b tagging)+(-2.0,0.7)$)
          {train \\ flavour \\ tagging};
        \draw[arr,tagging-red]
          (ex data ev)
          to[out=135,in=45] node[above] {apply \\ flavour tagging}
          (ex data jets);

        \begin{pgfonlayer}{unfolding}
          \draw[arr,unfolding-green]
            (th detector obs)
            -- node[right,align=left] {construct \\ unfolding}
            (th hadron obs);
          \draw[arr,unfolding-green]
            (ex detector obs)
            -- node[left,align=right] {apply \\ unfolding}
            (ex hadron obs);
        \end{pgfonlayer}

        \draw[arr,<->]
          (th hadron obs)
          -- node[above] {compare}
          (ex hadron obs);
        \draw[arr,<->,dashed]
          (th detector obs)
          -- node[above] {compare}
          (ex detector obs);
      \end{tikzpicture}
      \caption{Simplified overview of a generic jet-based analysis, illustrating both the main steps for calculating theoretical predictions and for analysing experimental data. The coloured arrows show the propagation of information about flavour. See the main text for details.}
      \label{fig:flowchart}
    \end{figure}
  \end{landscape}
}

It is useful to first describe the typical way in which theory and experiment are used to compare jet predictions and measurements.
The simplified picture of a generic jet analysis is illustrated in a flowchart given in \cref{fig:flowchart}. This chart displays how information is translated from first-principles theory and from low-level information recorded by the particle detectors through different stages of simulation and experimental analysis to quantities that can be compared directly.\footnote{The distinction between theory and experiment on the left- and right-hand sides of the diagram should not imply who is producing these predictions: experimental collaborations frequently use simulation tools to produce predictions, generate corrections, etc. Instead, the idea is to illustrate how information derived from either theory or from measurements is propagated between different levels.} 
In the following text, we use this simple chart as a common frame of reference to define certain terminology which we use throughout this article.

The left-hand side of \cref{fig:flowchart} sketches how theory predictions are obtained, starting from the description of events at the level of hard scattering. Events are successively translated to parton level by applying a parton shower, then to the hadron level through hadronisation\footnote{At hadron level one can further distinguish between hadrons produced from hadronising the partonic final state and their decay products.} and finally to the detector level by feeding the hadronic final state through a simulation of an experimental detector's response. In principle, we can analyse the events at any of these stages, construct jets by applying a jet algorithm to the events and calculate observables such as fiducial cross sections or distributions from these objects. Jets constructed from simulated events at parton or hadron levels are called \emph{truth jets}. From them we can calculate \emph{fixed-order}, \emph{parton-level}, or \emph{hadron-level observables}, respectively. Jets obtained from simulated detector-level events are called \emph{reconstructed jets}, and we calculate \emph{detector-level observables} from them.

On the right-hand side of \cref{fig:flowchart}, low-level detector objects (tracker hits, energy deposits in calorimeter cells, etc.) are used to reconstruct higher-level physics objects (tracks, topological cell clusters, displaced vertices, ParticleFlow objects, etc.). We call jets constructed from these physics objects \emph{data jets} and observables calculated from them \emph{measured observables}. There is therefore a level of parity between reconstructed jets and data jets, as well as detector-level observables and measured observables, which are at the same level but reconstructed using either simulation or experimental data, respectively.

The predictions for detector-level observables can be compared to measured observables (dashed black arrow in the flowchart), and this is done both in the context of measurements of Standard Model processes and in the context of searches for new physics. But since these quantities are detector specific, \emph{unfolding} is often used to translate detector-specific measurements to hadron level (green arrows in the flowchart), making them more universal. Unfolding requires constructing a response matrix by comparing predictions for hadron-level observables to detector-level observables calculated on the same simulated events.\footnote{Recently, methods to unfold events as a whole rather than differential distributions of observables are being studied, e.g., Refs.~\cite{Andreassen:2019cjw,Bellagente:2020piv,Butter:2024vbx,Huetsch:2024quz}.} This unfolding procedure is then applied to the measured observables to obtain \emph{detector-corrected observables}. These unfolded results can finally be compared to predictions for hadron-level observables, as well as to results from other experiments.

Given the focus of this study, it is useful to describe how flavour information is propagated both theoretically and experimentally. Here and in the following, we distinguish the procedure of \emph{flavour tagging} (red arrows in the flowchart) from \emph{flavour labelling} (blue arrows).
We define flavour tagging as what is used to assign flavour to \emph{reconstructed} or \emph{data} jets using just detector-level information as input.
On the other hand, by flavour labelling we refer to the procedure which assigns flavour based on truth-level flavour information available from simulation. Usually, this is applied  to \emph{truth} jets, in which case this is sometimes called ``truth-level flavour tagging'' in the literature. (Sometimes flavour labelling is also applied to reconstructed jets in simulation for the purpose of training flavour tagging algorithms.)
While this distinction in terminology is not used universally throughout the literature, we find it useful to keep the concepts separate in this paper.

For flavour labelling, a number of different strategies are used by experimentalists and theorists.
\begin{itemize}
  \item \emph{Naive/constituent labelling} (also called anti-\kt labelling in this paper):
        jets are first reconstructed without making reference to flavour. Flavour labels are then assigned based on the flavour of the jet constituents. This method requires the flavoured objects (quarks at parton level or undecayed heavy flavour hadrons at hadron level) to be in the list of inputs to the jet algorithm. It has been used in particular in theory calculations, but it is known to be IRC unsafe at higher orders in perturbation theory, as we will discuss below.
  \item \emph{Exclusive labelling}~\cite{UA1:1990mkp,CDF:1989gpa,LHCb:2012dgy,ALICE:2022wpn}:
        this can be thought of as an extension of naive labelling to situations where an exclusive decay mode of a particular flavoured hadron can be fully reconstructed. The flavour of this hadron can be used to assign flavour to a jet. Before jet reconstruction, the decay products of the flavoured hadron are removed from the list of inputs and replaced by the reconstructed hadron.
        In this case, flavour labelling and flavour tagging can be done using essentially identical procedures (up to standard detector corrections to account for particles which were missed or poorly reconstructed). However, experimentally this full-hadron reconstruction is typically only performed in a single decay channel for a single hadron, meaning that the hadronic selection is exclusive and necessarily misses most hadrons of the same flavour.
  \item \emph{Ghost labelling}~\cite{Cacciari:2005hq,Cacciari:2008gn}:
        in general, ``ghost association'' is a method for studying jet clustering in which approximately-zero-momentum ``ghost'' particles are introduced to an event, to see which jet they are clustered into by the jet algorithm.
        ``Ghost labelling'' applies this to flavour labelling by adding prompt, undecayed $b$ or $c$ hadrons from the truth record of the event to the input to the jet algorithm as ``ghosts''. These ghosts, with momentum scaled by a very small factor, e.g., $10^{-20}$, label the jet they are clustered into with the flavour they are carrying.
        When ghost labelling is applied at hadron level with undecayed hadrons using the anti-\kt algorithm, the method is equivalent to naive labelling.
  \item \emph{Cone labelling}~\cite{D0:2010zho} (or geometrical matching):
        after jet reconstruction, the truth record of the generator is searched for the undecayed $b$ or $c$ hadrons, potentially above a certain threshold in transverse momentum, and jets in the event are labelled as $b$ or $c$ jets if a previously unmatched flavoured hadron is found within a maximum $\Delta R$ of the jet axis.
  \item \emph{Flavoured jet algorithms}:
        dedicated jet algorithms that come with a prescription to assign flavour to the resulting jets in an IRC-safe way. This includes the flavour-\kt algorithm~\cite{Banfi:2006hf,Banfi:2007gu} as well as the newly developed ones that we will discuss in more detail below.
\end{itemize}
Exclusive, ghost and cone labelling are frequently used in experimental analyses,\footnote{Given that a $b$ hadron can decay to $c$ hadrons, a flavour definition priority must be assigned, with the usual convention that a $b$-labelled jet is not further labelled as a $c$ jet.} while naive labelling and flavoured jet algorithms have so far been predominantly applied in theoretical calculations.

The flavour of reconstructed or data jets is determined using \emph{flavour tagging} algorithms that rely, depending on the algorithm, on information about physics objects such as tracks or calorimeter clusters or also higher-level derived objects like fully-reconstructed heavy-flavour hadrons (when available), or secondary vertices to infer the flavour.
Many early measurements relied mostly on the impact parameter of tracks from decays of flavoured hadrons or the reconstruction of secondary vertices \cite{CDF:2003xlw,CMS:2012feb,ATLAS:2015thz,ALICE:2021wct}. Modern flavour tagging algorithms rely more and more on complex machine learning approaches in order to combine low-level information into powerful discriminants~\cite{ATLAS:2022qxm,CMS:2017wtu,LHCb-PAPER-2016-039}.

These algorithms must be validated or trained on simulated datasets where reconstructed jets have been labelled in some way. Assigning flavour to reconstructed jets for training or validation (dashed blue arrow) can be accomplished in several ways.
One possibility is to use a flavour labelling procedure on hadron-level truth jets and associate these jets to reconstructed jets at detector level on an event-by-event basis using geometric matching between the truth and reconstructed jets. The reconstructed jets then inherit the flavour label from the truth jet.
Another possibility is to include undecayed hadrons from the event record at hadron level as ghosts in the clustering of the event at detector level.
Finally, more recent approaches also include matching flavoured hadrons from hadron level to tracks after detector simulation and assigning these tracks to reconstructed jets via ghost association.
All of these approaches have in common that flavour labelling procedures influence flavour tagging algorithms via their training.

The tension between experimental and theoretical approaches to jet flavour labelling results in additional systematic uncertainties that hamper comparisons between high-precision calculations and experimental data.
In addition, it is well-known that flavour labelling based solely on the presence of flavoured quarks in the partonic jet is not IRC safe~\cite{Banfi:2006hf} for massless flavoured quarks.

\subsection{Infrared and collinear safety}\label{sec:IRC-safety}
When calculating the probability of a particular outcome of a measurement, quantum mechanics compels us to treat all degenerate states as equal and to sum over their contributions. Since we cannot distinguish states in which a massless particle splits into two collinear particles or where a massless particle radiates another massless particle with vanishing energy, we have to sum over those indistinguishable contributions even though they have different multiplicities. This is the basic tenet of the Kinoshita--Lee--Nauenberg (KLN) theorem~\cite{Kinoshita:1962ur,Lee:1964is}, which states that infrared singularities from soft or collinear massless particles that arise in real emission and virtual contributions cancel in ``sufficiently inclusive'' observables. If the cancellation holds true to all orders in perturbation theory for a given observable, it is called \emph{infrared and collinear safe}.

A necessary (but not sufficient) condition for IRC safety is that observables must be consistent between final states with different multiplicities in singular kinematic regions.\footnote{For a detailed discussion of this point see, e.g., Ref.~\cite{Banfi:2004yd}.} More concretely, this means, for example, that an observable must be the same for a configuration with a final state gluon as for a configuration where this gluon has been replaced by a collinear quark--anti-quark pair. If an observable does not obey these consistency conditions for all soft or collinear limits of massless particles the observable is \emph{IRC unsafe}. In particular, this implies that an unsafe observable cannot be computed in fixed-order perturbation theory.

When constructing an observable, one way to ensure IRC safety is to pay attention to whether it agrees across different multiplicities in all soft and collinear limits. Another way is to reduce the list of kinematic limits in which the matrix elements become singular. For example, this can be achieved by considering massive quarks.
This regulates collinear singularities involving this quark (giving rise to the well-known dead cone effect~\cite{Dokshitzer:1991fd,Dokshitzer:1995ev}). Therefore, if an observable does not agree on configurations in these limits, it is still technically IRC safe. However, the observable does pick up a logarithmic dependence on the ratio of the mass of the quark to a hard scale of the process. If we were to send the quark mass to zero, the logarithm would diverge, recovering the noncancellation of infrared singularities in the massless case. While the observables are IRC safe in this case, they become strongly sensitive to infrared scales, such as the quark mass, and they would diverge if we were to send the quark mass to zero. We call observables that exhibit this behaviour \emph{infrared sensitive}.\footnote{Note that some authors would classify infrared-sensitive observables as IRC unsafe. However, we think that making this distinction is useful in the context of this study.}

Besides introducing a regulator in the underlying theory, we can also make observables IRC safe by suitably excluding singular regions of phase space. Usually, the observables then also pick up a logarithmic dependence on the parameter controlling this exclusion. Thus, they become infrared sensitive and depend on low-energy details of the theory. Good examples would be the jet radius, a photon-isolation cut or the cut on transverse momentum in weak boson production.
Observables sensitive to these parameters can be computed in fixed-order perturbation theory. However, theoretical predictions are often supplemented with all-order resummation of infrared effects and, potentially, require the inclusion of nonperturbative contributions. 

Despite the theoretically motivated definitions of IRC safety and infrared sensitivity, these classifications are also relevant for experimental analyses. While IRC-unsafe observables can be still measured in experiments, the availability of theory predictions largely depends on the observables being IRC safe. Moreover, IRC-safe and infrared-insensitive observables tend to give a better handle on the underlying short-distance physics, whereas infrared-sensitive observables, as the name suggests, are influenced more by long-distance, low-energy details of the theory.

\subsection{Origin of the difficulties of defining flavoured jets}
This brings us to the origin of the challenges in defining flavoured jets in theoretical and experimental situations. Crucially, the object that carries flavour information and, therefore, distinguishes a flavoured from a nonflavoured jet is inherently of infrared origin. For example, at parton level, the mass of the quark is (much) smaller than the energy scales of the labelled jet and thus represents a potential infrared parameter.

Depending on the specifics of the jet-flavour definition, we can face two different situations. If the mass is needed to regulate infrared singularities, perturbative predictions will exhibit logarithmic corrections that need to be resummed. However, the resummation of these contributions is challenging because not all logarithms are of collinear origin and thus not accessible with DGLAP or LL/NLL parton shower methods. Further, such a definition would make strictly fixed-order perturbative computations, which typically achieve the highest precision, more complicated and less reliable due to the aforementioned logarithms. 
Another way to ensure the predictivity of the theory is to define flavoured jets such that the quark mass is not required as a regulator. This second avenue has the additional advantage of allowing a definition of jet flavour in massless QCD computations that is IRC safe and naturally removes the infrared sensitivity to the mass parameter, i.e., all associated logarithms.

Including flavour assignments in the observable definition can cause the observable to become IRC unsafe. This happens if the flavour assignment can be changed by particle splittings with soft or collinear kinematics. For concreteness, we pick $b$ flavour as the feature we want to track.
As examples, we discuss three cases:
\begin{enumerate}[(i),noitemsep,nolistsep]
    \item collinear $g \to b\bar{b}$ splitting,
    \item collinear $b \to b g$ splitting,
    \item soft $g \to b\bar{b}$ emissions,
\end{enumerate}
which illustrate the problem and point to different solutions that need to be combined.

If we assign flavour to jets based on the presence of flavoured jet constituents (partons or particles), the flavour assignment depends on whether a gluon in the jet splits collinearly into a $b \bar{b}$ quark pair. Given their collinearity, any sensible jet algorithm will cluster those quarks into the same jet, but whether or not this splitting happens will determine the presence of flavour in the jet. Since a collinear $g \to b \bar{b}$ splitting introduces a singularity for massless $b$ quarks, this makes the observable IRC unsafe. For massive $b$ quarks, the observable acquires a logarithmic dependence on the quark mass, i.e., it becomes infrared sensitive. To cure this problem, we can change the flavour assignment from being based on the presence of any flavoured jet constituent (\emph{any flavour} scheme) to being based on the net flavour number (\emph{net flavour} scheme): if a gluon splits into a $b \bar{b}$ pair, the net flavour number does not change. Alternatively, we can also count the number of flavoured jet constituents ($b$ and $\bar{b}$ quarks each count positively) and ask for an odd number of flavoured jet constituents (\emph{mod2 flavour} scheme). This might be closer to what is possible in experiments since measuring the charge of the flavoured particles can be challenging and $B$--$\bar{B}$ oscillations further blur the picture. Also the ``mod2 flavour'' number does not change under $g \to b \bar{b}$ splitting. Both with the ``net flavour'' and ``mod2 flavour'' schemes, a jet with two $b$-flavoured jet constituents would have to be regarded as unflavoured. This cures the IRC unsafety or infrared sensitivity that arises from collinear $g \to b \bar{b}$ splittings. Since the discussions below will frequently refer to the different flavour recombination schemes, we summarise them in \cref{tab:flavour-recombination}.

\begin{table}
  \centering
  \caption{Summary of the different flavour recombination schemes. The sum goes over all particles $i$ under consideration whose flavour should be combined. Flavoured particles count as $f_i = +1$, flavoured anti-particles count as $f_i = -1$. Unflavoured particles count as $f_i = 0$.}
  \label{tab:flavour-recombination}
  \begin{tabular}{ll}
    \toprule
    Scheme & Consider a set of particles flavoured if \dots \\
    \midrule
    any flavour  & $\sum_i |f_i| > 0$ \\
    net flavour  & $\sum_i f_i \neq 0$ \\
    mod2 flavour & $\sum_i |f_i| \equiv 1 \mod 2$ \\
    \bottomrule
  \end{tabular}
\end{table}

The second configuration concerns collinear-gluon emissions from flavoured quarks (i.e., $b \to b g$). A hard collinear emission from a flavoured quark might result in a quark that would fail transverse momentum constraints at particle or detector level. 
In a partonic calculation, if we try to implement this cut directly at parton-level, we will encounter a collinear singularity (in the case of massless quarks) or a mass logarithm for massive ones. In either case, the proper way to account for this is to introduce a fragmentation function that can resum this contribution through DGLAP evolution. This is a well-established formalism and we shall not discuss this further in this work. 

Finally, another problem arises for soft $g \to b \bar{b}$ splittings. If the splitting happens at wide angles, one of the $b$ quarks can end up in a hard jet (or both $b$ quarks can end up in separate jets), and thereby change the flavour assignment in those jets---even if the net or mod2 flavour scheme is applied. Since the gluon is soft, this induces IRC unsafety for massless quarks or infrared sensitivity for massive quarks. To remedy the issue, we have to change the flavour assignment from just being derived from the flavour of the jet constituents, to a jet algorithm that takes flavour into account.

The first such algorithm was the flavour-\kt algorithm~\cite{Banfi:2006hf}, where the metric that underlies the clustering was changed in order to prefer early recombinations of soft flavoured quark pairs. Being based on the \kt algorithm, the resulting jets do not have the same nice properties as jets based on the now ubiquitous anti-\kt algorithm.
To solve these issues, different anti-\kt-like jet algorithms have been proposed~\cite{Caletti:2022glq,Caletti:2022hnc,Czakon:2022wam,Gauld:2022lem,Caola:2023wpj}. Each of them introduces a new ``flavoured'' jet definition with different properties. The details of these algorithms are briefly reviewed in \cref{sec:new_alg}. 
Most of the novel flavoured jet algorithms need complete and detailed flavour and kinematical information about the events, and this is not achievable given the efficiencies and resolutions of realistic detectors. Therefore, they cannot be used for flavour tagging directly. Instead they are useful for flavour labelling, because flavour labelling is applied at the truth level where flavour information is available from the simulation tools for all final state particles. The study presented here employs the algorithms as flavour labelling strategies. We will briefly comment about possible implementations in experimental analyses in the Conclusions, \cref{sec:conclusions}.

\subsection{Goals of a jet flavour definition}
Given the fundamental differences in the way jet flavour is defined in theoretical and experimental situations, it is obvious that a sensible definition of jet flavour that captures most of the theoretical requirements but is also implementable in experiment will rely on finding common ground. This is very similar to the agreement of using the anti-\kt algorithm for (nonflavoured) jets. Adopting the anti-\kt algorithm, allows us to compare calculations and measurements at various stages of the event evolution and is one of the cornerstones of LHC phenomenology. We would argue that a good definition of flavoured jets would allow similarly consistent comparisons between theory and experiment.

There are several perspectives under which we can evaluate the new flavoured jet algorithms.
One open question is how similar these algorithms are to existing flavour labelling procedures that are currently used in experimental analyses and theoretical studies. This is helpful for understanding the size of the effect caused by switching from the current approaches to using one of the new algorithms. The similarity has been a design consideration for recent flavoured jet algorithms. In particular, they all strive to yield jets with kinematics that are as close as possible (if not identical) to nonflavoured anti-\kt jets. This is useful because the flavour tagging algorithms applied to detector-level objects mostly work with reconstructed or data jets that are constructed using the nonflavoured anti-\kt algorithm. Therefore, one could expect that a greater similarity between truth- and reconstructed-level jet flavour could reduce the size of uncertainties introduced by the unfolding procedure. It is tempting to use similarity to current flavour labelling procedures as the primary evaluation criterion. However, as we argue above, current flavour labelling strategies suffer from sensitivity to infrared effects. Using similarity as the sole evaluation criterion runs the risk of perpetuating these problems.

Another interesting aspect is the stability of predictions across different stages of the event description: how much do predictions change when comparing predictions for fixed-order, parton-level and hadron-level observables? Keeping corrections small when going from one level to another is advantageous because it lets us more easily compare predictions from fixed-order calculations, where usually the highest precision is available, to experimental measurements. Stability under ``event evolution'' is also tightly linked to infrared (in)sensitivity. Depending on the goals of a given analysis infrared insensitivity may be a major consideration. For example, if we want to study PDFs and fit them with perturbative input, we want to aim for an infrared-insensitive jet flavour definition, i.e., a definition which truthfully represents the short-distance process. If we are interested in the event evolution, for instance studying the dead-cone effect of a heavy quark/hadron, we might want to adopt a more infrared-sensitive definition. This choice also comes with an expectation on the modelling precision and accuracy. An insensitive definition should allow for precise computations within perturbation theory and should be stable under hadronisation and soft-physics modelling. Definitions that are more sensitive to fragmentation, hadronisation and soft-physics modelling allow us to validate or constrain them but of course also will suffer from the associated uncertainties. Over the course of the study, the stability of the predictions turned out to be a useful complementary evaluation criterion.

Further criteria include the computational efficiency of the implementations, their ease of use in existing experimental analysis chains, and the applicability of the algorithms to jet substructure analyses. While these criteria play a role in some parts of the study, they were not the main focus and we leave their evaluation to future work.

As discussed above, the introduction of new flavoured jet algorithms lays the foundations for infrared insensitive flavour labelling.
We note that IRC safety of these algorithms has been investigated through high perturbative orders in Ref.~\cite{Caola:2023wpj}. However, the aspect of infrared sensitivity and, in particular, the overall dependence of observables on the event evolution is less well studied. In addition, there are no detailed phenomenological comparisons between these algorithms. To fill this gap, the goal of this study is two-fold:
\begin{enumerate}
    \item compare the phenomenology of the different jet flavour definitions and study their infrared sensitivity, and
    \item investigate the impact of these modified jet definitions in experimental analyses at truth level.
\end{enumerate}

\section{A new generation of algorithms}\label{sec:new_alg}
In this section, we review the main features of the four new flavoured jet algorithms that have been recently introduced. We also describe their implementation in \fastjet~\cite{Cacciari:2011ma}. All algorithms are sequential clustering algorithms that are based on the anti-\kt algorithm, which performs two-to-one recombination steps of final state partons or particles. In the following, we use the term ``pseudojet'' to denote any object that results from the recombination ($i+j$), of particles, or previous pseudojets ($i$ and $j$), in the clustering sequence.  The clustering sequence of the anti-\kt algorithm is determined by a distance measure between the pseudojets ($d_{ij}$) as well as the beam ($d_{iB}$): 
\begin{align}
    d_{ij} &= \min(\ktof{i}^{-2},\ktof{j}^{-2}) \frac{\Delta R_{ij}^2}{R^2}
    \,, &
    d_{iB} &= \ktof{i}^{-2}
    \,.
\end{align}
The algorithm recombines the pair $i,j$ if their distance is the smallest among all pair- and beam-distances or removes pseudojet $i$ from the list if its beam distance is the smallest and declares it a jet. The sequence ends if all pseudojets are removed.

\subsection{SoftDrop Flavour (SDF)}\label{sec:new_alg_SD}
The two principles that guided the construction of the SoftDrop flavour algorithm~\cite{Caletti:2022hnc} can be summarised as follows.
First, one would like an IRC-safe jet flavour labelling procedure that does not alter the kinematics of the original jet, likely defined with the anti-\kt algorithm. Second, one would like to design a procedure as simple as possible, that makes use of tools and algorithms already used both in phenomenology and by the experimental collaborations. 

The idea of using a grooming procedure such as SoftDrop~\cite{Larkoski:2014wba} originates from the realisation that the configuration that naive flavour counting IRC unsafe at NNLO is analogous to the one responsible for nonglobal logarithms~\cite{Dasgupta:2001sh} in jet shape observables. In fact, similar to nonglobal logarithms, it is possible to eliminate the infrared ambiguities of jet flavour grooming.  There are, however, two restrictions or modifications to the SoftDrop algorithm that one must implement to ensure IRC safety of jet flavour through NNLO.  First, the angular exponent $\beta$ in the grooming constraint must be strictly positive.  Secondly, the $\textsc{Jade}$ clustering algorithm~\cite{JADE:1986kta,JADE:1988xlj} must be used as a reclustering algorithm to properly order and groom soft emissions that can render the jet flavour ambiguous. 

Thus, the SoftDrop flavour algorithm proceeds as follows:
\begin{enumerate}
  \item Recluster the jet with the \textsc{Jade} algorithm which has a metric $d_{ij}$ corresponding to the invariant mass of $i$ and $j$.
  \item At each stage of the clustering, require that particles $i$ and $j$ pass the SoftDrop grooming requirement, where:
  \begin{align}
    \label{eq:sdcrit}
    \frac{\min \left( p_{ti}, p_{tj}\right)}{p_{ti} + p_{tj}} > z_c\left( \frac{\Delta_{ij}}{R_0}\right)^\beta, \\ \nonumber \Delta_{ij}=\sqrt{(y_i-y_j)^2+(\phi_i-\phi_j)^2},
  \end{align}
  with the initial jet radius $R_0$,  the energy cut $z_c$, and  the angular exponent $\beta > 0$.
  \item If the stage in the clustering passes the grooming requirement, terminate and return  the flavour of the jet using a flavour recombination scheme on the jet's constituents that are left. If the grooming requirement fails, then remove the softer of the two branches, and continue to the next stage of the \textsc{Jade} clustering along the harder branch.
\end{enumerate}
The above procedure ensures IRC safety at NNLO, but it is known to fail beyond it. 

\subsection{Flavoured anti-\texorpdfstring{\kt}{kt} (CMP)}\label{sec:new_alg_CMP}

A proposal for a flavoured anti-\kt algorithm has been made in Ref.~\cite{Czakon:2022wam}, for the purpose of the comparisons in this article dubbed CMP. The guiding principle for achieving IRC safety is similar to the flavour-\kt algorithm, which introduces a modified distance measure used for jet clustering that captures soft and flavoured quark pairs. Starting from the standard anti-\kt distance measure~\cite{Cacciari:2008gp}, a suppression factor $\mathcal{S}_{ij}$ is introduced for flavoured quark or pseudojet pairs while all other distances are kept:
\begin{align}
  d_{ij}^{(\text{flavoured})} = d_{ij}^{(\text{standard})} \times
  \begin{cases}
    \mathcal{S}_{ij} \, , & \parbox{0.5\linewidth}{if both $i$ and $j$ have nonzero flavour of opposite sign and magnitude,} \\[0.5\baselineskip]
    1 \, , & \text{otherwise.}
  \end{cases}
  \label{eq:flavour-anti-kT}
\end{align}
where $\mathcal{S}_{ij}$ is supposed to vanish if the quark pair is soft compared to a hard process scale. While the flavour of a pseudojet might be decided by different flavour recombination schemes, i.e., net or mod2 flavour, the recommendation is to use net flavour. This modification has two implications: 1) it makes the algorithm identical to the anti-\kt algorithm in the absence of flavoured pairs, and 2) it leads to modifications in the kinematics of the clustered jets compared to nonflavoured anti-\kt jets. If $\mathcal{S}_{ij}$ is properly chosen, it will only modify the distances of pseudojets that are soft with respect to the relevant scales of the process, and therefore, the modification of the clustered jet kinematics is expected to be small. This distinguishes this approach from the other algorithms, which use nonflavoured anti-\kt jets and modify the flavour labels of the constituents.

The original proposal considered the following definition for the soft term:
\begin{align}
  \mathcal{S}_{ij} = 1-\theta\left(1-\kappa_{ij}\right)\cos\left(\frac{\pi}{2}\kappa_{ij}\right)
  \quad \text{with} \quad \kappa_{ij} \equiv \frac{1}{a} \, \frac{k_{t,i}^2+k_{t,j}^2}{2 k_{t,\text{max}}^2}\; .
  \label{eq:Sij}
\end{align}
Here $k_{t,\text{max}}$ is some (possibly dynamic) hard scale relevant to the process. To facilitate variations and the study of the sensitivity to the choice made, the parameter $a$ is introduced, which is expected to be of $\order{10^{-1}}$. The usage of $k_{t,i}$'s as a measure of the hardness of the quarks is motivated by boost invariance along the collision axis but leads to an unwanted suppression in the case of double collinear limits to the two incoming hadrons (i.e., each quark becomes collinear to a different initial-state parton). This has been pointed out in Ref.~\cite{Caola:2023wpj} together with a possible solution to this issue. Originally, the proposed solution involved the replacement (see discussion in \cref{sec:new_alg_IFN})
\begin{equation}
  \mathcal{S}_{ij} \to \bar{\mathcal{S}}_{ij} = \mathcal{S}_{ij} \frac{\Omega_{ij}^2}{\Delta R^2} \; , \quad \text{where} \quad
  \Omega_{ij}^2 = 2 \left[ \frac{1}{\omega^2}(\cosh(\omega \Delta y_{ij})-1) - (\cos \Delta \phi_{ij} -1 )\right]
  \; .
  \label{eq:Sij-fix}
\end{equation}
However, it turns out that this replacement leads to undesirable deviations between the plain anti-\kt algorithm and the CMP algorithm in regions of phase space where modifications are neither necessary nor expected. Instead, we found that the replacement\footnote{Such a modification was considered as an alternative to \cref{eq:Sij-fix} to restore IRC safety in the code underlying the numerical tests of Ref.~\cite{Caola:2023wpj}, as well.}
\begin{equation}
  \kappa_{ij} \to \bar{\kappa}_{ij} = \kappa_{ij} \sqrt{\frac{\Omega_{ij}^2}{\Delta R^2}}  \label{eq:kappaij-fix}
\end{equation}
shows better behaviour. We describe the observations and reasoning in more detail in \cref{app:HW}. We denote the variant of the CMP algorithm with the replacement in \cref{eq:kappaij-fix} as CMP$\Omega$ and employ this variant throughout this study. Unless stated explicitly, the label CMP also refers to this variant in the rest of the article.

To minimise the impact on the jet kinematics through the modified clustering, it is advised to consider only the flavour of interest, i.e., $b$ quarks if $b$ jets are to be studied, as ``flavoured'' and not to use the modification for several partonic flavours at once.

\subsection{Flavour dressing (GHS)}\label{sec:new_alg_GHS}

The flavour dressing algorithm~\cite{Gauld:2022lem} aims at a flavour assignment that is IRC safe in perturbation theory and can be combined with any definition of a jet. For the purpose of this study it will be dubbed GHS. In this algorithm, flavour assignment is \emph{factorised} from jet reconstruction: in a first step, flavour-agnostic jets are obtained with an IRC-safe jet algorithm (such as anti-\kt); in a second step, jets are \emph{dressed} with flavour quantum numbers according to flavour information about the event (e.g., flavoured particles in parton-level predictions or stable heavy-flavour hadrons in hadron-level predictions).
The set of flavour-agnostic jets is considered to be composed of truth or reconstructed jets that have passed a fiducial selection criterion, such as a minimum transverse momentum requirement, $p_{t,\text{cut}}$.
The flavour dressing algorithm then proceeds as follows:
\begin{enumerate}
  \item given the set of all particles in the event (that can be identified with the same input that enters the flavour-agnostic jet algorithm), populate a set of distances:
    \begin{itemize}
      \item if the particle $p_i$ is associated to jet $j_k$, add the distance measure $d_{p_i j_k}$. By \emph{association}, in parton-level predictions, we mean that a particle is a constituent of a jet (other options are possible, see Ref.~\cite{Gauld:2022lem}). In a hadron collider environment, the beam distances $d_{p_i B_{\pm}}$ should be added if $p_i$ is not associated to any jet;
      \item for each unordered pair of particles $p_i$ and $p_j$, add the distance measure $d_{p_ip_j}$ if either \emph{both particles are flavoured} or \emph{at least one particle is unflavoured and $p_i$ and $p_j$ are associated with the same jet} (the latter requirement guarantees collinear flavour safety when recombining particles close to the kinematic boundary of a jet);
    \end{itemize}
  \item select the pairing with the smallest distance measure:
    \begin{enumerate}
      \item $d_{p_i p_j}$ is the smallest:
        the two particles merge into a new particle $k_{ij}$ carrying the sum of the four-momenta and flavour. All entries that involve $p_i$ or $p_j$ are removed and new distances for $k_{ij}$ are added.
      \item \(d_{p_i j_k}\) is the smallest: assign the particle $p_i$ to the jet $j_k$ and remove all entries that involve $p_i$.
      \item \(d_{p_i B_{\pm}}\) is the smallest: discard particle $p_i$ and remove all entries that involve $p_i$.
    \end{enumerate}
  \item iterate until there are no more entries in the set of distances;
  \item the flavour assignment for jet $j_k$ is determined according to the accumulated flavours: one can either perform a net summation of flavours or mod2 summation.
\end{enumerate}

The distances adopted are inspired by the flavour-\kt algorithm~\cite{Banfi:2006hf}, with the generalisations of Ref.~\cite{Caola:2023wpj}.
For the distance $d_{ab}$ between any two final-state objects (particles or jets), we adopt \cref{eq:flktdist1,eq:flktdist2}.
For the distance between a particle $p_i$ and beam $B_{\pm}$ in direction of positive/negative rapidity, we adopt:
\begin{align}
  \label{eq:dist_fB}
  d_{p_i B_{\pm}} &=
  \max\bigl(p_{ti}^\alpha,p_{tB_{\pm}}^\alpha(y_{i}) \bigr)
  \min\bigl(p_{ti}^{2-\alpha},p_{tB_{\pm}}^{2-\alpha}(y_{i})\bigr) \,,
  \nonumber\\
  p_{tB_{\pm}}(y) &=
  \sum_{k=1}^m p_{t,j_k} \Bigl[
  \Theta( \pm \Delta y_{j_k}) +
  \Theta( \mp \Delta y_{j_k}) \; e^{\pm \Delta y_{j_k}}
  \Bigr] \,,
\end{align}
with $\Delta y_{j_k} = y_{j_k} - y$ and $\Theta(0) = \tfrac{1}{2}$. Hence, the flavour dressing algorithm depends on two parameters, $\alpha$ and $\omega$, the same parameters entering the IFN algorithm (see \cref{eq:flktdist1,eq:flktdist2}), with default choice $\left( \alpha, \omega \right) = (1, 2)$.

\subsection{Interleaved Flavour Neutralisation (IFN)}\label{sec:new_alg_IFN}

Interleaved Flavour Neutralisation (IFN)~\cite{Caola:2023wpj} takes a hybrid
approach with respect to the flavoured jet algorithms mentioned above. It
retains the set of distances from a given underlying generalised-\kt
algorithm (such as anti-\kt, or Cambridge/Aachen), but allows the possibility
of ``neutralising'' flavour at each stage of the kinematic clustering.

The IFN is triggered during the clustering of pseudojets $i$ and $j$ whenever
the softer pseudojet, $p_{ti} < p_{tj}$, carries flavour. Before performing the
clustering step, the algorithm runs a global search for potential
neutralisation candidates among a list $L$ of all current, flavoured
pseudojets, $k \in L$. The decision to neutralise flavour within $i$ from flavour in $k$ is based on
a separate set of distances, $u_{ik}$, which are taken to be
\begin{align}
    u_{ik} &= \max(p_{ti}, p_{tk})^\alpha \, \min(p_{ti},p_{tk})^{2-\alpha}  \; \Omega_{ik}^2\label{eq:flktdist1}\,,\\
    \Omega_{ik}^2 &= 2 \left[ \frac{1}{\omega^2} \left( \cosh (\omega \Delta y_{ik}) -1 \right)
                    - \left( \cos \Delta \phi_{ik} -1 \right)\right]\label{eq:flktdist2}\,,
\end{align}
with $\Delta y_{ik} = y_i - y_k$ and $\Delta \phi_{ik} = \phi_i - \phi_k$, and
parameters $0 < \alpha \leq 2$ and $\omega$.

\begin{figure}[ht]
    \centering
    \includegraphics[width=0.9\linewidth]{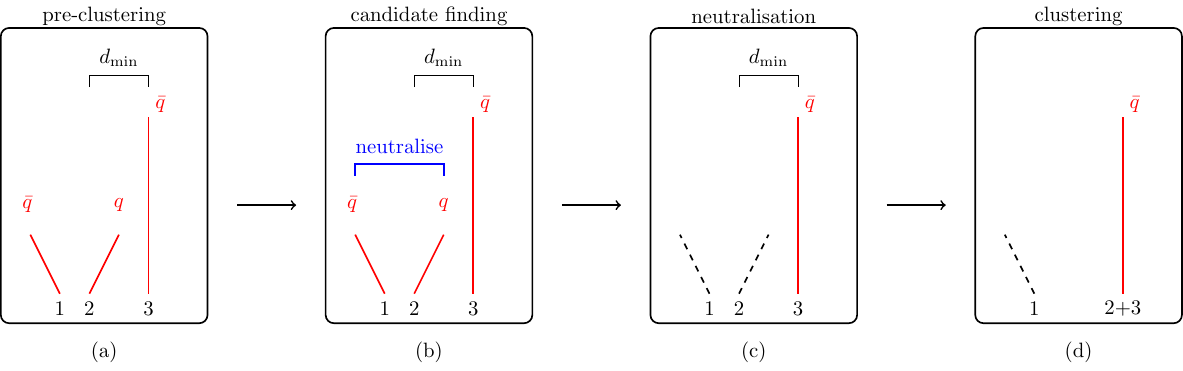}%
    \caption{Illustration of the IFN algorithm (see main text for description).}
    \label{fig:ifn}
\end{figure}
\Cref{fig:ifn} shows a schematic representation of the IFN algorithm for
the typical IRC-unsafe configuration at NNLO, with a soft gluon splitting to a
quark-antiquark pair (numbered $1$ and $2$), and a hard jet (here a $\bar q$,
numbered $3$). (a) The anti-\kt algorithm is about to cluster the soft quark
$2$ with the hard jet $3$, which would result in a change of flavour in the hard jet.  (b) Instead, with the IFN, a neutralisation candidate search is
     triggered (as the softer recombining pseudojet is flavoured). The algorithm
finds a neutralisation candidate, jet $1$ for the soft quark $2$. (c) In this
particular case, the algorithm decides to neutralise the flavours of $1$ ($\bar q$) and
$2$ ($q$). From this point onwards, pseudojets $1$ and $2$ have had their flavour
neutralised, and are to be considered flavourless (as indicated by the black
dashed lines). (d) As there is no more
flavour to neutralise globally in the event, the clustering step is now allowed to proceed
(with net flavour summation). Pseudojet $2$, which is now flavourless, recombines with
pseudojet $3$ and the final flavour of the jet $2+3$ is $\bar q$.

The crucial point is that the kinematics of the
underlying anti-\kt jets are left unchanged (the same pseudojets are combined
at each clustering step).  The (flavour) IRC unsafety of the typical NNLO
configuration is lifted, because soft quark-antiquark pairs coming from a gluon
splitting are neutralised before their flavour can contaminate harder jets.

The form of the $u_{ik}$ distances is critical. In particular the angular
component, $\Omega_{ik}^2$, is designed to eliminate IRC divergences stemming
from the interplay of initial-state radiation with soft flavour at large angle
(similar configurations plague other flavoured algorithms, and IRC unsafety
issues are resolved by the same $\Omega_{ik}^2$ choice, instead of the standard $\Delta R_{ik}^2$ angular factor).  In the IFN, IRC
safety requires a recursive definition of the neutralisation search, as well as
\begin{equation}
    \omega > 2 - \alpha
\end{equation}
for the choice of the algorithm's parameters. We usually choose $\left( \alpha,
\omega \right) = (1, 2)$ or $(2,1)$.

\subsection{Winner-Take-All flavour (WTA)}
\label{subsec:WTA_flav}

An alternative approach to defining jet flavour, first presented in Ref.~\cite{Caletti:2022glq}, uses the Winner-Take-All (WTA) axis~\cite{Bertolini_2014, Larkoski:2014uqa} to define jet flavour in the infrared limit. The procedure is as follows:
\begin{enumerate}
    \item reconstruct the jet using any desired jet reconstruction algorithm, for example standard anti-\kt using $E$-scheme recombination and any desired resolution (radius) parameter;
    \item recluster the jet using any IRC-safe jet clustering algorithm and recursively follow the hardest splitting until reaching the outermost branch in the reclustered jet tree (WTA axis);\footnote{This follows the original formulation of the algorithm~\cite{Caletti:2022glq} whereas more recent developments that focus on high-accuracy resummation~\cite{Larkoski:2023upz, Larkoski:2024nub} explicitly adopt the \kt algorithm for the reclustering step.}
    \item define the flavour of the jet by the sum of flavours which are collinear along the resulting WTA axis.
\end{enumerate}

Since the WTA axis is completely insensitive to soft radiation, defining jet flavour using the WTA algorithm is also infrared safe. However, the WTA flavour definition is not collinear safe. This collinear unsafety is handled by introducing a ``flavour fragmentation function,'' which is calculable to any perturbative order~\cite{Larkoski:2023upz}.

Due to its collinear unsafety and its fundamentally different approach from the four IRC-safe algorithms described above, this algorithm is not included in the majority of studies reported in this article. However, the algorithm also presents experimental benefits in the case when a subset of heavy-flavoured hadrons can be fully reconstructed experimentally, since the flavour definition does not require all flavoured particles in the jet to be known, only whether or not a single flavoured particle is collinear with the WTA axis. Therefore, some experimentally-motivated studies are performed and their findings are presented below in~\cref{sec:findings}, as well as in the corresponding \cref{app:lhcb}.

\subsection{\texorpdfstring{\fastjet}{FastJet} implementations}
\label{sec:fj-implementations}

The four IRC-safe flavoured jet algorithms presented above are available in the form of \fastjet~\cite{Cacciari:2011ma} plugins 
(\texttt{SDFPlugin}, \texttt{CMPPlugin}, \texttt{GHSAlgo}, and \texttt{IFNPlugin} respectively), and can be found as part of the \fastjet contrib repository at
\begin{center}
    \url{https://fastjet.hepforge.org/contrib/}\,.
\end{center}
To compile the plugins one needs \texttt{fastjet-config} in the environment path. (Due to its comparatively simple implementation using existing \fastjet tools, namely the absence of a new flavour reclustering scheme, no discrete plugin for WTA flavour is provided.)
The flavour machinery relies on \fastjet's \texttt{PseudoJet::UserInfo} extra user information 
holder, which we initialise with a \texttt{FlavInfo} class instance (or the more generic
class \texttt{FlavHistory}, for cases where one is interested in the flavour
information at different stages of the clustering), which is part of the \texttt{IFNPlugin} installation.
The \texttt{IFNPlugin} requires at least \fastjet v.~3.4.1.
The \texttt{SDFPlugin}, \texttt{CMPPlugin} and \texttt{GHSAlgo} packages all depend on the \texttt{IFNPlugin} to handle flavour information,
through the already mentioned flavour infrastructure.
Furthermore, the \texttt{SDFPlugin} also requires the \texttt{RecursiveTools}~\cite{Dasgupta:2013ihk,Larkoski:2014wba,Frye:2017yrw}
library to be installed since it exploits its SoftDrop and \textsc{Jade} clustering implementations. 

An example file (\texttt{example.cc}) is provided in each algorithm's directory, which demonstrates how to set up the \texttt{JetDefinition} and run the jet clustering on a given event (an input vector of \texttt{PseudoJet} objects).
The main steps are summarised for each algorithm in \cref{app:alg-setup}.
The example files show how to set up the flavour information of the input particles, and access the flavour of the resulting jets.
An input particle's extra \texttt{UserInfo} can be initialised, e.g., by
its PDG id,
\begin{verbatim}
    PseudoJet particle(px, py, pz, E);
    particle.set_user_info(new FlavHistory(pdg_id));
    event.push_back(particle);
\end{verbatim}
and the flavour of the final recombined jet can be accessed by
\begin{verbatim}
    vector<PseudoJet> jets = jet_def(event);
    for (const auto &j : jets) {
        string flav = FlavHistory::current_flavour_of(j).description();
        cout << "Jet flavour: " << flav << endl;
    }
\end{verbatim}
to print each jet's flavour in human-readable format, or to label flavour in a
jet according to the \fastjet setting (``net'' or ``mod2'', set through the \texttt{FlavRecombiner} class):
\begin{verbatim}
    bool is_blabelled = FlavHistory::current_flavour_of(j)[5] != 0;
\end{verbatim}

\section{Findings}\label{sec:findings}

The main objective of this study is to compare the algorithms presented in various LHC-based scenarios and to provide a comprehensive discussion of the phenomenological differences. To this end, individual studies for different final states or phase-space regions have been performed. 
The details of the setup and results for each individual study are collected in dedicated appendices. 
In this section, we condense the most relevant findings to provide the reader with a compact overview. We refer interested readers to the extensive discussions in \crefrange{app:HW}{app:atlas}. 
We will see that, despite the different kinematic configurations, selection procedures, and details in each analysis, a coherent picture emerges. 

All algorithms we compare introduce new parameters. We identify the following baseline setup for the new algorithms:
\begin{itemize}
    \item SDF: The SoftDrop algorithm described in \cref{sec:new_alg_SD} with parameters $z_c=0.1$ and $\beta=1$, or $\beta=2$ (where the value of $\beta$ is indicated explicitly in the various studies).
    \item CMP$\Omega$: The flavour anti-\kt algorithm described in \cref{sec:new_alg_CMP} with parameters $a = 0.1$, $\omega=2$ and the choice of $k_{t,\text{max}}$ as the hardest transverse momentum of a pseudojet at each clustering step.
    \item GHS: The flavour dressing algorithm described in \cref{sec:new_alg_GHS} with $\alpha = 1$, $\omega = 2$ and $p_{t,\text{cut}} = \SI{15}{\giga\eV}$.\footnote{Natural units with $\hbar=c=1$ are used throughout.}
    \item IFN: The interleaved flavour neutralisation algorithm from \cref{sec:new_alg_IFN} with $\alpha = 2$ and $\omega = 1$.
\end{itemize}
For all algorithms, we choose either the ``mod2'' or ``net-flavour'' recombination scheme. In addition to flavoured jet algorithms, we show results for the plain anti-\kt algorithm equipped with the ``mod2-flavour'' recombination scheme in parton-shower computations. This algorithm is infrared sensitive due to the missing treatment of soft $b\bar{b}$ pairs, but we use it to gauge the impact of the corresponding modifications in the flavoured jet algorithms.

\subsection{Impact of flavour recombination schemes}

\begin{figure}
    \centering
    \includegraphics[width=0.5\linewidth,page=3]{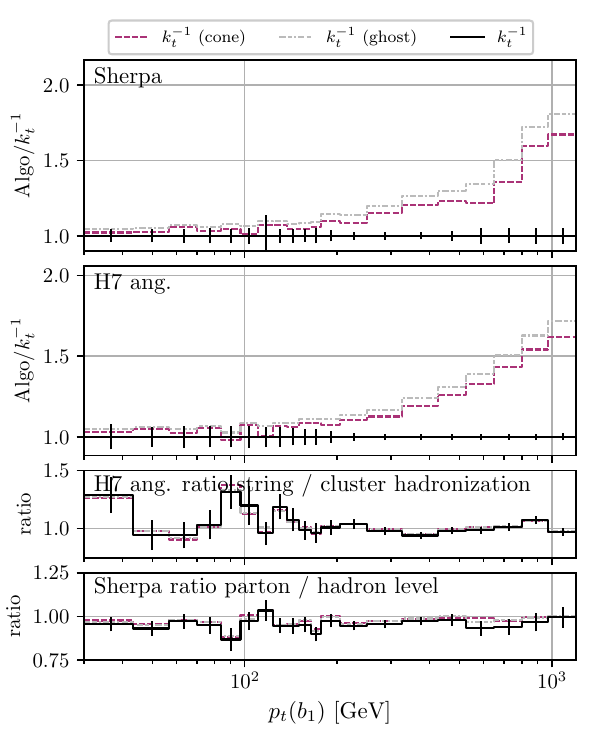}%
    \includegraphics[width=0.5\linewidth,page=3]{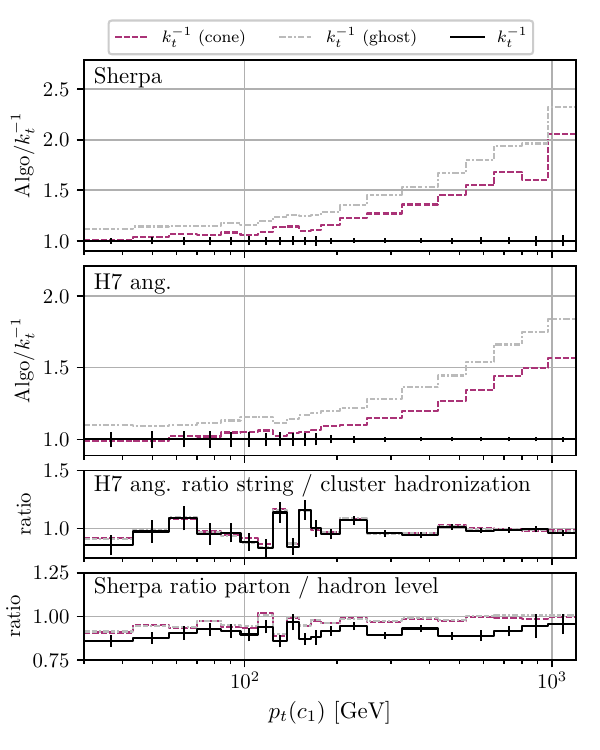}
    \caption{NLO+PS predictions from \sherpa and \herwig for $pp \to \Z + b$ (left) and $pp \to \Z + c$ (right) in central kinematics. Different experimentally inspired jet labelling algorithms are compared to the anti-\kt algorithms with mod2 recombination for the transverse momentum of the leading $b$/$c$ jet. Additionally, differences between the cluster and string hadronisation model for \herwig, as well as parton and hadron level for \sherpa, are shown.}
    \label{fig:summary_ppzj_nlops_bottom_exp}
\end{figure}

Before discussing the new flavoured jet algorithms, we point out the impact of the flavour recombination scheme. As discussed above, experimental analyses often correspond to the ``any-flavour'' scheme, where one or more flavoured particles with the correct kinematical properties are required to assign the jet flavour. To ensure IRC safety from NLO QCD onwards, ``mod2-flavour'' or ``net-flavour'' must be employed as discussed in \cref{sec:objectives}. The phenomenological impact of these schemes is profound because they remove the flavour that arises from the splitting of gluons ($g\to b\bar{b}$ or $g \to c\bar{c}$) within a jet. To assess the size of this effect, we compare the anti-\kt algorithm using the ``mod2-flavour'' scheme to the anti-\kt algorithm with cone and ghost labelling methods used in some experimental analyses for central $Z+\text{jet}$ production (see \cref{app:Zjnnlonlops} for details). In the simulation, the decay of all $b$ and $c$ hadrons is deactivated, and the anti-\kt algorithm with the labelling of the ``any-flavour'' scheme would be the same as ghost labelling. In \cref{fig:summary_ppzj_nlops_bottom_exp} we present the transverse momentum spectrum for $b$- and $c$-labelled jets. Focusing on the second and third panels from the top, we see that the three methods agree at low \pt, but deviate increasingly at high \pt. The effect at a \pt of about $\SI{300}{\GeV}$ is $\sim 20\%$ and reaches up to $50\%$ at $\SI{1}{\TeV}$. The interpretation of this observation is that the cone and ghost labelling strategies are picking up additional $b$/$c$-labelled jets from $g\to b\bar{b}$ splittings and are therefore infrared sensitive. We want to stress that this observation is well known (see, for example, Refs.~\cite{CMS:2023aim, ATLAS:2024tnr}), and phenomenological analyses typically take these into account by Monte Carlo and unfolding corrections. These corrections are however inherently infrared sensitive since they are computed with algorithms that are not infrared flavour safe.

Differences between ``net-flavour'' and ``mod2-flavour'' arise due to different flavour assignments in the case of jets with $bb$ or $\bar{b}\bar{b}$ content. Phenomenologically, these effects are comparatively small, see for example the discussion in \cref{app:Zjnnlonlops}, because such configurations do not correspond to infrared-enhanced regions of the cross section.

\subsection{Fixed-order comparisons}
The algorithms can be compared at different stages of the event evolution: the hard process (i.e.~in fixed-order computations), the parton level, or the particle level (i.e.~in matched parton shower predictions with and without hadronisation). We will first discuss the scenarios accurately describing the highest energy scales, i.e., fixed-order predictions, and move then to the more complete simulations of the event evolution to lower scales. This will allow us to judge the stability of the flavour assignment across energy scales.

Fixed-order computations have been performed through NNLO QCD for the final states $H(\to b\bar{b})W$ and $Z+b/c$-jet. 
Details of $HW$ studies are collected in \cref{app:HW}.
The $Z+b/c$-jet final state has instead been studied in a central phase space selection, inspired by ATLAS and CMS analyses, and at forward rapidities, as appropriate for \lhcb.
Details on the phase-space selection cuts can be found in \cref{app:Zjnnlonlops} and \cref{app:ppzcharm_nnlo}.

\begin{figure}
    \centering
    \includegraphics[width=0.49\linewidth]{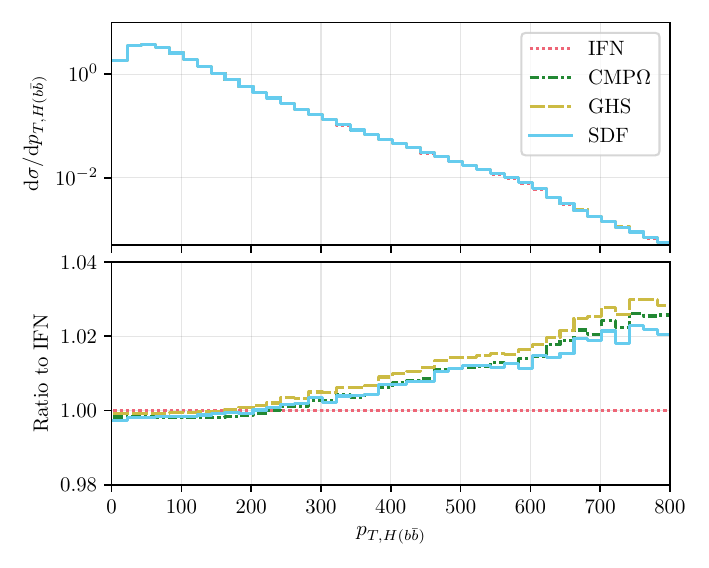}%
    \includegraphics[width=0.48\linewidth,page=51]{figures/nnlo_ppzc_LHCb/ZcLHCb-crop.pdf}
    \caption{Comparison of the flavoured jet algorithms at fixed-order. The \pt spectrum of the Higgs boson, reconstructed from two $b$-labelled jets in $pp\to WH$ production is shown on the left. The right shows the \Z rapidity spectrum in a LHCb $Z+c$ selection.}
    \label{fig:summary_pQCD_1}
\end{figure}

We start with the $HW$ final state and show on the left-hand side of \cref{fig:summary_pQCD_1} the transverse momentum spectrum of the Higgs boson reconstructed from two identified $b$ jets. The upper panel shows the absolute spectra, while the lower panel shows the ratio to the IFN algorithm. Generally, the predictions agree with each other across the phase space, with only a slight shape difference between IFN and all the other flavoured algorithms that reaches $2\%$ at $\SI{800}{\GeV}$. Noticeably, similar observations hold for a rather different process, with different selection cuts, namely the $Z+c$-jet final state in the \lhcb kinematic regions. Here, we show on the right-hand side of \cref{fig:summary_pQCD_1} the rapidity spectrum of the \Z boson. We find qualitative agreement between the different algorithms, with some shape differences in the forward region between GHS/IFN and SDF/CMP. The same holds for central phase space selections, see \cref{app:Zjnnlonlops}.

From the fixed-order studies, we can conclude that the differences produced by the algorithm on NNLO QCD calculations are largely negligible, and the predictions are consistent. The low sensitivity to the jet algorithms in these fixed-order scenarios originates from the low flavour multiplicity. Up to NNLO QCD, there are at most two additional ($b$ or $c$) flavoured partons in the final state on which the algorithms act. The corresponding channels are also colour-suppressed compared to the same process with $g\to gg$ splitting, and therefore have small numerical contributions to the cross sections. Although not explicitly shown here, at NLO QCD, all algorithms almost identically reproduce the plain anti-\kt algorithm (equipped with ``mod2-flavour'' recombination for IRC-safety).

\subsection{Effects from parton shower and hadronisation}

The phenomenological differences become more apparent when looking at full Monte Carlo predictions that include the parton shower and nonperturbative effects. To guide this discussion, we consider studies performed for vector boson plus jet final states, in the central phase-space region.
We also study the impact of labelling
schemes in the development of flavour tagging algorithms.

\begin{figure}
    \centering
    \includegraphics[width=0.5\linewidth,page=3]{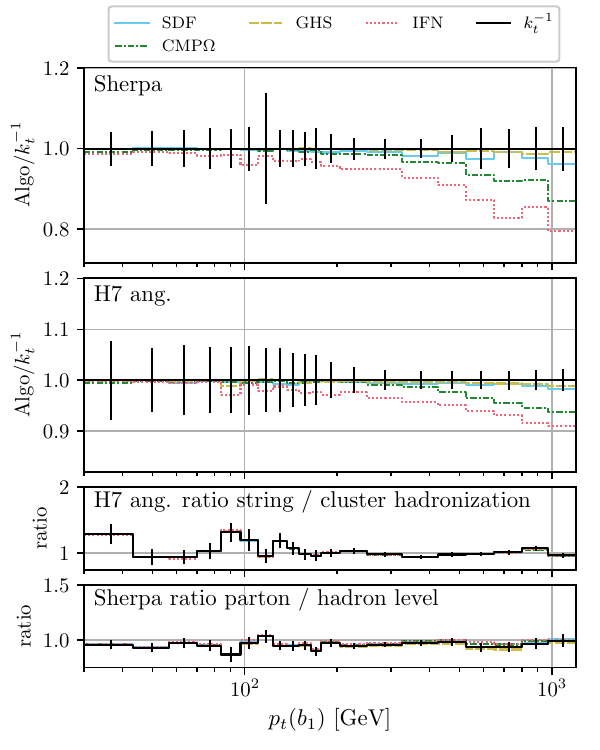}%
    \includegraphics[width=0.5\linewidth,page=3]{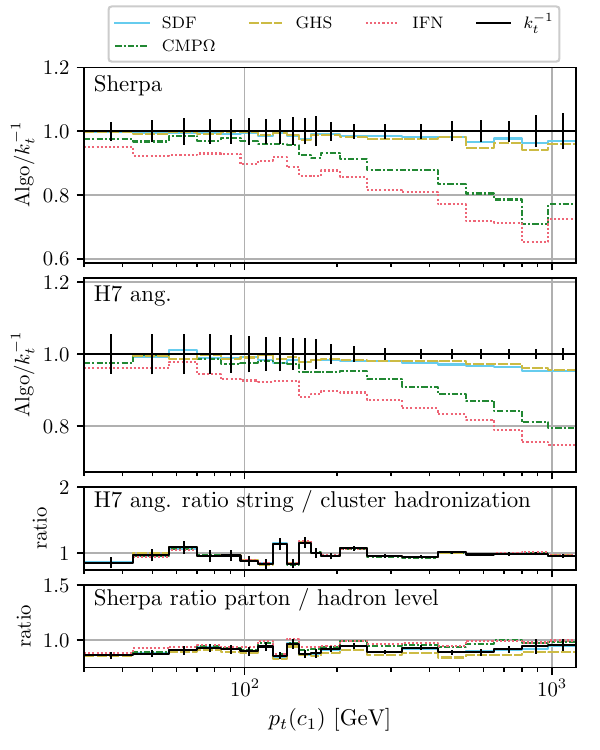}
    \caption{NLO+PS predictions from \sherpa and \herwig for $pp \to \Z + b$ (left) and $pp \to \Z + c$ (right) in central kinematics. Different jet algorithms (CMP$\Omega$ -- green, SDF -- light blue, GHS -- yellow, IFN -- red) are compared for the transverse momentum of the leading $b$/$c$ jet. Additionally, differences between the cluster and string hadronisation model for \herwig, as well as parton and hadron level for \sherpa, are shown. The vertical bars indicate statistical uncertainties.}
    \label{fig:summary_ppzj_nlops_bottom}
\end{figure}

In \cref{fig:summary_ppzj_nlops_bottom}, we summarise the findings of the $Z+\text{jet}$ studies in the central phase space. We focus on the transverse momentum spectrum of $b$-labelled jets (left-hand side) and $c$-labelled jets (right-hand side), but similar conclusions essentially hold for all energy-dependent observables. The statistical uncertainties are in some cases large compared to the observed differences between the algorithms. Still, the observed differences are meaningful since the analysis has been performed on the same sample, which leads to a strong statistical correlation.

The first two panels from the top in \cref{fig:summary_ppzj_nlops_bottom} compare different jet algorithms with respect to the anti-\kt algorithm (with ``mod2-flavour'' recombination) at the particle level for \sherpa and \herwig, respectively. The differences between the flavoured jet algorithms are smaller than what we observed for the flavour-recombination-scheme dependence, but also become more pronounced for high transverse momentum. The SDF and GHS algorithms reproduce the plain anti-\kt behaviour throughout the spectrum, while the CMP and IFN algorithms give reduced rates at high transverse momentum. The effect is more pronounced for the $c$-label case, i.e., the case of smaller mass, for which we also expect a larger flavour multiplicity (i.e., number of flavoured quarks/hadrons) in the final state, see \cref{app:Zjnnlonlops}.

In the lowest two panels in \cref{fig:summary_ppzj_nlops_bottom}, we show the ratio of two different hadronisation models implemented in \herwig and the ratio between the hadron and parton level for \sherpa, for the various algorithms. The difference between the hadronisation models is very minor and within the statistical errors. The differences between parton and hadron levels for the $b$-label case are also rather small, as are the differences between the responses of the jet algorithms. However, for the $c$-label case, the hadronisation effects are more significant, in particular for a small transverse momentum. The algorithms' responses are comparable. Still, differences arise again in the tail of the distribution, where we can observe that the CMP, IFN, and experimental labelling techniques for anti-\kt jets lead to smaller hadronisation corrections than the rest.

\begin{figure}
    \centering
    \includegraphics[width=0.49\linewidth,page=2]{figures/lops_ppzjet/pythia-Zj-flav-algs.pdf}
    \includegraphics[width=0.49\linewidth]{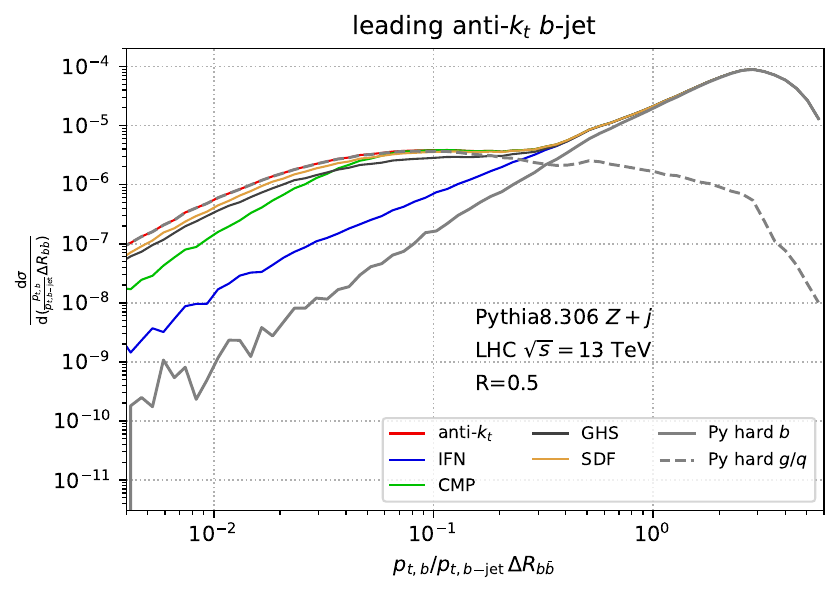}
    \caption{LO+PS $Z+j$. Left: The ratio, to net-flavour anti-\kt (in red), of IFN ($\alpha=1$ and $\alpha=2$, blue), CMP ($a=0.1$, green), GHS ($\alpha=1$, $\omega=2$, black) and SDF ($\beta=2$, $z_{\cut}=0.1$, yellow), for the transverse momentum of the leading $b$-labelled jet $p_{t,b\mathrm{-jet}}$. Right: Distribution of $\frac{p_{t,b}}{p_{t,b\mathrm{-jet}}} \Delta R_{b\bar b}$, in events with two $b$ quarks at parton level, compared to classification of events according to the \pythia hard matrix element (hard $b$, solid grey, vs.~hard $g$ or light $q$, dashed grey).}
    \label{fig:summary_lops_ppzjet}
\end{figure}

Before we interpret these results, it is instructive to view the problem from the perspective that the most robust algorithm is the one that most reliably reproduces the correct flavour of the hard-scattering process after shower evolution. This has been studied in detail for $Z+\text{jet}$ events in \cref{app:ppzb_kinematics}. First, this study confirms the qualitative behaviour of the algorithms as a function of \pt as seen on the left-hand side in \cref{fig:summary_lops_ppzjet}, where the same quantity is shown for a LO \pythia setup (see \cref{app:ppzb_kinematics} for details).

For a selection of events with at least two $b$ quarks in the final state, the right-hand side of \cref{fig:summary_lops_ppzjet} shows the quantity $p_{t,b}/p_{t,b\text{-jet}} \Delta R_{b\bar{b}}$, where $p_{t,b\text{-jet}}$ is the transverse momentum of the $b$ jet, $p_{t,b}$ denotes the transverse momentum of the hardest $b$ quark in the jet and $\Delta R_{b \bar{b}}$ in the angular distance between the two $b$ quarks in the event. This quantity separates the hard-$b$-initiated events (solid grey line) from the $g/q$-initiated events (dashed grey line) in a clean way; see \cref{app:ppzb_kinematics} for more details. The $b$-labelled jets originating from light flavours are likely to have obtained their label from a soft $b$ pair (small $p_{t,b}/p_{t,b\text{-jet}}$) while $g$-initiated jets are changed into $b$-labelled jets through collinear $g\to b\bar{b}$ splittings (small $\Delta R_{b\bar{b}}$). We can find qualitatively different behaviours by looking at the $b$-labelled rates for this observable provided by the various algorithms. The cleanest labelling, i.e., the labelling that is least contaminated by $q/g$-initiated events, is provided by IFN. The other algorithms pick up a significant fraction of the $q/g$-initiated events. These events populate the differences observed in the \pt spectrum, i.e., the differences arise from mislabelled jets.

\begin{figure}
    \centering
    \includegraphics[width=0.5\linewidth,page=3]{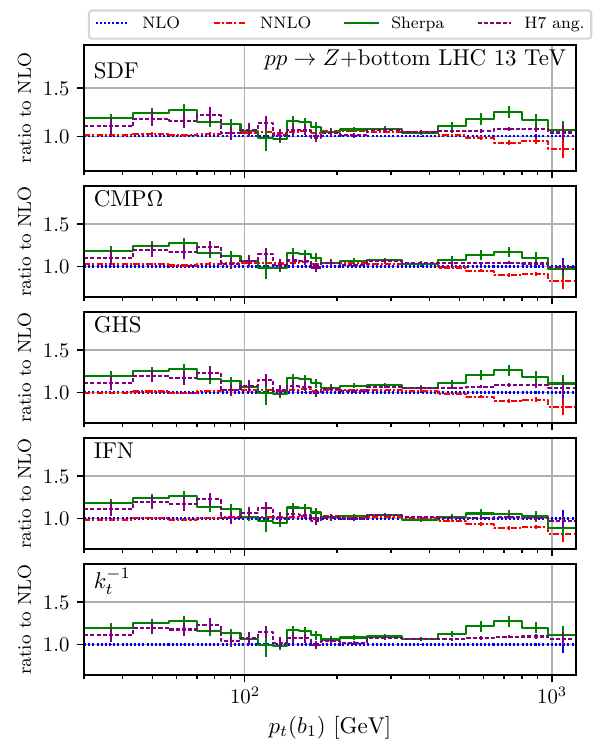}%
    \includegraphics[width=0.5\linewidth,page=3]{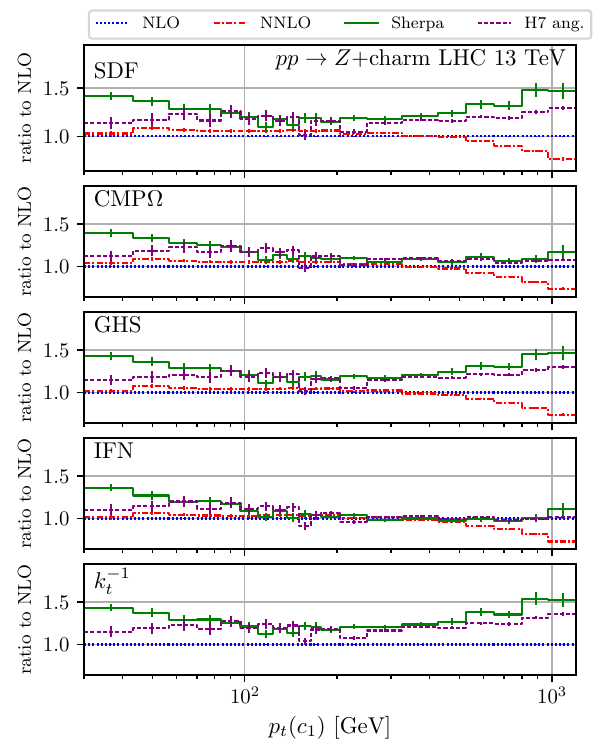}
    \caption{Comparison between fixed-order and NLO+PS predictions for $pp \to \Z + b/c$ in central kinematics. Different jet algorithms are compared for the transverse momentum of the leading $b$/$c$ jet in the bulk and tail of the phase space.}
    \label{fig:summary_ppzj_fo_vs_nlops_bottom}
\end{figure}

This conclusion can be confirmed by comparing the showered and hadronised results with the fixed-order predictions for $Z+\text{jet}$ from \cref{app:Zjnnlonlops}. For this purpose, we plot the ratio of the various predictions to the fixed-order NLO QCD results\footnote{The motivation is that all NLO QCD results, regardless of the algorithm, are very similar for the observables considered here (see, for example, \cref{fig:nnlo_ppzj_algos_pt} in \cref{app:Zjnnlonlops}).} in \cref{fig:summary_ppzj_fo_vs_nlops_bottom}. Several observations can be made. First, we see a visible parton-shower effect in the threshold region, which can be traced back to the matching and shower scale dependence as the same differences occur for nonflavoured jets (compare to the discussion in \cref{app:Zjnnlonlops}). Focusing on the $c$-label case, we find different behaviour of the various algorithms when looking into the tail of the distribution. SDF and GHS start to have more significant corrections from the shower at higher \pt, comparable to the effects observed for nonflavoured anti-\kt jets. The IFN algorithm receives minor corrections that approach unity at a very high \pt, while the CMP algorithm is somewhat between these extremes. A similar behaviour, though smaller in scale, can be observed for the $b$-label case (the \sherpa predictions' statistical uncertainty is significant here, but for \herwigv{7}, this observation is statistically significant).

The comparison of fixed-order with showered cross sections in the high-\pt limit indicates the infrared sensitivity of the observable definition provided by a given algorithm. The smaller the corrections, the less sensitive the algorithm is to infrared physics. We observe that IFN delivers the most robust definition of infrared-insensitive jet flavour from that measure, followed by CMP, GHS and SDF.

\begin{figure}
    \centering
    \includegraphics[width=0.49\linewidth]{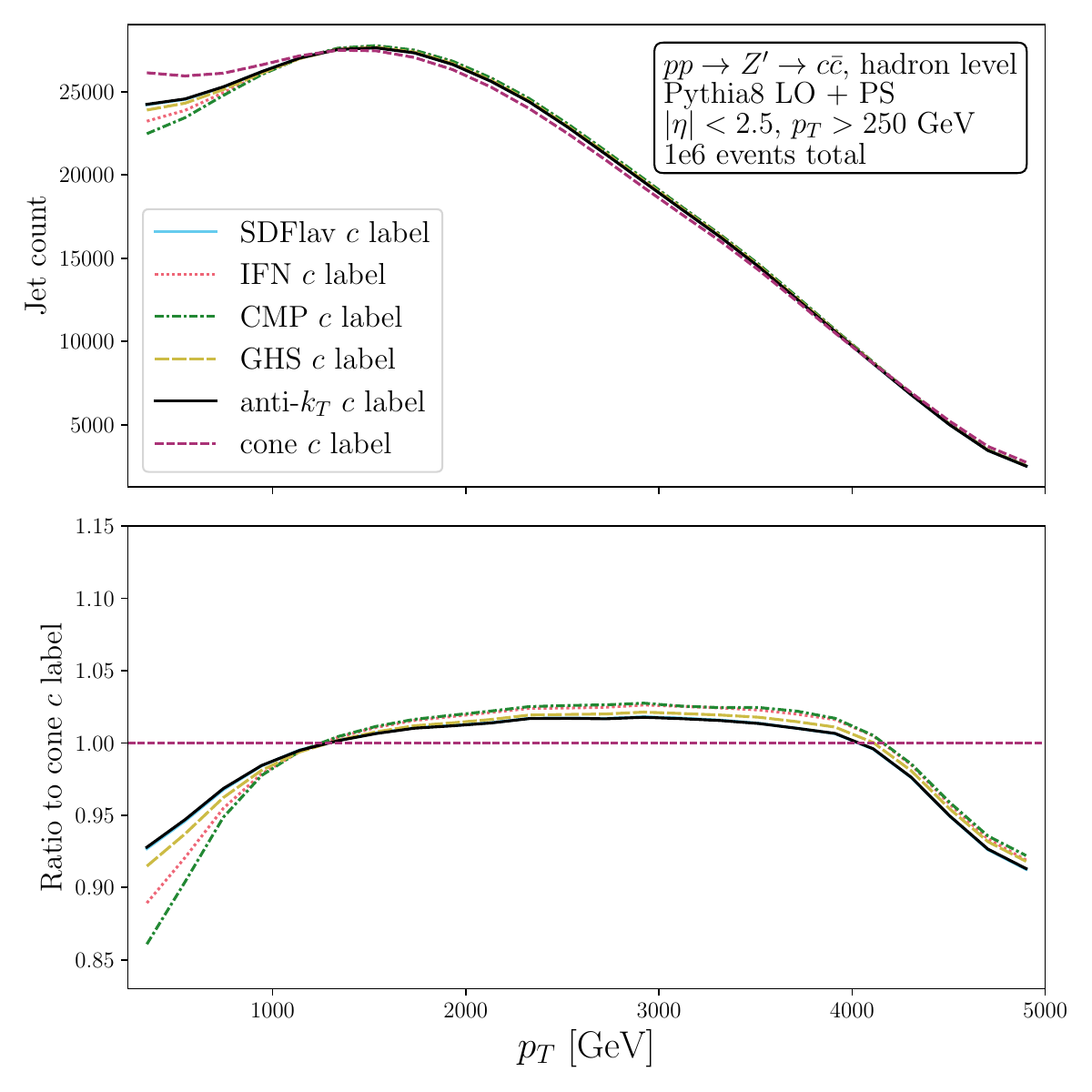}
    \includegraphics[width=0.49\linewidth]{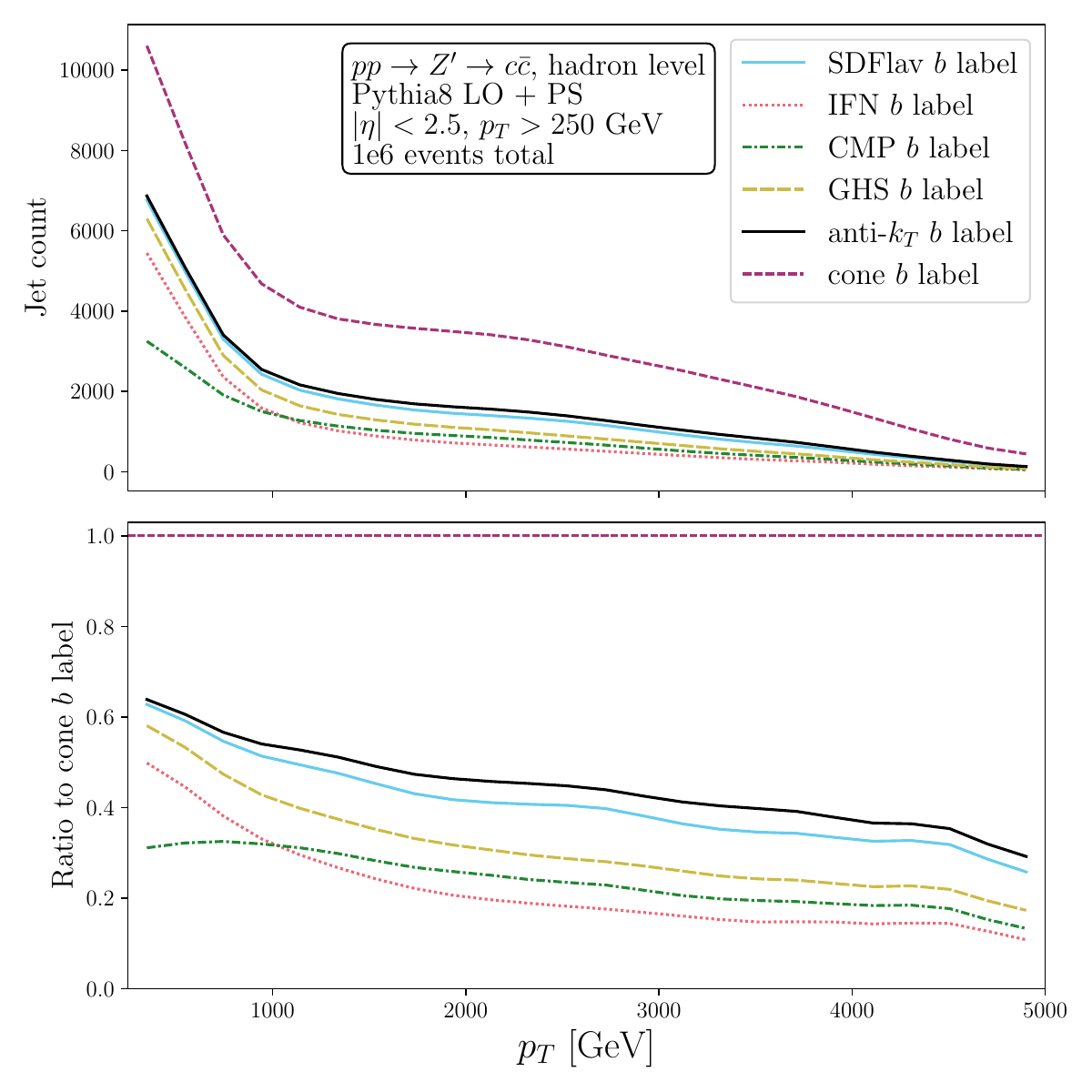} 
    
    \caption{The \pt spectra of $c$ and $b$ jets found in $pp \rightarrow Z' \rightarrow c \bar{c}$ events, simulated at LO and showered with \pythiav{8}. All labelling schemes and algorithms shown use the  ``net-flavour''  recombination scheme, except for the cone label, which uses the ``any-flavour'' scheme.}
    \label{fig:summary_ATLAS-FTAG-comparison}
\end{figure}

Finally, it is also interesting to investigate the impact of the different flavour labelling schemes in the development of multivariate flavour tagging algorithms. In particular, \cref{app:atlas_ftag} reports flavour labelling studies on simulated $pp\to Z'\to c\bar{c}$ events, generated with \pythiav{8} LO+PS with $m_{Z'}=\SI{4}{\TeV}$, because such a MC sample is used by the ATLAS Collaboration for the training of multivariate flavour tagging classifiers at the jet-\pt scale of $\SI{1}{\TeV}$. The sample is used, together with other MC simulations for the other flavours and phase space regions, in order to create a training dataset with well-separated $c$-jet signatures in the detector. However, \cref{fig:summary_ATLAS-FTAG-comparison} shows that such an assumption is not robust: a jet produced with \pt far from the bulk of the events is likely to be polluted by gluon splitting into quark flavours different from the $c$ quarks present in the hard-scattering matrix element. The effect is even more striking considering the right part of \cref{fig:summary_ATLAS-FTAG-comparison}, where the experimental flavour labelling algorithms find a much larger contamination of $b$ jets in a sample that has no $b$ quarks from the hard scattering with respect to the IRC-safe algorithms. The pollution of the training sample of the multivariate flavour tagging classifiers is considered as one of the possible reason for a sizeable systematic uncertainty due to model dependence in the efficiency calibration of the flavour tagging algorithm~\cite{ATLAS:2022qxm,ATLAS:2019bwq,CMS:2017wtu}. In this respect, the use of IRC-safe algorithms in the definition of new training samples would surely improve this bias.

\subsection{Impact of hadron reconstruction  on flavoured jet definitions}

Studies presented so far have focused on the impact of an IRC-safe flavour definition on inclusive jet observables. However, we realized that additional effects arise when considering flavour in comparisons with experimental measurements handling the treatment of heavy-flavour hadrons in different ways. We consider two examples: the inclusion of the decay products of $b$ and $c$ hadrons in inclusive jet observables, as well as an example of an analysis that tackles jet flavour from a different perspective, namely using the WTA flavour algorithm.

\begin{figure}
    \centering
    \includegraphics[width=0.49\linewidth]{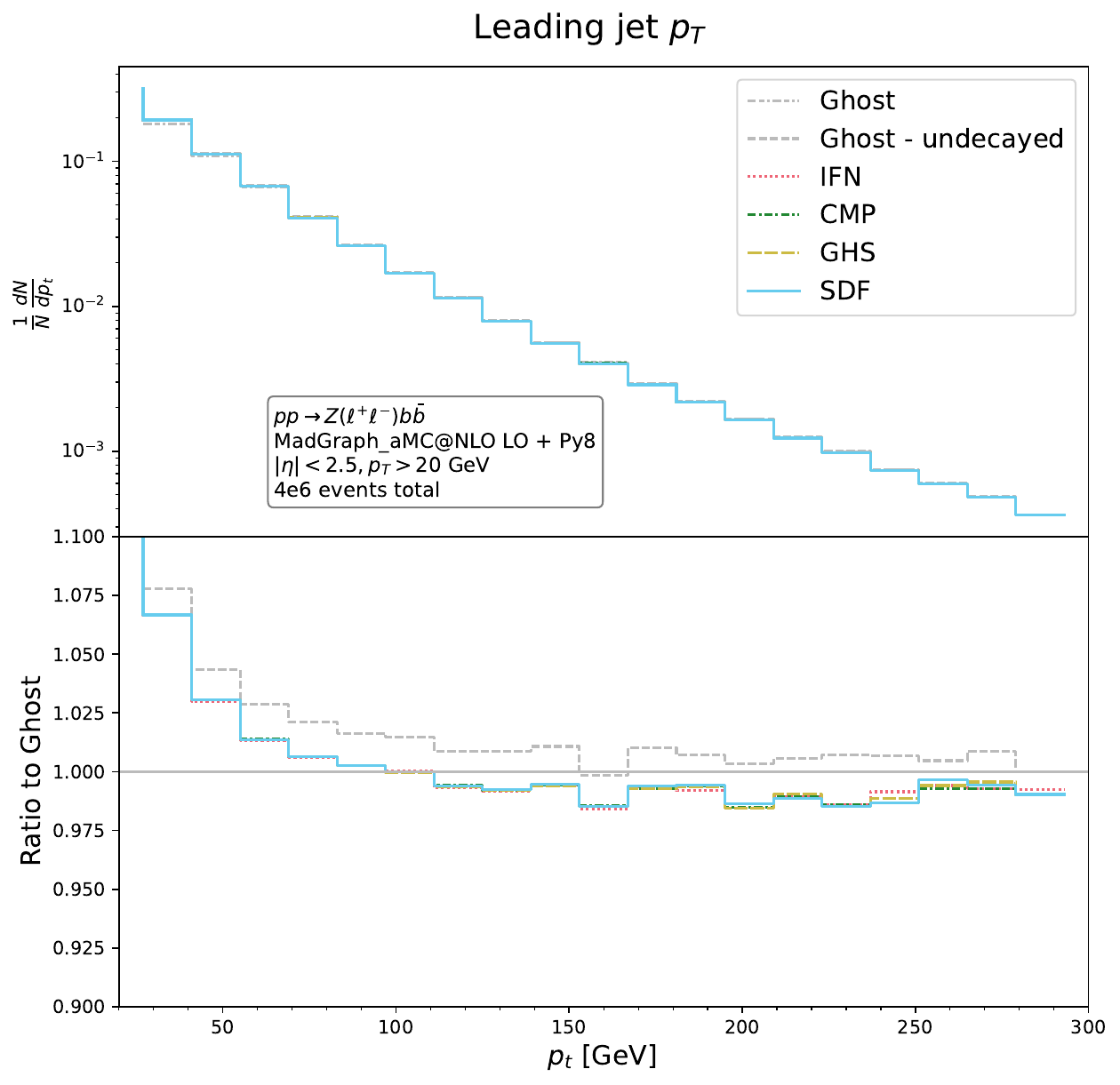}
    \includegraphics[width=0.49\linewidth]{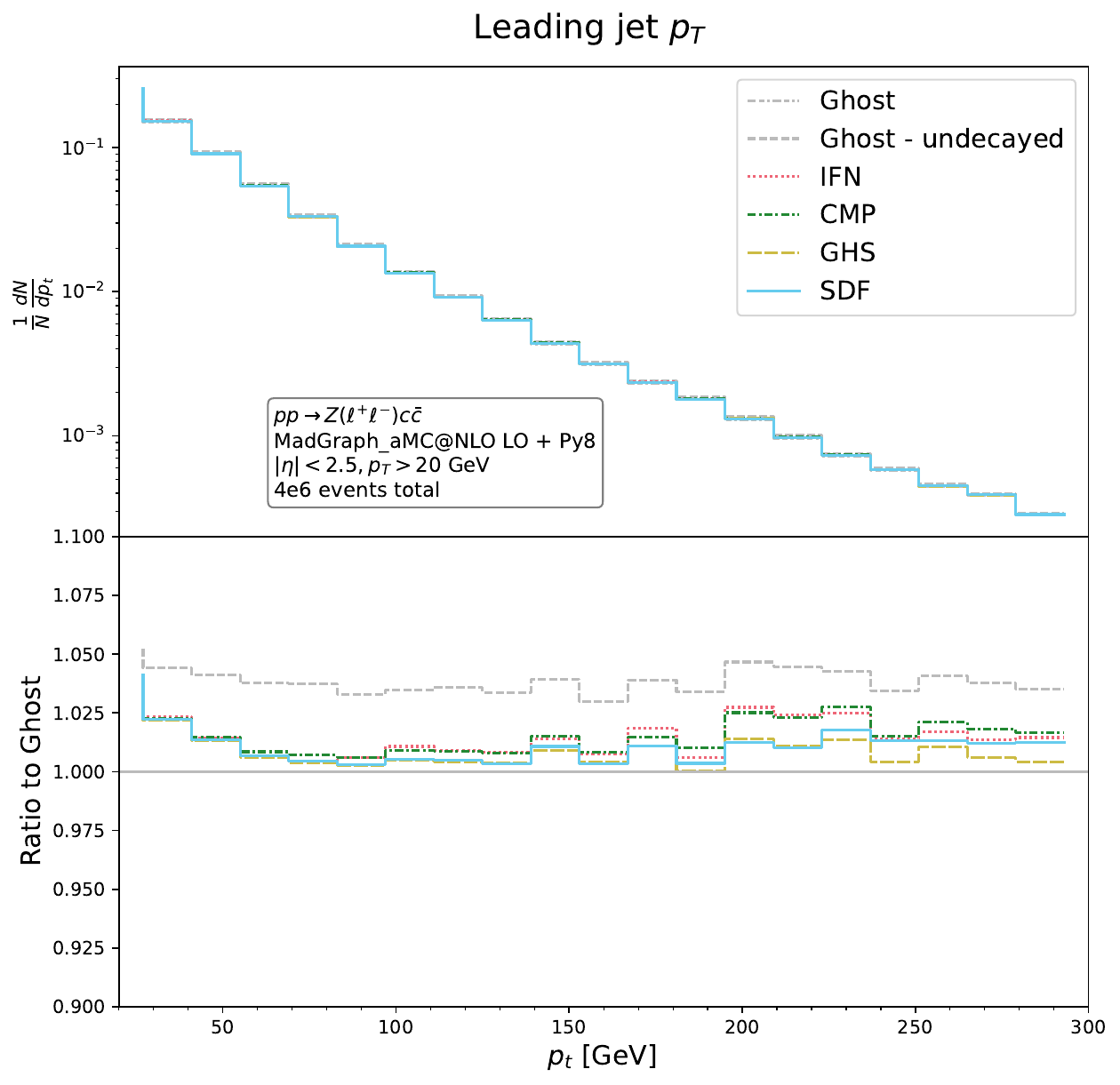} 
    \caption{The \pt distribution of leading $b$-jets (left) and leading $c$-jet (right) identified with the ghost and flavoured jet algorithms labels using final state particles.
    Ratios are compared to the distribution for jets obtained with ghost labelling.}
    \label{fig:decayed_cmp}
\end{figure}

\paragraph{Including heavy flavour decays.} For what concerns the first, a comparison of the different flavour labelling algorithms has been performed using  $pp \to \Z + bb$ and $pp \to \Z + cc$ events and selected within a fiducial phase with a di-lepton mass cut and jets of $\pt> \SI{20}{\GeV}$, similar to ``standard'' measurements of a \Z boson produced in association with flavoured jets in the ATLAS or CMS experiments~\cite{ATLAS:2020juj,CMS:2021pcj,ATLAS:2024tnr}.
Detailed results are reported in \cref{app:atlas}. The comparison, visible in \cref{fig:decayed_cmp}, shows a sizeable discrepancy in the low jet \pt spectrum when heavy flavour jets are selected in the analysis phase space using the experimental labelling scheme---namely ghost labelling---versus the new labelling algorithms---namely SDF, CMP, GHS, and IFN.
The simple, yet initially unexpected, reason behind this comes from the need to define experimental flavour labelling schemes using final state particles, i.e., including the hadrons originating from heavy flavour decay chains, while the new labelling algorithms need to rely on undecayed heavy-flavour hadrons in order to be correctly evaluated. This introduces a difference in the kinematics of the clustered jet, stronger for $b$ jets than for $c$ jets, according to the larger decay pattern multiplicity of the first. The same effect was already reported in a recent ATLAS \Z + heavy-flavour measurement~\cite{ATLAS:2024tnr}, signalling the need of a second unfolding step to pass from phase space definition using final state particles to a phase space defined using undecayed heavy hadrons.
In order to investigate smaller differences between the various labelling schemes, the analysis has been repeated switching off the heavy-flavour decay in the events, as visible in \cref{fig:decayed_cmp} with the label ``ghost-undecayed''. Percent-level differences in the selection of $b$ jets appear over the full jet \pt range; while for $c$ jets the effect starts at the jet \pt threshold, of $\SI{20}{\GeV}$, and it grows to about 5\% at higher \pt.
This is consistent with an increasing $g\to q\bar{q}$ contribution at higher jet \pt, with two heavy-flavour quarks (produced by the hard ME or the parton shower) winding up in the same jet: in such cases all the IRC-safe labelling algorithm---with tiny differences between all of them---are classifying the jets as flavourless, while the current experimental algorithms assign a flavour to the jets.

\paragraph{Exclusive $b$ tagging with WTA.}  As expected, more significant differences between the currently available experimental labelling algorithms and the IRC safe ones arise in special region of the phase space where soft or collinear emission is more abundant. This prompted further investigation using simulated samples used in experiments and where these effects may be enhanced. 
Other LO+PS studies were carried out in approximate \lhcb kinematics, as documented in \cref{app:lhcb}. 
The \lhcb experiment has a forward detector geometry which is currently unique at the LHC, covering roughly the pseudorapidity range $2 < \eta < 5$~\cite{LHCb-DP-2022-002}.
Because of its access to forward kinematics, flavoured hadrons are produced with a larger longitudinal boost than at mid-rapidity, providing a longer distance between the interaction point of colliding hadrons and the subsequent decay vertex of heavy-flavour hadrons.
This makes it easier to separate and identify boosted decays of single hadrons in LHCb kinematics.
These benefits, coupled with the ability to identify charged hadron identities across a vast range of momenta~\cite{LHCb-DP-2012-003}, enables \lhcb to exclusively reconstruct specific heavy-flavour hadrons at detector level before jet reconstruction.
In this case, the WTA flavour algorithm (see \cref{subsec:WTA_flav}) offers an advantage over other algorithms, as knowledge of all flavoured particles in the event is not required; it is only important whether or not the reconstructed flavoured hadron is collinear with the WTA axis.

Although WTA flavour tagging (as discussed above) follows a fundamentally different approach to the flavour tagging problem, similar behaviour is observed between WTA flavour and other new flavoured jet algorithms.
These results confirm the dominance of flavour recombination effects over different algorithms and show that while WTA flavour tagging does not use any modified flavour recombination scheme, it similarly tends to reject gluon splitting at a rate comparable to other algorithms, in the case of $b$-jets.

\begin{figure}
    \centering
    \includegraphics[width=0.475\linewidth]{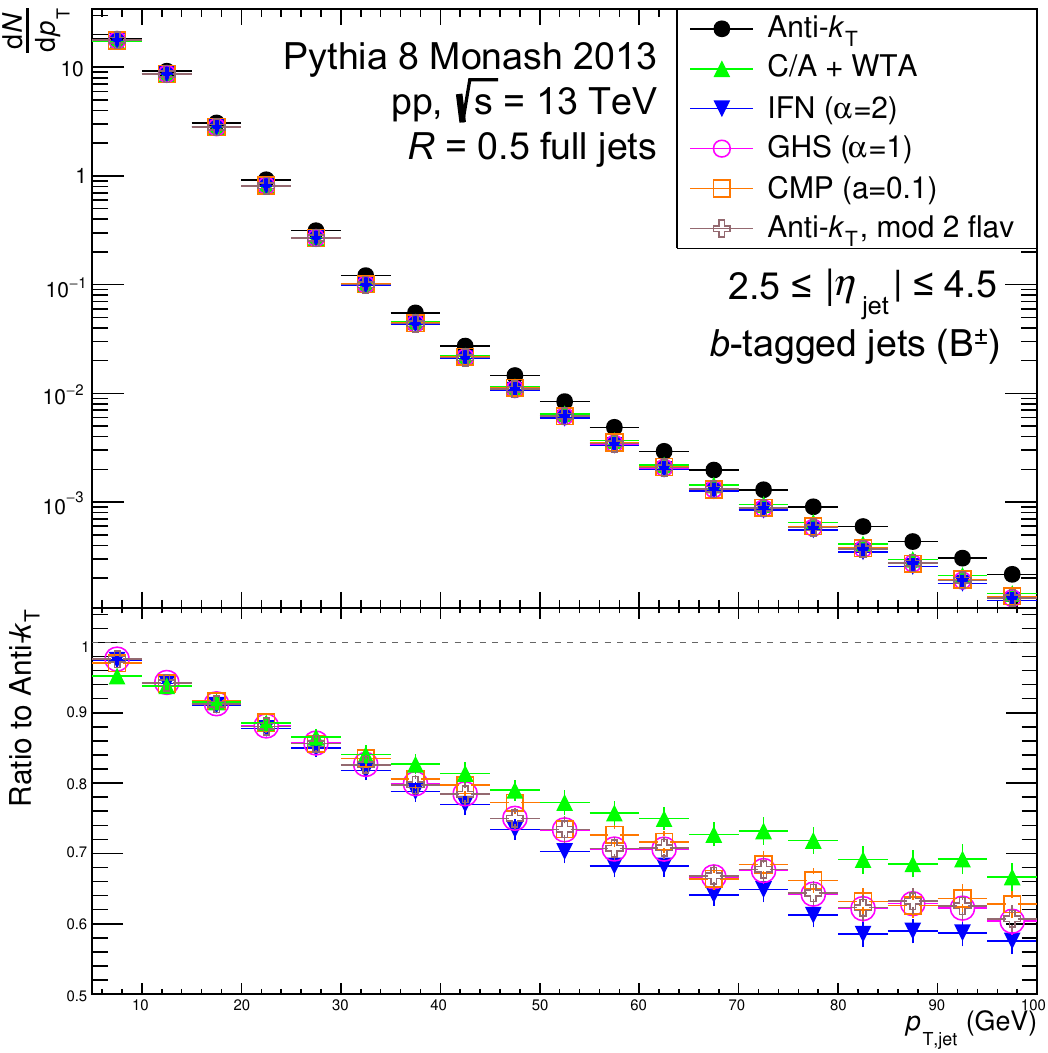}
    \includegraphics[width=0.48\linewidth]{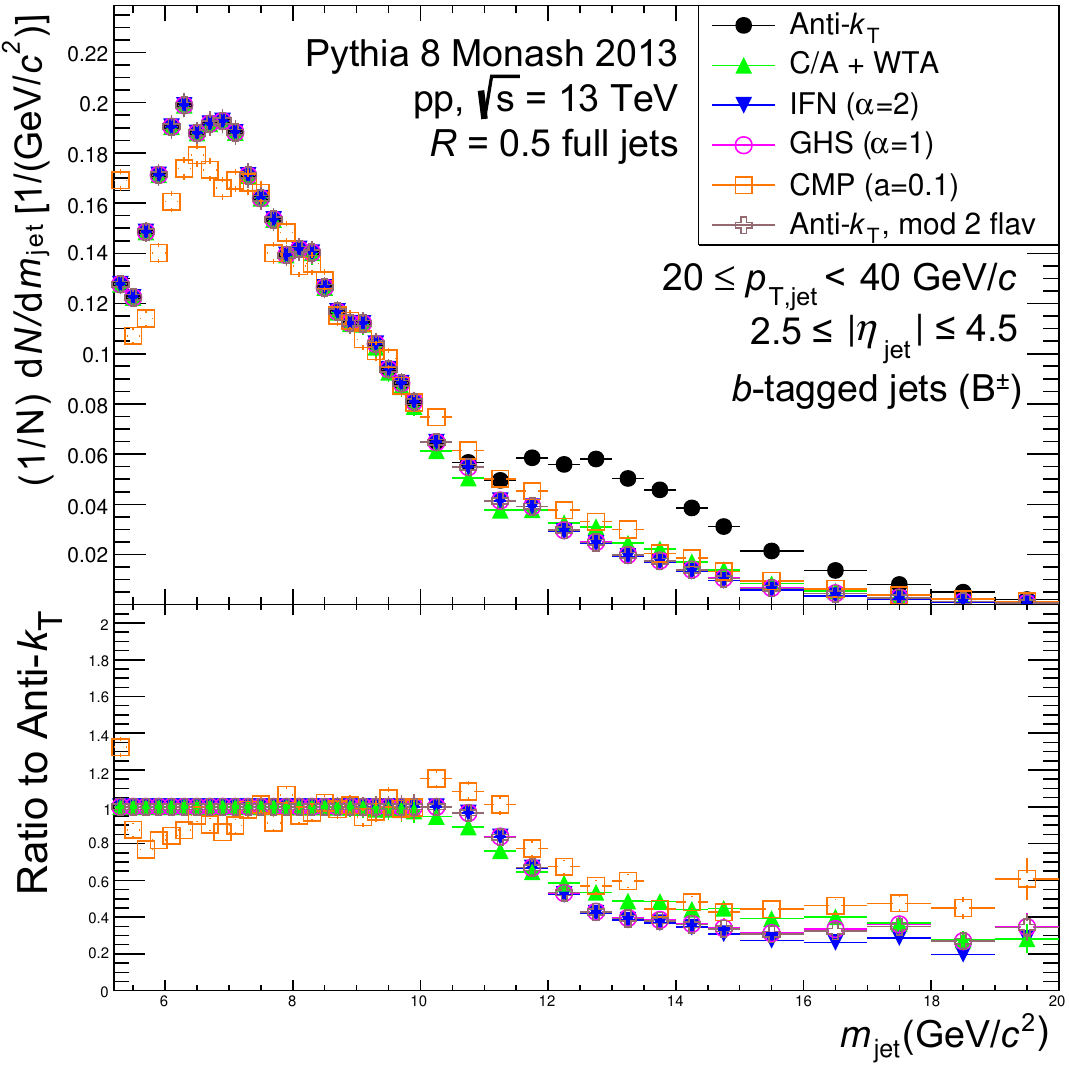}
    \caption{LO+PS predictions for an exclusive sample of $b$ jets containing a fully-reconstructed \Bpm-meson, comparing flavour defined using ``any-flavour'' anti-\kt (black) with the new algorithms and flavour recombination schemes. \emph{Left:} $b$-jet cross section. \emph{Right:} $b$-jet invariant mass. For more details, see \cref{app:lhcb}.}
    \label{fig:LOPS_LHCb_bjets}
\end{figure}

\Cref{fig:LOPS_LHCb_bjets} shows the $b$-tagged jet cross section for jets containing an explicitly reconstructed \Bpm-meson for the different tagging algorithms as compared to an ``any flavour'' anti-\kt baseline. The tagging fraction is observed to fall as a function of \pTjet for all algorithms, which is expected due to the growing gluon splitting contribution at higher \pTjet. The same figure also shows the $b$-jet invariant mass \mjet distribution for the same sample of jets. A secondary peak is observed in the tail of the distribution at the kinematic limit of $\mjet = 2 \cdot m_{\Bpm}$, where a significant proportion of jets with $b$ flavour originating from gluon splitting populate the spectrum. All new algorithms are observed to suppress this secondary peak with little-to-no modification of the primary mass peak, with the exception of CMP, which has small modifications at low \mjet due to the slightly modified kinematics of the clustering redefinition. 

\FloatBarrier
\section{Concluding remarks}\label{sec:conclusions}

In this paper, we have performed a first comprehensive phenomenological study that focuses on the challenges and opportunities presented by new definitions of jet flavour in the context of high-precision collider physics. 
After a short introduction, we have reviewed the concepts of flavour labelling and flavour tagging in the context of a typical workflow of an experimental analysis in particle physics, highlighting tensions with an IRC-safe definition of jet flavour from a theory viewpoint.

The new generation of algorithms---SoftDrop Flavour (SDF), Flavoured anti-\kt (CMP), Flavour Dressing (GHS), and Interleaved Flavour Neutralisation (IFN)---represent a significant advancement in jet flavour labelling, offering IRC-safe definitions applicable at fixed order and throughout the simulation chain. Their key role is to define jet flavour labels that are resilient against the contamination from collinear splittings and soft emissions, particularly $g \to b \bar b$, which traditionally hampers our ability to compare high-precision calculations to flavour-tagged measurements.
In this context, we have stressed the importance of adopting a flavour recombination scheme such as ``net-flavour'' or ``mod2-flavour'', rather than ``any-flavour'', which is the current standard.

We have performed several studies for LHC phenomenology, which are detailed in the appendices, exploiting theory predictions computed with different tools and levels of precision. In particular, we have considered fixed-order predictions at NNLO, Monte Carlo parton showers, both at hadron and parton level, as well as their matching to NLO matrix elements. 
In the bulk of the phase space, all new algorithms give compatible results and most differences with respect to current experimental strategies can be traced back to the use of different flavour recombination schemes.
However, as we move to regions of phase space where the hierarchy of scales is more pronounced, e.g., smaller quark masses or higher transverse momenta, soft and collinear emissions are abundant, and the algorithms start to differ from each other.
Among these, IFN stands out for its robustness against soft and collinear effects, delivering distributions that change little through the simulation chain. 

In this study, we have focused on the role played by the novel algorithms on flavour labelling, which is performed on Monte Carlo pseudodata (truth level), for which full flavour information is available. However, the new flavour algorithms may impact other components of experimental workflows.
We can identify three main aspects:
\begin{itemize}
  \item definition of appropriate samples for background estimate and/or signal yield evaluation in new physics searches or measurements;
  \item accurate estimate of detector efficiencies and selection phase space used to unfold data events to a fiducial phase space of a measurement with flavoured final states;
  \item optimisation and training of flavour tagging algorithms to be used in flavour identification in collision data events.
\end{itemize}
Indeed, using new flavour definitions as truth-level labels could improve the fidelity of unfolding procedures, providing a more stable baseline for correcting detector effects and modelling uncertainties. Since unfolding is typically performed separately for each analysis, this could be the easiest first step toward adoption.
A critical open question is whether a mismatch between the labelling schemes used in unfolding and flavour tagging, which is performed using actual experimental inputs, where flavour knowledge is limited, significantly degrades analysis performance. While definitive answers require more study, the large effects observed when switching from any-flavour to mod2 or net-flavour recombination schemes---despite potential efficiency losses---strongly motivate further experimental investigations.

Despite their formal appeal, these algorithms face practical limitations due to detector resolution and reconstruction capabilities. For example, their ability to identify the correct initiating flavour may be hindered in the presence of $B$- and $D$-hadron decays. 
However, we can envision the use of IRC-safe flavour labels to train flavour taggers. Because these are typically based on rather complex neural networks, making a switch in labelling schemes to re-train them is likely more laborious and deserves more investigation. Still, these algorithms offer a path toward defining more accurate training targets and mitigating model-induced bias due to poorly modelled infrared physics. 
Notably, training mismatches between \pythia and \herwig are already observed and may originate from such sensitivity~\cite{ATLAS:2022qxm,CMS:2017wtu,LHCb-PAPER-2016-039}.
Furthermore, with improved particle identification capabilities in LHCb and ALICE, and the advent of new vertex reconstruction algorithms like the GN2 tagger in ATLAS~\cite{ATLAS:2025dkv} or CMS’s ML-based $b$-hadron momentum estimators~\cite{CMS:2024gds}, some of these algorithms could soon be implemented experimentally for select final states.

Our findings suggest the best approach to a successful implementation may be gradual: using these algorithms first at truth level, to improve Monte Carlo-based corrections, and progressively moving toward direct application to reconstructed data as experimental techniques mature. Several open directions remain. Firstly, deeper studies are needed on the modelling of gluon splitting into $b \bar b$ pairs, possibly through double $b$-tagged jets. Secondly, validations of the algorithms’ performance under different hadronisation and parton shower models (e.g., string vs.~cluster models in \herwig and \sherpa) must continue. These variations are typically small, but still non-negligible in some regimes, especially for charm.

\acknowledgments
 
First and foremost, we are grateful to the organisers of Les Houches 2023 PhysTeV workshop and of the subsequent IPPP `Flavoured Jets at the LHC' workshop.
A.B.\ and A.S.\ thank the CERN Theoretical Physics Department for
hospitality while part of this research was being carried out.
The work of S.M.\ and of F.S.\ is supported is supported by the Italian Ministry of University and Research (MUR) grant PRIN 2022SNA23K funded by the European Union -- Next Generation EU, Mission 4, Component 2, CUP D53D23002880006 and by ICSC Spoke 2 under grant BOODINI, CUP I57G21000110007.
L.S.\ is supported by an ARC Discovery Early Career Researcher Award (project number DE230100867). 
The work of A.S.\ was supported by the OpenMAPP project via National
Science Centre, Poland under CHIST-ERA programme (grant No.\ NCN
2022/04/Y/ST2/00186)
and by the Priority Research Area Digiworld under the program
`Excellence Initiative -- Research University' at the Jagiellonian
University in Kraków.
J.W.\ gratefully acknowledges the support of the Polish National Science Centre
grant 2019/34/E/ST2/00457, and
the Polish high-performance computing infrastructure
PLGrid (HPC Centre: ACK Cyfronet AGH) for providing computing facilities
and support within computational grants PLG/2024/017081 and PLG/2025/018389.
A.Hinzmann\ acknowledges support from DESY (Hamburg, Germany), a member
of the Helmholtz Association HGF, and support by the Deutsche Forschungsgemeinschaft
(DFG, German Research Foundation) under Germany’s Excellence Strategy – EXC 2121
"Quantum Universe" – 390833306.
G.P.S.\ is supported by a Royal Society Research Professorship
(RP$\backslash$R1$\backslash$23100) and by the Science and Technology Facilities
Council (STFC, grant ST/X000761/1).
D.R.\ is supported by the European Union under the HORIZON program in Marie Sk{\l}odowska-Curie project No. 101153541.
D.R.\ was further supported by the STFC under grant agreement ST/P006744/1
during the initial stages of this work.
R.P.\ acknowledges the support from the Cambridge Service for Data Driven Discovery (CSD3), part of which is operated by the University of Cambridge Research Computing on behalf of the STFC DiRAC HPC Facility (www.dirac.ac.uk). The DiRAC component of CSD3 was supported by STFC grants ST/P002307/1, ST/R002452/1 and ST/R00689X/1. R.P.\ acknowledges further support from the Leverhulme Trust and the Isaac Newton Trust in the initial stages of the project.
S.C.\ acknowledges the support
of the Swiss National Supercomputing Centre (CSCS) under project ID ETH5f.
R.G.\ is supported by an STFC studentship (project ST/X508664/1), the Clarendon Scholarship, and the Wolfson Harrison UK Research Council Physics Scholarship.
R.X.\ acknowledges support from the Homer A.\ and Donna J.\ Neal Endowment Fund of the University of Michigan.

\appendix
\settocdepth{subsection}

\section{Flavoured jet algorithms in \texorpdfstring{\fastjet}{FastJet}}
\label{app:alg-setup}

The following code snippets allow the user to construct the resulting jets
(starting from a collection of input particles \texttt{vector<PseudoJet> event}, set up with flavour information as explained in~\cref{sec:fj-implementations}). Note that one can consider all possible flavour at once, or treat flavours separately, one at a time (e.g., $b$ and $c$).\footnote{For instance in the case of the IFN algorithm this choice can modify the neutralisation sequence. Also for CMP the choice of which flavours to consider has an impact on the clustering sequence and the general recommendation is to only consider one flavour at a time.} To consider a single flavour, one can reset all flavours but the one of interest; for example, when setting up flavour from the input particles, to consider \emph{only} $b$ flavour:

\begin{verbatim}
    int filter_flav = 5;
    auto flav_info = new FlavInfo(pdg_id);
    flav_info->reset_all_but_flav(filter_flav);
\end{verbatim}

In all cases below we set the jet radius to $R=0.4$, and we assume that the code has
access to the \texttt{fastjet} and \texttt{fastjet::contrib} namespaces.
The algorithms make use (directly or indirectly) of the \texttt{FlavRecombiner} class, which implements the flavour recombination scheme (either ``mod2'' or ``net'') during the clustering of the jets. Each algorithm returns a list of jets whose flavour can be accessed, as described at the end of \cref{sec:fj-implementations}, with the \texttt{FlavHistory} class. The return value inherits the choice of the recombination scheme. The structure of the code, however, allows for a different criterion at the stage of the label assignment. In particular, it is possible to use ``net'' flavour recombination during clustering and assign the label based on ``mod2'' flavour:\footnote{If ``mod2'' is used during clustering, ``net'' and ``mod2'' yield the same result during assigning.}
\begin{verbatim}
   bool is_blabelled_mod2 = abs(FlavHistory::current_flavour_of(j)[5])%2==1;
\end{verbatim}
The following examples demonstrate the basic usage for each algorithm.

\bigskip
\noindent
\textbf{IFN}:
\begin{verbatim}
    // jet radius parameter
    double R = 0.4;
    // create a base jetdef (anti-kt or C/A)
    JetDefinition base_jet_def(antikt_algorithm, R);
    // flavour recombiner (default is net flavour)
    FlavRecombiner flav_recombiner;
    base_jet_def.set_recombiner(&flav_recombiner);
    double alpha = 2.0;
    // IFN plugin (sets the parameter omega to 3-alpha by default)
    JetDefinition jet_def(new IFNPlugin(base_jet_def, alpha));
    // get the IFN jets
    vector<PseudoJet> IFN_jets = jet_def(event);
\end{verbatim}
\textbf{CMP}:
\begin{verbatim}
    // jet radius parameter
    double R = 0.4;
    // CMP parameters
    double a = 0.1;
    CorrectionType CMP_corr = CorrectionType::SqrtCoshyCosPhiArgument_a2;
    ClusteringType CMP_clust = ClusteringType::DynamicKtMax;
    // CMP plugin
    JetDefinition jet_def(new CMPPlugin(R, a, CMP_corr, CMP_clust));
    // flavour recombiner (default is net flavour)
    FlavRecombiner flav_recombiner;
    jet_def.set_recombiner(&flav_recombiner);
    // get the CMP jets
    vector<PseudoJet> CMP_jets = jet_def(event);
\end{verbatim}
\textbf{GHS}:
\begin{verbatim}
    // jet radius parameter
    double R = 0.4;
    // anti-kt jet definition
    JetDefinition base_jet_def(antikt_algorithm, R);
    // flavour recombiner (default is net flavour)
    FlavRecombiner flav_recombiner;
    base_jet_def.set_recombiner(&flav_recombiner);
    // run the jet clustering with the base jet definition
    vector<PseudoJet> base_jets = base_jet_def(event);

    // run the flavour dressing (requires some minimal pt cut)
    double ptcut = 20.0;
    double alpha = 1.0;
    double omega = 2.0;
    // get the GHS jets
    vector<PseudoJet> GHS_jets = run_GHS(base_jets, ptcut, alpha,
        omega, flav_recombiner);
\end{verbatim}
\textbf{SDF}:
\begin{verbatim}
    // jet radius parameter
    double R = 0.4;
    // anti-kt jet definition
    JetDefinition base_jet_def(antikt_algorithm, R);
    // flavour recombiner (default is net flavour)
    FlavRecombiner flav_recombiner;
    base_jet_def.set_recombiner(&flav_recombiner);
    
    // SoftDrop parameters
    const double beta = 2.0;
    const double zcut = 0.1;
    // initialise the SoftDrop flavour algorithm
    SDFlavourCalc sdFlavCalc(beta, zcut, R);
    
    // run the jet clustering with the base jet definition
    vector<PseudoJet> SDF_jets = base_jet_def(event);
    // run the SDFlav algorithm on the base jets
    sdFlavCalc(SDF_jets);
\end{verbatim}

\sectionAuthor{NNLO fixed-order \texorpdfstring{$p p \to WH(b\bar{b})$}{p p -> W H(bb~)}}{Arnd Behring}
\label{app:HW}

We adapt the calculation from Refs.~\cite{Caola:2017xuq,Behring:2020uzq} for the process
\begin{align}
  p p \to W^+(\to e^+ \nu_e) H(\to b\bar{b})
\end{align}
to study differences between the flavoured jet algorithms. The calculation is performed in the nested-soft collinear subtraction scheme~\cite{Caola:2017dug,Caola:2019nzf,Caola:2019pfz,Asteriadis:2019dte}, which is closely related to the sector-improved residue subtraction scheme~\cite{Czakon:2010td, Czakon:2011ve, Czakon:2014oma, Czakon:2019tmo}. We treat the decays of the Higgs and \W-boson in the narrow-width approximation. The decay $W^+ \to e^+ \nu_e$ is treated at LO, but both the production subprocess $p p \to H W^+$ and the decay subprocess $H \to b \bar{b}$ are calculated through NNLO in QCD. In this study we treat the $b$ quark as massless.

\subsection{Setup of the calculation}
The calculation closely follows the setup from Ref.~\cite{Behring:2020uzq}, but we collect a few details below for the convenience of the reader.
\paragraph{Input parameters}
For the input parameters of the calculation we use the values
\begin{align}
  m_H &= \SI{125}{\giga\eV}
  \,, &
  m_W &= \SI{80.399}{\giga\eV}
  \,, &
  \Gamma_W &= \SI{2.1054}{\giga\eV}
  \,, \\
  m_t^\text{pole} &= \SI{173.2}{\giga\eV}
  \,, &
  m_b^\text{pole} &= \SI{4.78}{\giga\eV}
  \,, &
  G_F &= \SI{1.16639e-5}{\giga\eV^{-2}}
  \,, \\
  \sin^2 \theta_W &= \num{0.2226459}
  \,.
\end{align}
The value of the $b$-quark mass $m_b^\text{pole}$ is only used to calculate the $b$-quark Yukawa coupling. For details about this and about the treatment of the narrow width approximation we refer to Ref.~\cite{Behring:2020uzq}. We use PDFs in the $n_f=5$ flavour scheme (\texttt{NNPDF31\_nnlo\_as\_0118} \cite{NNPDF:2017mvq}), as provided through the \lhapdf library~\cite{Buckley:2014ana}. The running of the strong coupling $\alpha_s$ is treated in the $\overline{\text{MS}}$ scheme with $n_f=5$ active flavours.

\paragraph{Jet algorithms}
We use the four new IRC-safe flavoured jet algorithms (IFN, CMP$\Omega$, GHS, and SDF) via their implementations in \fastjet~\cite{Cacciari:2011ma}. For reference, we also include the flavour-\kt algorithm. In all algorithms we use the net flavour recombination scheme and a jet radius of $R=0.4$. The remaining parameters that are specific to each jet algorithm are chosen as listed in the baseline setup in \cref{sec:findings}, except for the minimal jet transverse momentum $p_{t,\text{cut}}$ in the GHS algorithm, where we use $p_{t,\text{cut}} = \SI{25}{\GeV}$ to match the fiducial cuts, and we use $\beta = 2$ in the SDF algorithm.

\paragraph{Fiducial cuts}
We consider the following cuts on jets. In order to be counted as a $b$ jet, the jet in question has to have a transverse momentum of at least $\pt > \SI{25}{\giga\electronvolt}$ and a rapidity of at most $|\eta| < 2.5$. Similarly, flavourless jets also have to have at least $\pt > \SI{25}{\giga\electronvolt}$ of transverse momentum. We require at least two $b$-labelled jets.
The positron from the decay of the \W boson has to fulfil $\pt > \SI{15}{\giga\electronvolt}$ and $|\eta| < 2.5$.

For certain observables, we reconstruct a candidate for the Higgs boson by selecting among all $b$-labelled jets the pair whose invariant mass is closest to the invariant mass of the Higgs boson. The sum of momenta of these two $b$ jets is then treated as the reconstructed momentum of the Higgs boson, from which further observables (transverse momentum, rapidity, invariant mass, etc.) can be derived. We will label those observables with a subscript $H(b\bar{b})$.

\subsection{Results}
\begin{figure}
  \centering
  \includegraphics[width=0.49\linewidth]{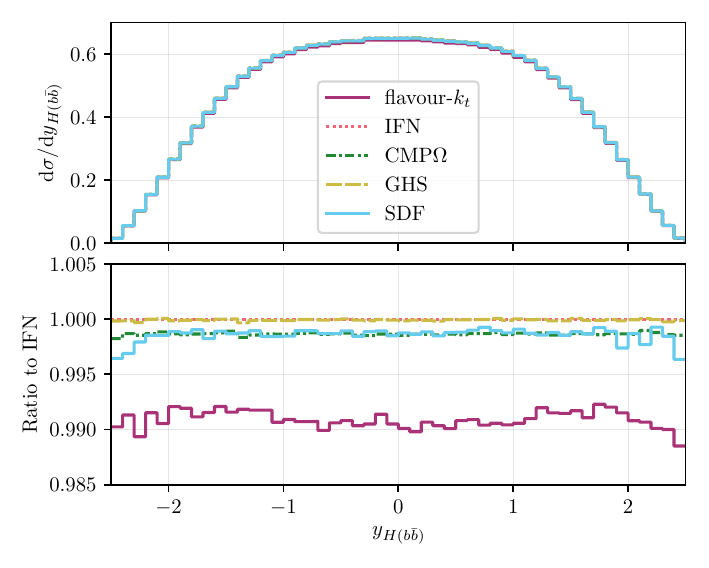}
  \includegraphics[width=0.49\linewidth]{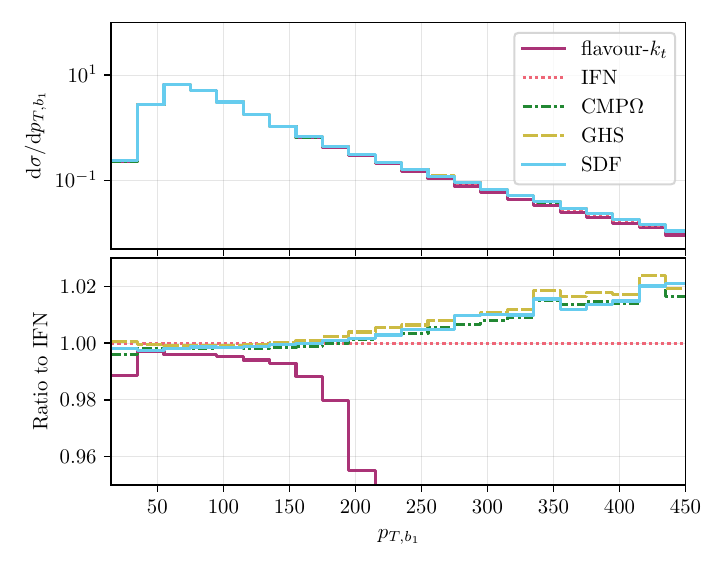}
  \caption{Fixed-order NNLO QCD predictions for the rapidity of the reconstructed Higgs boson in $p p \to W H(b\bar{b})$ (left) and for the transverse momentum of the leading $b$ jet for different jet algorithms. The distributions for different jet algorithms are calculated on the same events. The lower panes show the ratio of the distributions for the respective algorithms to that for IFN.}
  \label{fig:whbb0-yH-pTb1}
\end{figure}
For many of the observables one can define for this process, the differences between IFN, CMP$\Omega$, GHS, and SDF, are very small, at or below the percent level. Examples for this behaviour are the rapidity distribution of the reconstructed Higgs momentum, $y_{H(b\bar{b})}$ (\cref{fig:whbb0-yH-pTb1}, left), or the transverse momentum of the leading $b$ jet, $p_{t,b_1}$ (\cref{fig:whbb0-yH-pTb1}, right). Larger differences are observed between the new flavoured jet algorithms and the flavour-\kt algorithm, which is to be expected given that it is based on the \kt instead of the anti-\kt algorithm.

\begin{figure}
  \centering
  \includegraphics[width=0.49\linewidth]{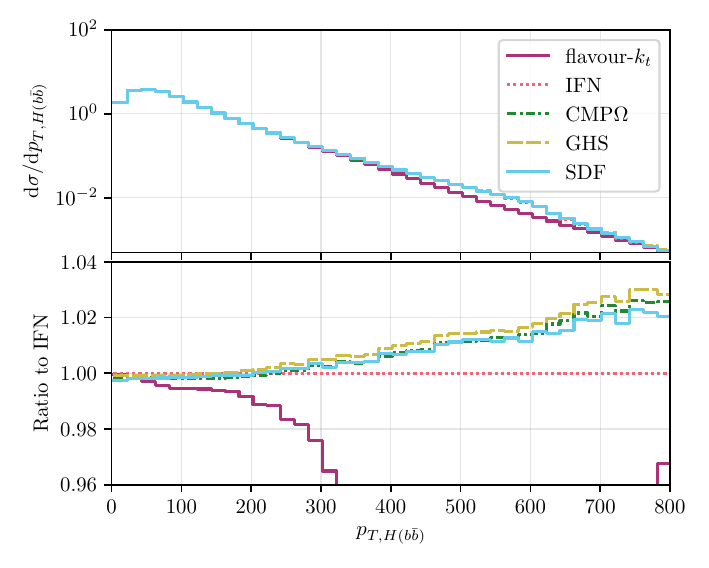}
  \includegraphics[width=0.49\linewidth]{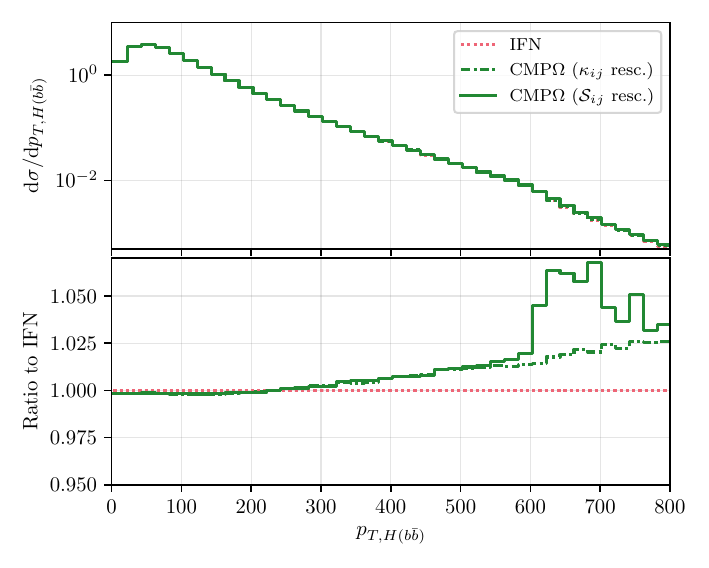}
  \caption{Predictions for the transverse momentum distribution of the reconstructed Higgs boson momentum, $p_{t,H(b\bar{b})}$, at NNLO in QCD. The distributions for the different jet algorithms were calculated using the same events. The lower pane shows the ratio of the distributions for the respective algorithms to that for IFN. Left: comparison of different flavour jet algorithms with each other; right: comparison of the different modifications of the CMP$\Omega$ algorithm discussed in \cref{eq:Sij-fix,eq:kappaij-fix}, see main text for details.}
  \label{fig:whbb0-pTH}
\end{figure}
Somewhat more pronounced effects can be observed in the tail of the transverse momentum distribution of the reconstructed Higgs, $p_{t,H(b\bar{b})}$, see \cref{fig:whbb0-pTH} (left). There is a stark difference between the flavour-\kt algorithm and the novel flavoured jet algorithms, starting from around $p_{t,H(b\bar{b})} \sim \SI{200}{\giga\eV}$. A similar effect was already observed in Ref.~\cite{Behring:2020uzq} when comparing the massless calculation using the flavour-\kt algorithm to the massive calculation using the plain anti-\kt algorithm. There, it was interpreted to be due to the fact that the flavour-\kt algorithm start to cluster the $b$ quarks coming from the Higgs boson decay into a single flavourless jet for much less boosted configurations compared to the anti-\kt algorithm. Since we require at least two $b$ jets, many of those events will not pass this cut.
Interestingly, also the distributions for the anti-\kt-based algorithms seem to change slope around $p_{t,H(b\bar{b})} \sim \SI{600}{\giga\eV}$ and towards higher transverse momenta the curves start to approach the curve of the flavour-\kt algorithm. Around this value of $p_{t,H(b\bar{b})}$, also the anti-\kt-based algorithms start to predominantly cluster the $b$ quarks from the Higgs boson decay together into a flavourless jet, as was shown analytically in Ref.~\cite{Caola:2023wpj}.

During the study it was observed that the CMP$\Omega$ algorithm with the rescaling of $\mathcal{S}_{ij}$ shown in \cref{eq:Sij-fix} shows a somewhat surprising behaviour above $p_{t,H(b\bar{b})} \gtrsim \SI{600}{\giga\eV}$ (shown as the solid green line in \cref{fig:whbb0-pTH} (right)).
Compared to the other algorithms, the CMP$\Omega$ algorithm yields about $\SIrange{2}{4}{\percent}$ larger results, with a very sudden onset. It turns out that this effect is already present at LO and the position of the onset scales linearly in $R$: when the jet radius is chosen twice as large, the jump in the distribution moves to about $\SI{300}{\GeV}$. This indicates that the issue is connected to configurations where the $b$ quarks from the Higgs decay are boosted in such a way that they are on the verge of being clustered into a single jet. Detailed investigations show that the effect is in fact due to the rescaling of $\mathcal{S}_{ij}$ by $\frac{\Omega_{ij}^2}{\Delta R^2}$. This modifies the metric compared to the anti-\kt algorithm also away from singular configurations. In particular, it can modify whether boosted Higgs decay events are clustered into a single jet. Since the three other flavoured jet algorithms obey exact anti-\kt kinematics, this modification shows up in the ratio plot. If we instead rescale $\kappa_{ij}$ by $\frac{\Omega_{ij}^2}{\Delta R^2}$, as described in \cref{eq:kappaij-fix} and shown as the dash-dotted green line in \cref{fig:whbb0-pTH} (right), the modification of the metric is confined to fairly soft configurations, due to the step function in $\mathcal{S}_{ij}$. This ensures that close to the edges of the jet the metric agrees with the anti-\kt metric. Switching to this alternative fix removes the sudden jump in the $p_{t,H(b\bar{b})}$ distribution for the CMP$\Omega$ algorithm.

\sectionAuthors{\texorpdfstring{$p p \to \Z + b/c$}{p p -> Z + b/c}  in central
  kinematics: NNLO and NLO+PS}{Davide Napoletano, Rene Poncelet, Daniel Reichelt, Andrzej Si\'odmok, and James Whitehead} 
\label{app:Zjnnlonlops}

In this appendix, we report on our study of $\Z$+jet events, at fixed order up to NNLO QCD and with matched NLO-plus-parton-shower calculations from event generators \herwig and \sherpa, with and without hadronisation.

\subsection{Details of the simulations}
\label{subsec:Zjnnlonlops_setup}

We consider the process
\begin{align}
  p p \to \mu^+\mu^- + \text{jet} + X\;.
\end{align}
Loosely inspired by the selection criteria used in Ref.~\cite{CMS:2016gmz}, we identify events corresponding to central nearly-on-shell \Z bosons through the requirements:
\begin{align}
    \label{eq:app_nnlonlops_analysis_cuts}
    \centering
	p_{\mathrm{T}}^{\mu} &> \SI{20}{\GeV} \,, & \left\vert y^{\mu} \right\vert &< 2.4, \\
    p_{\mathrm{T}}^{\mu\mu} &> \SI{20}{\GeV}  \,, & m_{\mu\mu} &\in  \SIrange[range-units=brackets, range-phrase={,}]{77}{111}{\GeV}.
\end{align}
Each accepted event contains exactly one $\mu^+\mu^-$ pair fulfilling these conditions.

We use \fastjet~\cite{Cacciari:2011ma} to implement the various jet algorithms as described in \cref{app:alg-setup}. We include all flavoured jet algorithms listed in \cref{sec:findings} and the standard anti-\kt with a jet radius of $R=0.5$, applied to all visible particles except the identified muon pair. For the flavoured jet algorithms, we use the net-flavour combination scheme during clustering and initialise only the flavour in question, e.g., either the $b$ or $c$ flavours, see \cref{app:alg-setup}. The labelling of the clustered jets is performed with the mod2-flavour scheme in the case of flavoured jet algorithms, while we employ the mod2, ghost and cone labelling strategies for the anti-\kt algorithm. The labelling is performed separately for $b$- and $c$-labelled jets, i.e., we do not consider multiple labels at a time. We identify as flavour-labelled jets those passing the selection cuts
\begin{align}
    \label{eq:app_nnlonlops_analysis_jetcuts}
    \centering
	p_{t}^{j_{b,c}} &> \SI{30}{\GeV} \,, & \left\vert y^{j_{b,c}} \right\vert &< 2.4.
\end{align}
For all predictions, we use the \texttt{PDF4LHC21} PDF set~\cite{PDF4LHCWorkingGroup:2022cjn} within the \lhapdf library~\cite{Buckley:2014ana}, and as factorisation and renormalisation scales, we adopt the transverse mass of the muon pair,
\begin{equation}
  \muf^2 = \mur^2 = M_{\rT,W}^2 \equiv p_{{\rT},\mu\mu}^2 + M_{\mu\mu}^2~.
\end{equation}
For the hard process, we are working in QCD with $n_f = n_l = 5$ flavours, and the electromagnetic coupling is obtained in the $G_\mu$ scheme using the Fermi constant
\begin{align}
  \alpha = \frac{\sqrt{2}}{\pi} G_\mu m_W^2 \left( 1 - \frac{m_W^2}{m_Z^2} \right)  \qquad \text{with}  \qquad   {G_{\mu} = \SI{1.16638e-5}{\GeV^{-2}}}.
\end{align}
The values of the masses and widths used for the numerical simulations read:
\begin{align}
  m_Z &=  \SI{91.1535}{\GeV},      & \Gamma_Z &= \SI{2.4943}{\GeV},  \nonumber \\
  m_W &=  \SI{80.352}{\GeV},       & \Gamma_W &= \SI{2.084}{\GeV}.
\end{align}
The fixed-order LO, NLO and NNLO computations have been performed with \textsc{Stripper}, a \texttt{C++} implementation of the four-dimensional formulation of the sector-improved residue subtraction scheme~\cite{Czakon:2010td, Czakon:2011ve, Czakon:2014oma, Czakon:2019tmo} which has been used for various processes involving flavoured jets~\cite{Czakon:2020coa, Hartanto:2022qhh, Czakon:2022wam, Hartanto:2022ypo, Czakon:2022khx}.  Conventional 7-point scale variations are used to estimate uncertainties from missing higher-order corrections.

For the NLO-plus-parton shower predictions we generate large event samples for the process $pp \to Z/\gamma^*[\to\mu^+\mu^-]+\text{jet}$ at NLO accuracy, using the \sherpa~\cite{Sherpa:2024mfk} and \herwigv{7} (H7)~\cite{Bellm:2015jjp} event generators (versions 3.0.0 and 7.3.0~\cite{Bewick:2023tfi}, respectively). We use the \mcatnlo~\cite{Frixione:2002ik} method for matching the perturbative hard process with the parton shower, as implemented in \sherpa~\cite{Hoeche:2011fd} and \herwig \matchbox~\cite{Platzer:2011bc} respectively. In anticipation of the analysis cuts (given in \cref{eq:app_nnlonlops_analysis_cuts}) we apply generator cuts:
\begin{align}
    \centering
	p_{\mathrm{T}}^{\mu} &> \SI{20}{\GeV} \,, & \left\vert y^{\mu} \right\vert &< 2.4, \\
    p_{\mathrm{T}}^{\mu\mu} &> \SI{20}{\GeV}  \,, & m_{\mu\mu} &\in \SIrange[range-units=brackets, range-phrase={,}]{77}{111}{\GeV}.
\end{align}
Hadron decays are disabled, such that stable $B$ hadrons can be used as unambiguous tags of flavour. A single sample is generated for each generator set-up, which is used to study both $b$- and $c$-tagged jets. The renormalisation scale $\mur$ is also used as the dipole-shower starting scale.

Within \sherpa, matrix elements at tree level are provided by the internal generators \amegic~\cite{Krauss:2001iv} and \comix~\cite{Gleisberg:2008fv}. Virtual corrections for this process are likewise included internally. The events are showered with \sherpa's default \csshower~\cite{Schumann:2007mg} based on Catani--Seymour dipoles and hadronised using the cluster hadronisation model~\cite{Webber:1983if} as implemented in \sherpa~\cite{Winter:2003tt,Chahal:2022rid} with the parameter tune from Refs.~\cite{Knobbe:2023njd,Knobbe:2023ehi}. The underlying event is simulated in \sherpa following the Sj\"ostrand--Zijl model~\cite{Sjostrand:1987su}.

Within \herwig, events are showered using either the \herwig dipole shower~\cite{Platzer:2009jq} or the \herwig angular-ordered shower~\cite{Gieseke:2003rz}. Within the \herwig dipole shower, by default, heavy quarks are treated as massless within the shower, as in the perturbative hard process; their physical masses are restored prior to hadronisation. In the angular-ordered shower, heavy quarks are treated as massless within the perturbative hard process only, and as massive within the shower. For hadronisation, we employ the \herwig cluster model~\cite{Webber:1983if,Gieseke:2017clv} and, for comparison, the string model~\cite{Andersson:1983ia} from \pythiav{8.308}~\cite{Bierlich:2022pfr} with colour-reconnection model~\cite{Christiansen:2015yqa} interfaced to \herwig via the \ThePI{} \texttt{C++} package~\cite{PIeight}. The Lund string model, together with the angular-ordered parton-shower model, have been tuned to LEP and LHC data sets as described in Ref.~\cite{H7stringTune}. 

In summary, we consider mainly the following event generator setups:
\begin{itemize}
    \item \sherpa: \pt-ordered dipole shower with massive $b$ and $c$ quarks with a cluster hadronisation model,
    \item H7 ang.: angular-ordered shower with massive $b$ and $c$ quarks with a cluster hadronisation model and with the string hadronisation model,
    \item H7 dip.: \pt-ordered dipole shower with massless $b$ and $c$ quarks with a cluster hadronisation model.
\end{itemize}

The fiducial phase-space selection described above, and routines filling the histograms displayed elsewhere in this appendix, have been implemented in a \rivet~\cite{Bierlich:2019rhm} analysis which was used for the \sherpa and \herwig results presented below.%
\footnote{The analysis code is available at \url{https://github.com/DReichelt/LH23FlavAlgs}.}
The code is based on the Rivet analysis \texttt{CMS\_2017\_I1499471} \cite{CMS:2016gmz}. In the \sherpa sample, we use the code developed in Ref.~\cite{Reichelt:2021svh} for the extraction of nonperturbative corrections to simultaneously access the parton and hadron level of a given event.

\subsection{Results}

\begin{figure}
    \centering
    \includegraphics[width=0.5\linewidth,page=1]{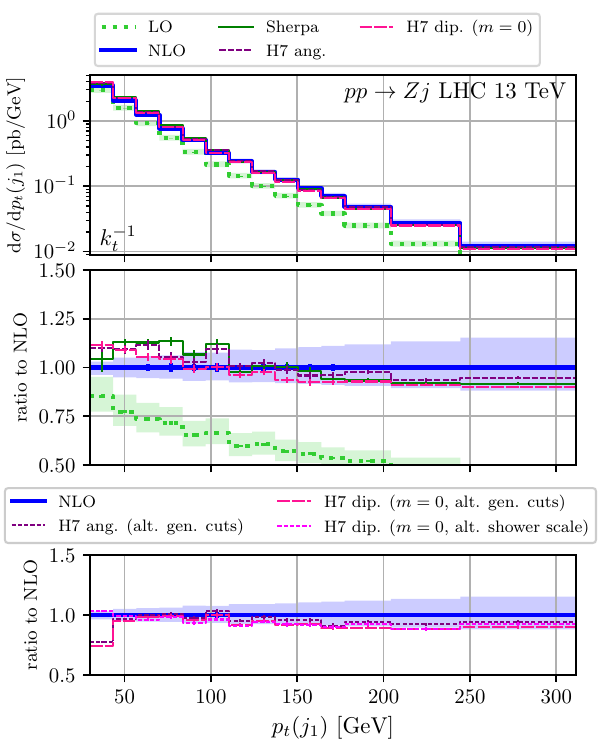}%
    \includegraphics[width=0.5\linewidth,page=2]{figures/ppzj_nnlo_nlops/ppzj_noflv_fo_vs_nlops_comparisons.pdf}
    \caption{The inclusive leading jet transverse momentum spectrum at low (left) and high \pt (right) in central $pp \to \Z + j$ kinematics. The comparison shows \sherpa, H7 dipole, H7 angular-ordered and FO QCD.}
    \label{fig:nnlo_nlops_jetpt}
\end{figure}

We compare the different algorithms at each of the three possible levels at which a jet or clustering algorithm can be applied:
\begin{enumerate}[(i)]
    \item to the hard-process final state, at fixed-order LO, NLO and NNLO;
    \item to the partonic final-state obtained from a parton shower (matched to NLO with \mcatnlo), using the \sherpa and \herwig dipole showers, and the \herwig angular-ordered shower;
    \item to the hadronic final-state obtained from (ii) with a hadronisation model, using the \sherpa and \herwig cluster models and the \pythia string model.
\end{enumerate}
The set-ups described in \cref{subsec:Zjnnlonlops_setup} have been chosen to maximise the consistency, and hence comparability, of the predictions between different combinations of codes and algorithms. We studied various jet observables without flavour labelling to validate the consistency. A comparison of the leading jet transverse momentum $p_t(j_1)$ can be found in \cref{fig:nnlo_nlops_jetpt}, the left-hand side shows the low-\pt and the right-hand side the high-\pt regime. The top panel displays the absolute results, and the middle panel shows the ratio with respect to fixed-order NLO QCD. We find reasonable agreement between the different event generators and good agreement with fixed-order NLO QCD within the theory uncertainties. The largest differences appear near the phase space threshold at low \pt, a region which is, however, sensitive to the details of the matching and shower scale. We demonstrate this sensitivity in the lowest panel, which contains \herwig predictions with altered shower starting scale or generation cuts.

\subsubsection{Sensitivity to perturbative order}

\begin{figure}
    \centering
    \includegraphics[width=0.45\linewidth,page=3]{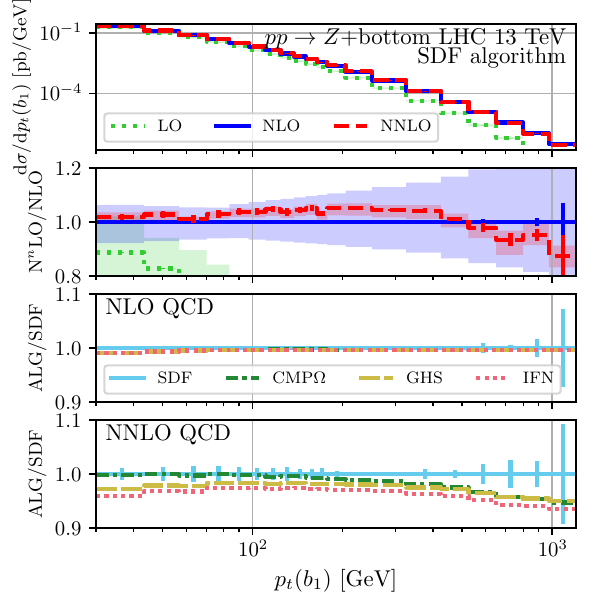}%
    \includegraphics[width=0.45\linewidth,page=3]{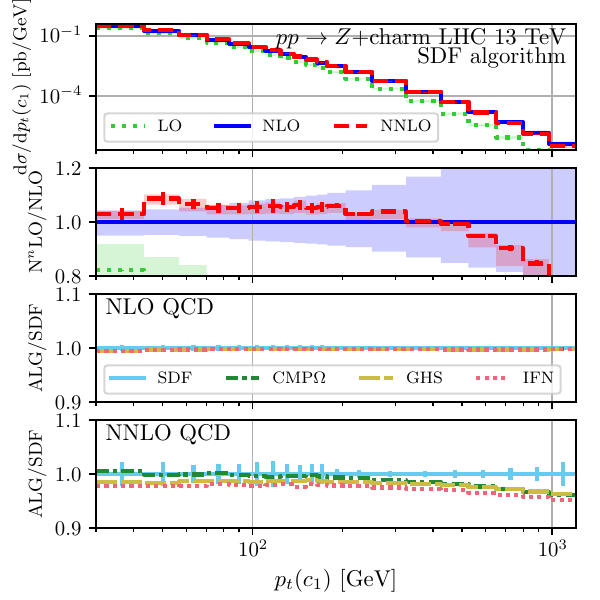}
    \caption{Perturbative predictions for the transverse momentum spectrum of the leading flavoured jet in central $pp \to \Z + j$ kinematics. The left column shows bottom-labelled jets and the right column charm-labelled jets. The upper two panels show the differential cross section obtained with the SDF algorithm through different perturbative orders (LO -- light green, NLO -- blue, NNLO -- red), absolute and relative to NLO QCD, respectively. The envelope of the 7-point scale uncertainties is shown as coloured bands. Different jet algorithms (CMP$\Omega$ -- green, SDF -- light blue, GHS -- yellow, IFN -- light red) are compared at NLO and NNLO QCD in the lower two panels. The Monte Carlo errors indicated by the vertical lines are almost 100\% correlated among the different algorithms and are only shown for the central prediction.}
    \label{fig:nnlo_ppzj_algos_pt}
\end{figure}

\begin{figure}
    \centering
    \includegraphics[width=0.45\linewidth,page=4]{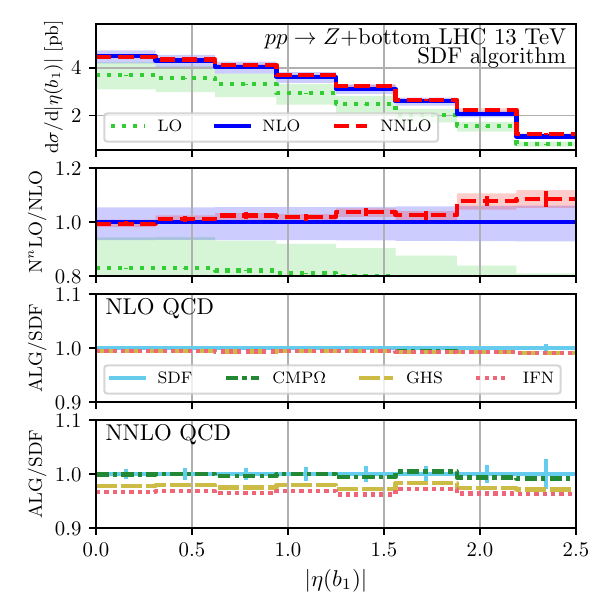}%
    \includegraphics[width=0.45\linewidth,page=4]{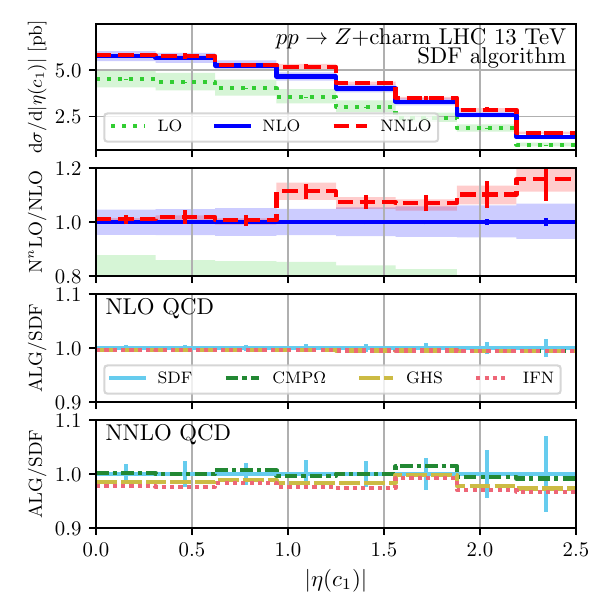}
    \caption{Same as \cref{fig:nnlo_ppzj_algos_pt} but for the rapidity distribution of the leading flavoured jet.}
    \label{fig:nnlo_ppzj_algos_eta}
\end{figure}

\Cref{fig:nnlo_ppzj_algos_pt} shows the perturbative predictions through LO (light green), NLO (blue) and NNLO QCD (red) for the SDF algorithm in the transverse momentum spectrum of the leading $b$/$c$ jet ($\pt(b_1)/\pt(c_1)$) in the upper two panels in absolute and relative to NLO QCD. The perturbative corrections are very similar for all the other algorithms and show the known stabilisation of the perturbatively predicted cross section at NNLO QCD and a significant reduction of the scale dependence. The $|\eta(b_1)|$ ($|\eta(c_1)|$) distributions in \cref{fig:nnlo_ppzj_algos_eta} exhibit similar behaviour.

The different algorithms are compared directly in the lower two panels in \cref{fig:nnlo_ppzj_algos_pt,fig:nnlo_ppzj_algos_eta} at NLO and NNLO QCD, respectively. The NNLO QCD predictions for the four algorithms are comparable within a few percent, and the differences are similar to the uncertainties of missing higher-order corrections. Focusing on the $\pt(b_1)/\pt(c_1)$ distributions, there are smaller shape differences at low and high $\pt(b_1)$/$\pt(c_1)$ of up to 4\%. The IFN, GHS, and CMP$\Omega$ algorithms seem to approach each other at higher $\pt(b_1)$ ($\pt(c_1)$).

\subsubsection{Sensitivity to parton showering}

We investigate various jet distribution observables. As outlined earlier, this analysis is conducted using both $b$- and $c$-flavoured objects. Our primary observation is that the qualitative effects are similar in both cases, though they tend to be more pronounced for charm flavour due to the smaller mass of the $c$ quark. It turned out that the sensitivity to the different flavour labelling algorithms depends significantly on the mass treatment within the parton shower.

\begin{figure}
    \centering
    \includegraphics[width=0.5\linewidth,page=3]{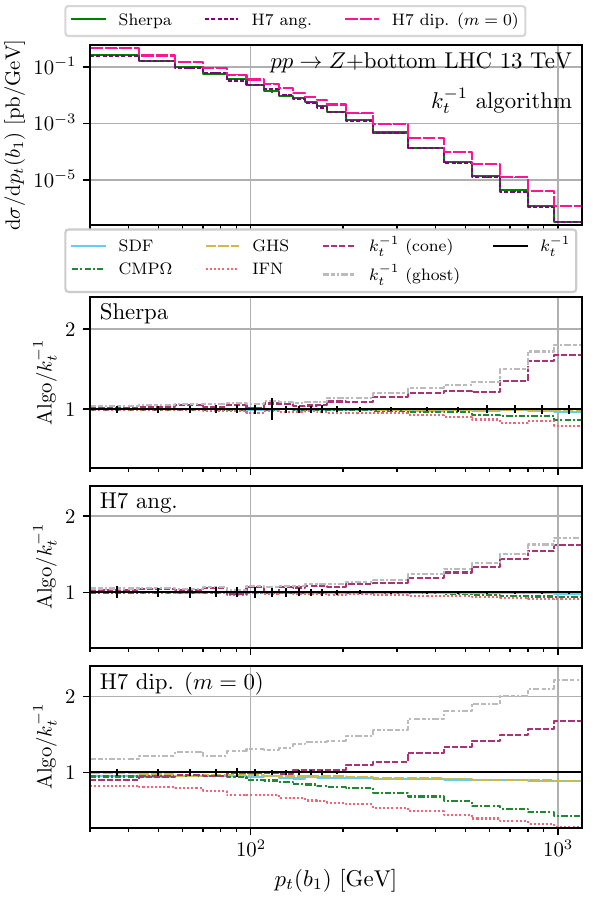}%
    \includegraphics[width=0.5\linewidth,page=3]{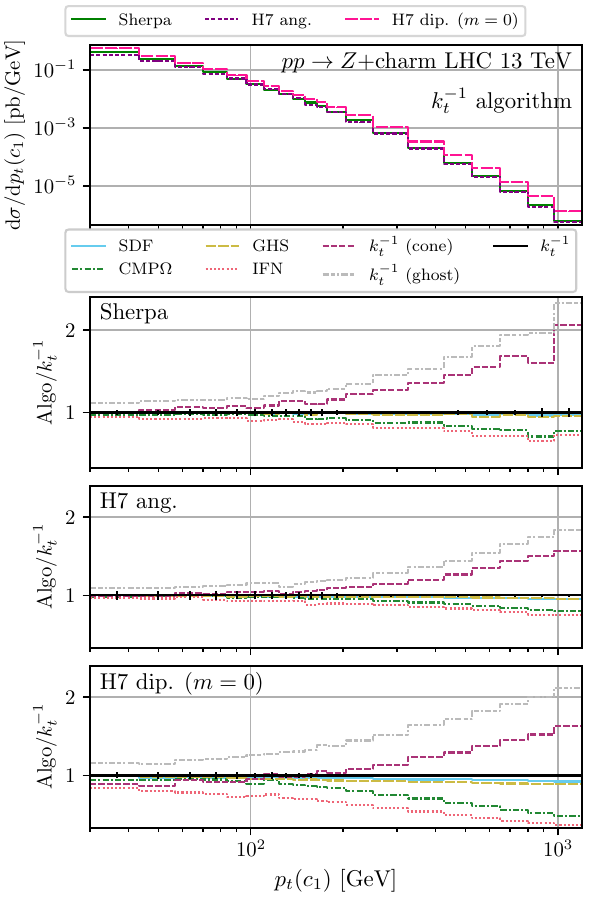}
    \caption{NLO+PS predictions from \sherpa and \herwig for the transverse momentum of the leading flavoured jet in central $pp \to \Z + j$ kinematics. The left column shows bottom-labelled jets and the right column charm-labelled jets. The top panel shows the differential cross section obtained with the standard anti-\kt algorithms with mod2-flavour combination. Different jet algorithms (CMP$\Omega$ -- green, SDF -- light blue, GHS -- yellow, IFN -- light red, as well as anti-\kt with different labelling: mod2 (black), ghost (grey) and cone (wine-red)) are compared for each generator in the lower three panels. The Monte Carlo errors indicated by the vertical lines are almost 100\% correlated among the different algorithms and are only shown for the central prediction.}
    \label{fig:nlops_ppzjet_pTb}
\end{figure}

We start by considering the transverse momentum of the leading tagged jet, $\pt(b_1)$ and $\pt(c_1)$ for the two flavour selections, in \cref{fig:nlops_ppzjet_pTb}.\footnote{Many more observables have been computed and can be found in the PDF files attached to this submission.} The left column of plots shows results for $b$-labelled jets, while the right column corresponds to the $c$-flavour selection. The top panel displays the absolute results for the different event generators at the hadron level, using the applied standard anti-\kt algorithm with mod2-flavour labelling. Below, we present a comparison of the different algorithms for each event generator in three panels.

The predictions from \sherpa and \herwig angular-ordered agree very well with each other across the spectrum. Strikingly, the \herwig dipole prediction is much higher than the other two, which can be traced back to an enhanced production of flavoured particles if the $b$/$c$ quarks are treated massless in the parton shower. We discuss this effect in more detail below.

We now turn our attention to the comparison of the jet algorithms. First, we can observe, by looking at the predictions with the standard anti-\kt algorithm, that there is a growing sensitivity to the different labelling strategies, i.e., mod2, cone and ghost. We want to note that since the decays of hadrons are deactivated in the samples used for this study, the ghost labelling and any-flavour scheme are the same because the ghosts will always be exactly collinear to the hadrons, i.e., zero clustering distance, and thus clustered immediately. Therefore, we can conclude that the difference between the ghost and mod2 flavour originates from the removal of $g\to q\bar{q}$ splittings of the latter. Furthermore, we can observe that differences also appear at high \pt when comparing the actual IRC-safe algorithms with anti-\kt. Comparing the $b$ and $c$ results, we see that in the $c$-label case the sensitivity is larger, reaching up to 10--20\% at $\SI{1}{\TeV}$. We find consistent results between the angular-ordered \herwig and \sherpa event samples and a significantly enhanced sensitivity for the dipole shower. This enhancement is again due to the increased flavour production within this sample, which naturally increases the sensitivity to the algorithms.

\begin{figure}
    \centering
    \includegraphics[width=0.5\linewidth,page=1]{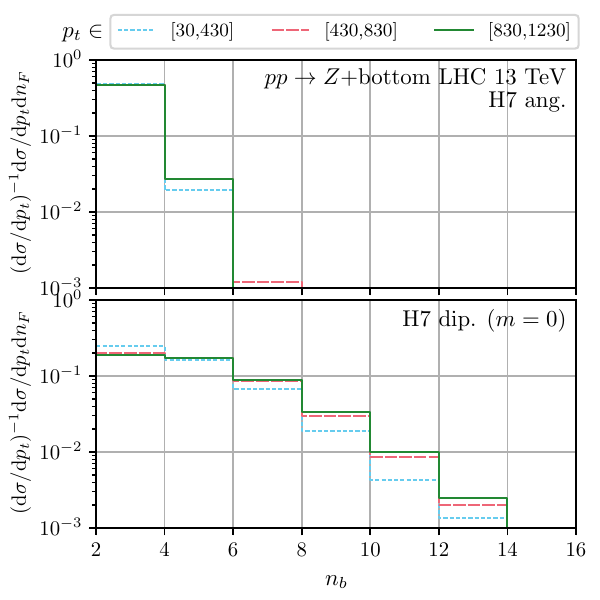}%
    \includegraphics[width=0.5\linewidth,page=1]{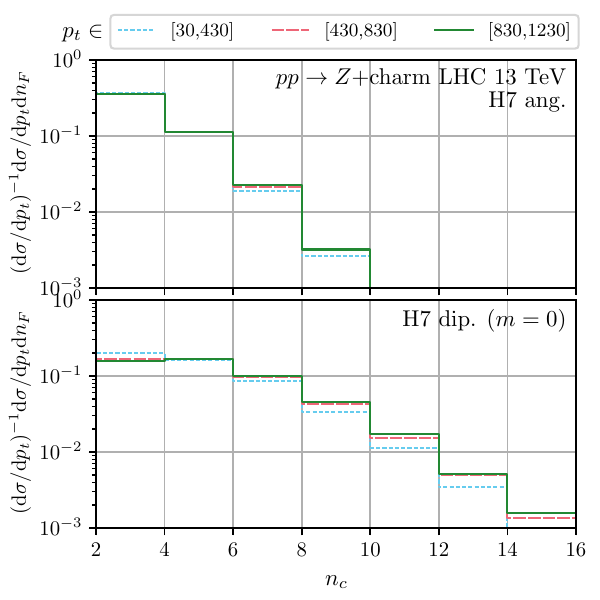}
    \caption{The normalised differential cross section as a function of the flavour multiplicity (the number of hadrons with the corresponding flavour) predicted by a massive (\herwig angular-ordered -- top) and massless (\herwig dipole -- bottom) parton shower, in three transverse momentum bins of the leading flavoured jet. The $b$ case is shown on the left and the $c$ case on the right.}
    \label{fig:nlops_ppzjet_pTb_mass}
\end{figure}

The striking feature that the massless dipole \herwig sample predicts a much larger cross section and is much more sensitive to the jet algorithms can be traced back to the amount of flavoured hadrons produced. In \cref{fig:nlops_ppzjet_pTb_mass} we show the fraction of the differential cross section as a function of the flavour multiplicity, i.e., the number of flavoured hadrons. We show results for \herwig angular (with massive quarks in the shower) and for the \herwig dipole shower (massless quarks in the shower) for the $b$ and $c$ case separately. We find a significantly enhanced hadron multiplicity in the case of the massless treatment within the shower. Additionally, we observe that the multiplicities for the massive quarks in the shower are larger for the charm case than for the bottom case, as one would expect naively. In contrast, the difference predicted by the shower with massless treatment is significantly smaller. Consequently, the differences between massless and massive treatments are larger for bottom than for charm.

\begin{figure}
    \centering
    \includegraphics[width=0.5\linewidth,page=3]{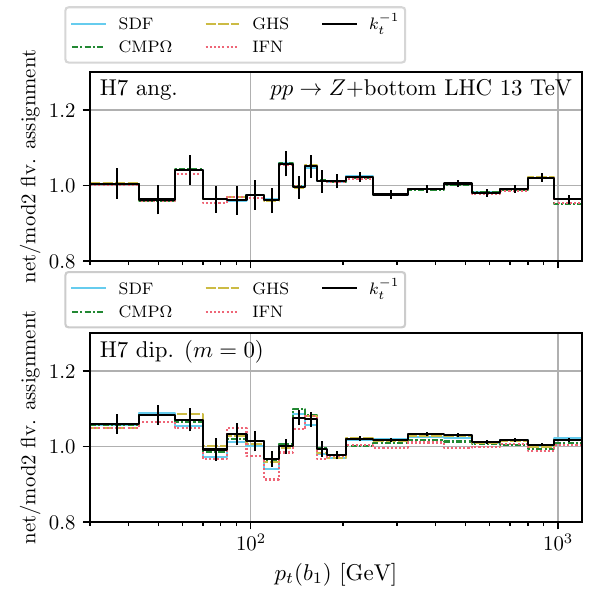}%
    \includegraphics[width=0.5\linewidth,page=3]{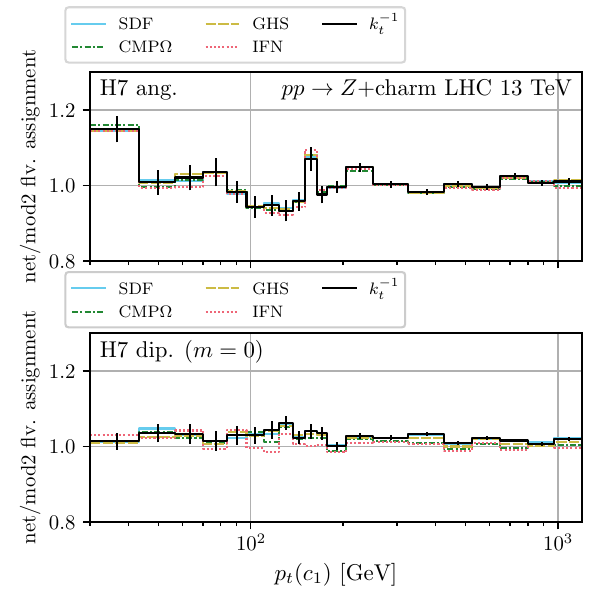}
    \caption{Comparison between the net and mod2 flavour label assignment on jets which are recombined with the net-flavour recombination scheme in case of the massive \herwig angular shower (top row) and the massless \herwig dipole shower (bottom row). The left column shows the bottom case, and the right column shows the charm case. The ratio is shown for a collection of algorithms (CMP$\Omega$ -- green, SDF -- light blue, GHS -- yellow, IFN -- light red, anti-\kt{} -- black).}
    \label{fig:nlops_ppzjet_pTb_flvassign}
\end{figure}

As discussed in \cref{app:alg-setup}, the different flavoured jet algorithms can have a different flavour assignment on the clustered jets during the clustering sequence itself. The impact of such a choice is shown in \cref{fig:nlops_ppzjet_pTb_flvassign} where we directly compare the mod2- and net-flavour assignment for the massive \herwig angular and massless \herwig dipole sample. We want to stress that in both cases, the clustering has been performed with the net-flavour recombination scheme. Overall, the ratio is close to 1, indicating only minor differences in the predicted flavoured jet rates. In most cases, the statistical uncertainties are of comparable size. In particular, in the high-\pt region, we observe a negligible difference between the two flavour assignments. A systematic difference can be seen for the massless \herwig dipole sample at low \pt of about $5\%$, which is, however, independent of the jet algorithm used. The most relevant potential source of differences can come from jets containing two particles of flavour $f$, which in the case of mod2 will be labelled as unflavoured and as flavoured with net-flavour. These cases are not infrared enhanced at NNLO since an $ff$ flavour cannot arise from a gluon splitting. As such, we do not expect significant differences and, if at all, of geometric nature (i.e., proportional to the area of the jet) due to accidental wide-angle radiation of a flavour into another flavoured jet.

\subsubsection{Sensitivity to hadronisation}

\begin{figure}
    \centering
    \includegraphics[width=0.5\linewidth,page=3]{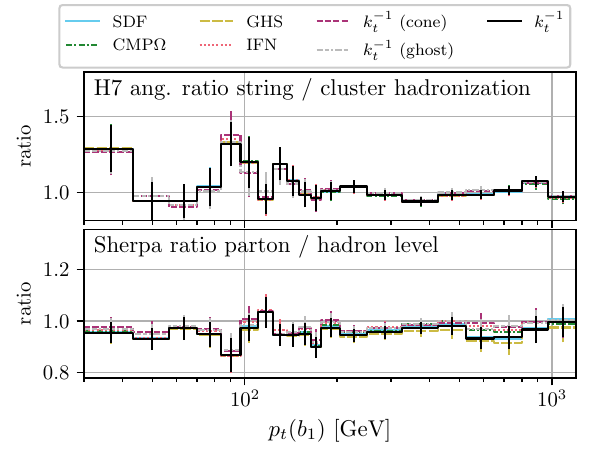}%
    \includegraphics[width=0.5\linewidth,page=3]{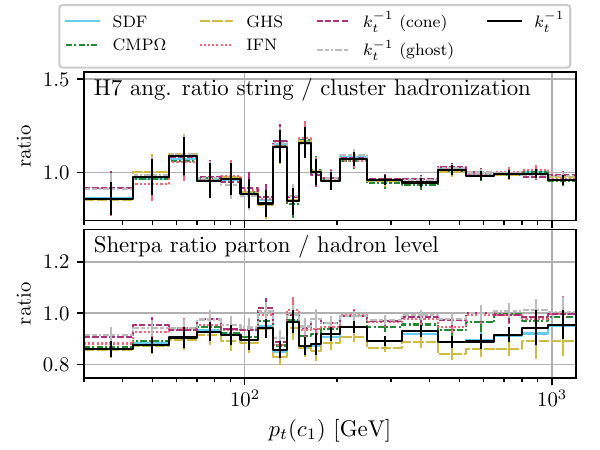}
    \caption{The top row shows a comparison between the string and cluster hadronisation in \herwig for the transverse momentum spectrum of the leading $b$-labelled jet on the left and $c$-labelled jet on the right. The lower row shows the ratio between parton level and hadron level in the \sherpa sample.}
    \label{fig:nlops_ppzjet_pTb_had}
\end{figure}

To investigate the sensitivity of the jet algorithms to hadronisation, we conducted two additional comparisons. First, we investigated the impact of the hadronisation model in \herwig on the results, which we show differentially in the leading flavoured jet \pt in the top row in \cref{fig:nlops_ppzjet_pTb_had}. We compare the cluster and string models, and we observe only minor effects, which are generally within the statistical uncertainties. The second comparison is between the parton and hadron level within the \sherpa sample, which we show in the lower row in \cref{fig:nlops_ppzjet_pTb_had}. For the charm case, we observe some minor effects, differentiating the algorithms in particular at high \pt.

\subsubsection{Comparison of fixed-order and NLO+PS}

Looking at the high-\pt end of the $\pt(b_1)$/$\pt(c_1)$ spectrum (see \cref{fig:nnlo_ppzj_algos_pt} and \cref{fig:nlops_ppzjet_pTb}), it appears that the NNLO fixed-order prediction and the NLO+PS predictions behave differently. Qualitatively, all algorithms appear similar in the case of fixed-order prediction, while they diverge (in the sense that they differ more and more) for NLO+PS. The scale of the differences is also very different. We see at most a few percent for the fixed-order calculation; for the NLO+PS, we see differences of $\order{10\%}$. When directly comparing the NLO+PS and fixed-order predictions for a fixed jet algorithm, which we show in 
\cref{fig:summary_ppzj_fo_vs_nlops_bottom}%
, we can observe that there is generally a fair agreement in the sense that the NLO+PS is (almost) within the NLO QCD uncertainty. However, we observe shape differences in the high-\pt region for the SDF, GHS and CMP algorithms, while the IFN results are the same using NLO FO and NLO+PS. The shape differences are consistent with the spread we see when comparing the algorithms at NLO+PS and FO. The effect is more clearly visible for the $c$-jet case than for the $b$-jet case, again due to the larger sensitivity in the $c$ case.

In other words, the agreement between NLO+PS and FO for SDF, GHS, and CMP depends on the mass parameter of the flavoured quark; the smaller the mass, the worse the agreement. The reason for this observation is likely to be found in the multiplicity of flavoured quarks, which is higher for smaller masses, leading to more significant differences between the algorithms. IFN shows a much smaller dependence and therefore provides a more robust prediction.

\subsubsection{Jet substructure observables}

\begin{figure}
    \centering
    \includegraphics[width=0.5\linewidth,page=16]{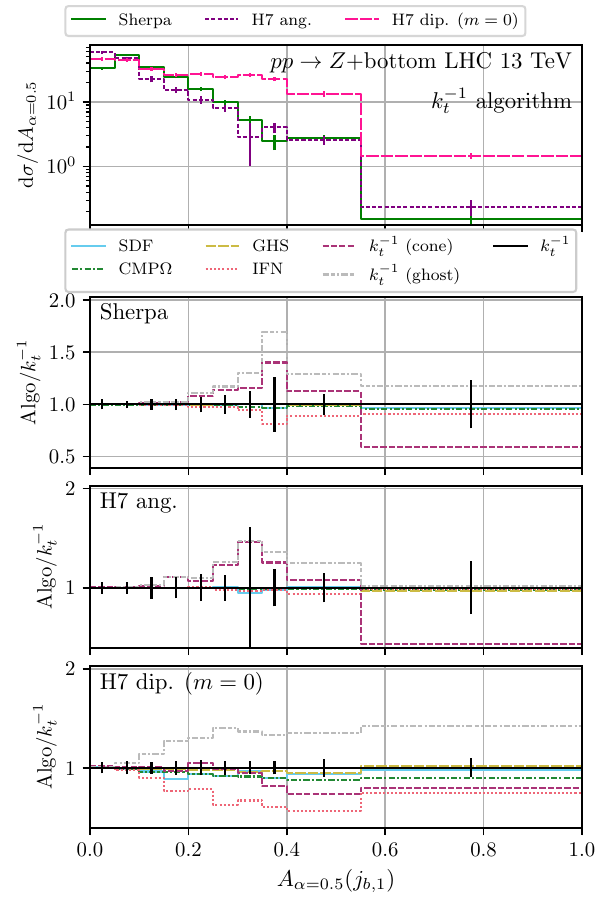}%
    \includegraphics[width=0.5\linewidth,page=16]{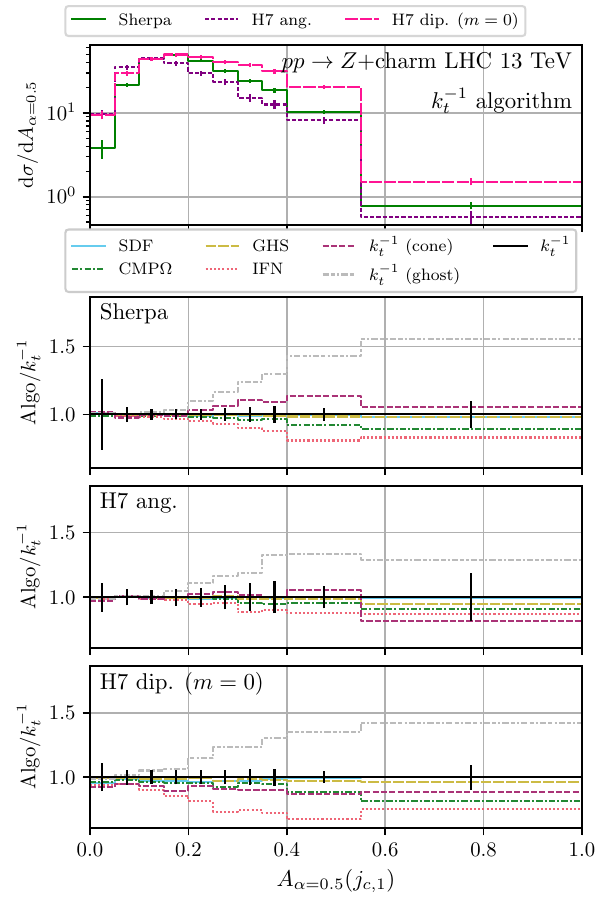}
    \caption{Same structure as \cref{fig:nlops_ppzjet_pTb} but for the Les Houches angularity of the leading flavoured jet,  $A_{\alpha = 0.5} (j_{b/c,1})$.}
    \label{fig:nlops_ppzjet_jss_ang}
\end{figure}

\begin{figure}
    \centering
    \includegraphics[width=0.5\linewidth,page=19]{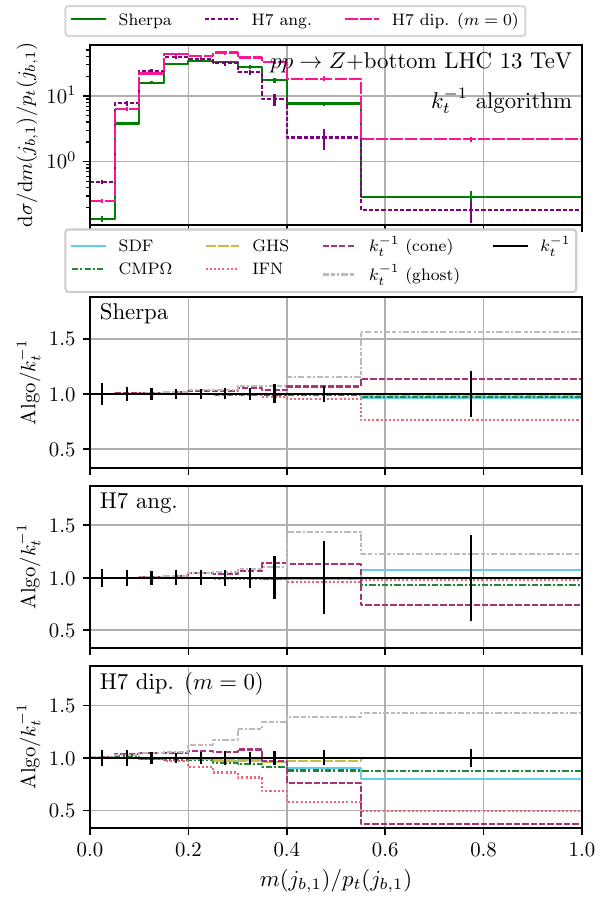}%
    \includegraphics[width=0.5\linewidth,page=19]{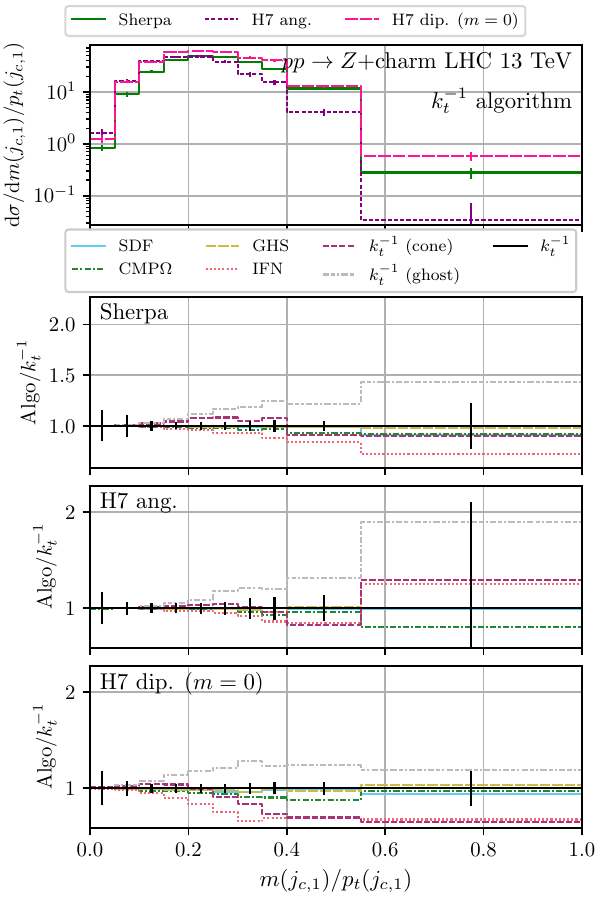}
    \caption{Same structure as \cref{fig:nlops_ppzjet_pTb} but for the normalised jet mass of the leading flavoured jet, $m(j_{b/c,1}) \, / \, {\pt}(j_{b/c,1})$.}
    \label{fig:nlops_ppzjet_jss_mass}
\end{figure}

We also take a first look at the jet substructure of the tagged jets. There are several subtleties to consider when discussing the jet substructure of jets tagged with IRC-safe jet algorithms. We here only take a simple look at jet substructure observables defined on all kinematical constituents of the anti-\kt jets that are labelled as $b$ flavoured by the respective algorithm, i.e., we ignore any information about particles outside in the jet that the flavour algorithm identified to belong to the jet in addition to the raw anti-\kt constituents (and vice versa, we keep constituents even if they did not contribute to the jet flavour label). The only exception is the CMP algorithm that slightly modifies the jet kinematics and hence defines jets that can not necessarily be associated with the raw anti-\kt jets. We concentrate on the angularity ($\alpha=0.5$, Les Houches angularity (LHA)) in \cref{fig:nlops_ppzjet_jss_ang} and the jet mass normalised to the transverse momentum of the jet in \cref{fig:nlops_ppzjet_jss_mass}.

We illustrate the general features of the distributions in the respective top panels of the figures, for the case of anti-\kt jets tagged with the mod2 prescription. \sherpa and \herwig's angular-ordered shower (using massive splitting functions) agree reasonably well, while the distribution is significantly shifted to larger values of jet mass and angularity, indication the presence of additional or harder radiation inside the jet. The differences between the showers are generally more pronounced for the LHA compared to the jet mass.

The different algorithms for each simulation are compared in the three lower panels. Comparing $b$ (left plot) and $c$ jets (right), we first note that, as expected, the behaviour is the same when considering the \herwig dipole shower that treats both as equally massless quarks. For the other showers the differences between the algorithms are significantly enhanced for the $c$-jet case, while they represent a minor correction for $b$ jets, consistent with our earlier observations. 

Focusing on the larger visible differences in $c$ jets, we observe consistent qualitative results between the different showers. SDF, GHS, and CMP behave roughly similarly to anti-\kt (and hence each other). This should not be surprising, since they behave similarly when considering the $p_t$ spectrum, indicating that they label roughly the same jets as flavoured, and we simply evaluate the substructure observables of those jets. We can hence mainly learn something about the characteristics of jets that are labelled in some algorithm but not the other. for example, the excess $b$-jet cross section predicted by the ghost labelling technique appears to be associated with a large jet mass and angularity. The reduced cross section in the IFN algorithm relative to anti-\kt on the other hand appears to be largely based on removing jets with large mass and angularities. This reinforces the idea that differences are mainly caused by configurations with a lot of radiation (and hence additional flavoured particles) in the jets. The cone-labelled jets tend to have a similar spectrum to anti-\kt and the algorithms behaving similarly, tentatively shifted towards the soft end of the distribution.

Having taken a first look at the jet substructure of flavour-labelled jets with the new generation of algorithms, we leave a detailed study, exploring the effects observed above and going beyond this simple procedure, to future studies.

\FloatBarrier\clearpage

\sectionAuthors{\texorpdfstring{$pp \to Z+b$}{p p -> Z+b} across kinematic regions}{Gavin P. Salam, and Ludovic Scyboz}\label{app:ppzb_kinematics}

Here we investigate further the results of the last section, in particular deviations of the
various flavoured jet algorithms from anti-\kt spectra (see, e.g., the
transverse momentum of the $b$-labelled jet in
\cref{fig:nlops_ppzjet_pTb}). For this purpose we generate a
sample of $Z+j$ events at parton level (i.e., without hadronisation) and follow
a roughly similar selection to \cref{app:Zjnnlonlops} in
our analysis.
The events are produced by \pythiav{8.306} at leading order,
at a centre-of-mass energy of $\sqrt{s}=\SI{13}{\TeV}$, with no flavour
constraints on the Born $Z+j$ process.
We require, as before, a \Z boson to be reconstructed in an invariant mass range $m_{\mu^+ \mu^-} \in [\SI{71}{\GeV}, \SI{111}{\GeV}]$, where the muons have to satisfy the set of cuts ($p_{t,\mu} > \SI{20}{\GeV}$, $|\eta_\mu| < 2.4$), and we cluster jets with a radius $R=0.5$.

\begin{figure}
    \centering
    \includegraphics[width=0.49\linewidth,page=1]{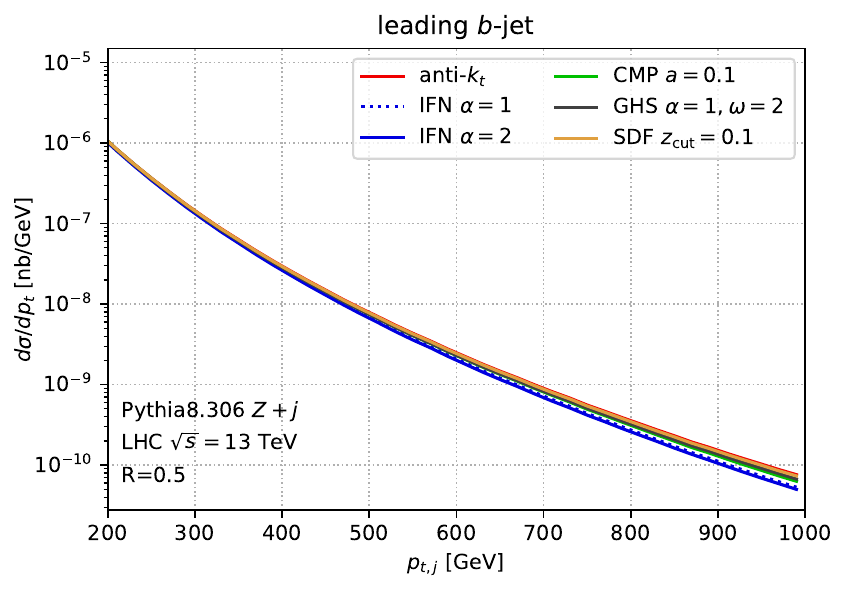}
    \includegraphics[width=0.49\linewidth,page=2]{figures/lops_ppzjet/pythia-Zj-flav-algs.pdf}
    \caption{Left: The distribution of the transverse momentum of the leading $b$-labelled jet $p_{t,b}$ for net-flavour anti-\kt (red), IFN ($\alpha=1$ and $\alpha=2$, blue), CMP ($a=0.1$, green), GHS ($\alpha=1$, $\omega=2$, black) and SDF ($\beta=2$, $z_{\cut}=0.1$, yellow). Right: the ratio to net-flavour anti-\kt.}
    \label{fig:lops_ppzjet_pTb_Py}
\end{figure}

In \cref{fig:lops_ppzjet_pTb_Py} we consider the distribution of the transverse momentum of the leading identified $b$ jet, focusing on a similarly high range of $p_{t,b}$ (from $\SI{200}{\GeV}$ to $\SI{1}{\TeV}$), as in \cref{fig:nlops_ppzjet_pTb}.
As can be seen in particular in the ratio to net-flavour anti-\kt (\cref{fig:lops_ppzjet_pTb_Py}, right), while the various flavoured jet algorithms differ from the anti-\kt distribution by only a couple of percent at $\pt \sim \SI{300}{\GeV}$ (and by $\sim 8\%$ for IFN $\alpha=2$), all algorithms start to develop larger differences at higher values of \pt.

\begin{figure}
    \centering
    \includegraphics[width=0.49\linewidth,page=3]{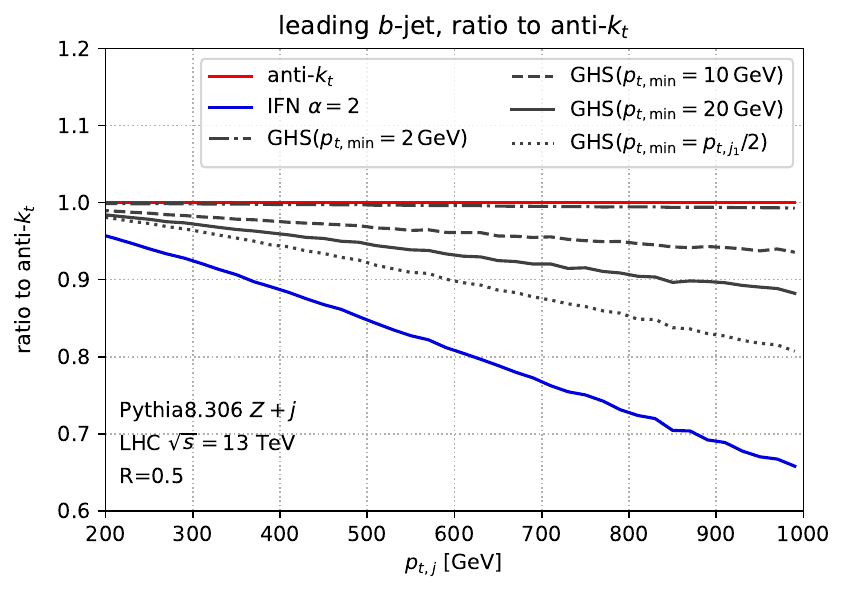}
    \includegraphics[width=0.49\linewidth,page=4]{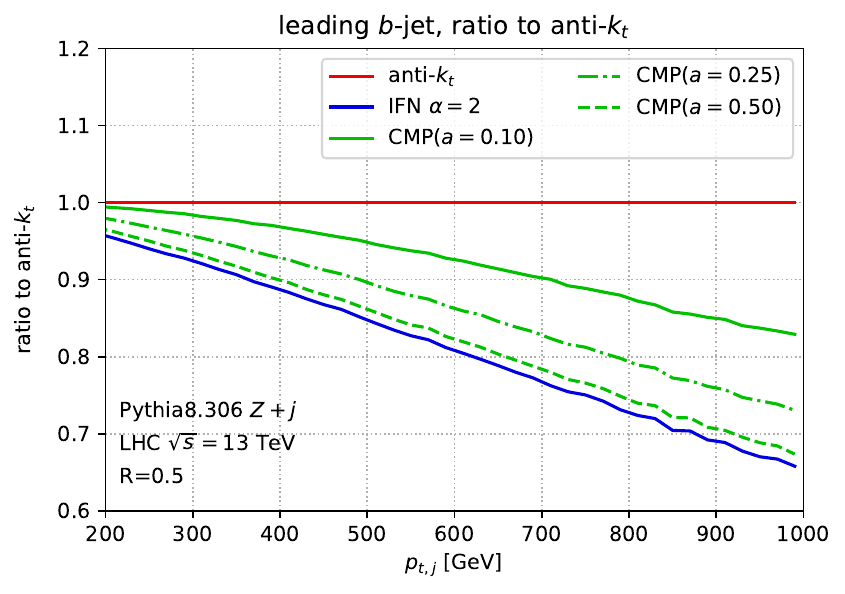}\\
    \includegraphics[width=0.49\linewidth,page=5]{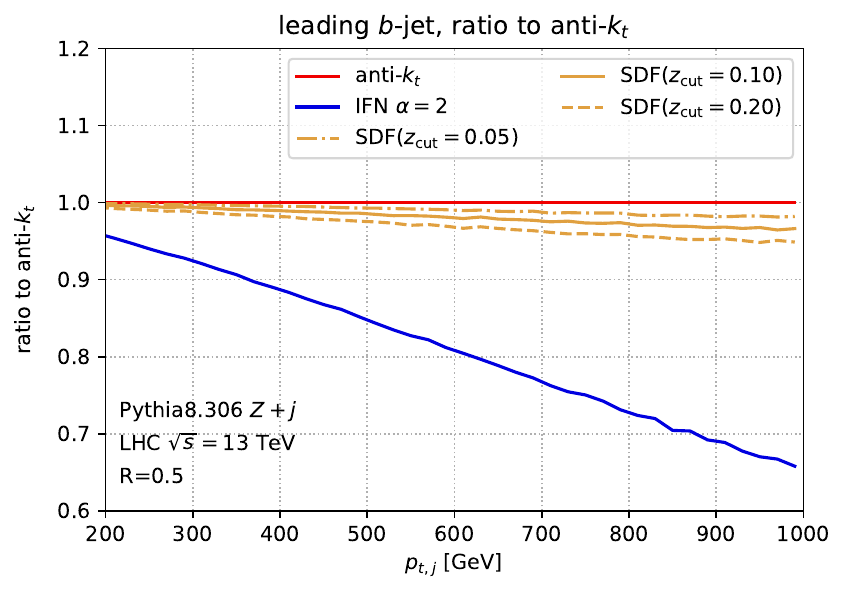}
    \caption{Same as \cref{fig:lops_ppzjet_pTb_Py}, right, for variations of the flavoured algorithms' parameters, as explained in the text. Top left: GHS, top right: CMP, bottom: SDF.}
    \label{fig:lops_ppzjet_pTb_Py_parms}
\end{figure}

Furthermore, this observation quantitatively depends on the actual set of parameters used by the
flavoured algorithms. To illustrate this, we show the same ratio to the anti-\kt result in \cref{fig:lops_ppzjet_pTb_Py_parms}, for variations of the IR-regularising parameters: the
minimum transverse momentum $p_{t,\text{min}}$ for the input jets in GHS (top left),
the parameter $a$ for CMP (top right), and the SoftDrop cut
$z_{\text{cut}}$ for SDF (bottom). The default values used in
\cref{fig:lops_ppzjet_pTb_Py} are pictured as full lines, and the
curves corresponding to variations of the parameters are shown as
dashed/dotted lines. In particular the CMP and GHS algorithms tend to the anti-\kt distribution in a limit where their defining parameters go to zero (in
the exact limit, those algorithms would become IRC unsafe).

The fact that the flavoured jet algorithms differ in assigning flavour labels is
not surprising.
As an aid to understand whether the $b$-label assignment is sensible,
however, we investigate a set of individual $Z+j$ events, again generated
with \pythia, with no flavour constraints in the generation.
We select events which have at least two $b$ quarks at parton level, and
where the leading anti-\kt jet  satisfies $p_{t,j} > \SI{200}{\GeV}$
and is classified as a $b$ jet by the net-flavour anti-\kt algorithm.

\begin{figure}
    \centering
    \includegraphics[width=0.49\linewidth,page=1]{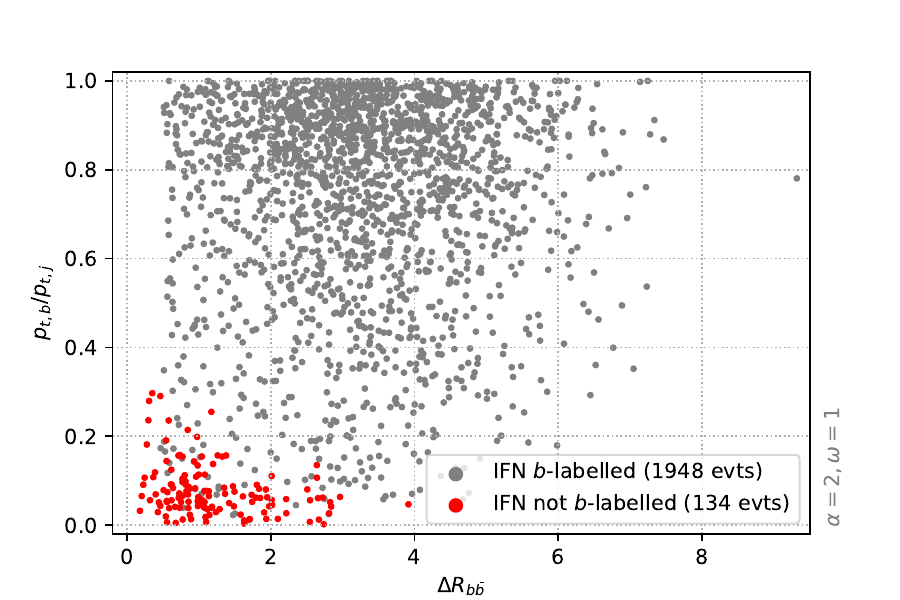}
    \includegraphics[width=0.49\linewidth,page=2]{figures/lops_ppzjet/pythia-Zj-scatter.pdf}\\
    \includegraphics[width=0.49\linewidth,page=3]{figures/lops_ppzjet/pythia-Zj-scatter.pdf}
    \includegraphics[width=0.49\linewidth,page=4]{figures/lops_ppzjet/pythia-Zj-scatter.pdf}
    \caption{$Z+j$ events with identified $b$ jets (grey: labelled as $b$ by
      both net-flavour anti-\kt and the algorithm under
      consideration, red: labelled as $b$ by anti-\kt, but not by the
      flavoured algorithm), as a function of $\Delta R_{b\bar{b}}$,
      $p_{t,b} / p_{t,b\text{-jet}}$.
      Top left: IFN, top right: CMP, bottom left: GHS, bottom right: SDF.}
    \label{fig:lops_ppzjet_scatter_all}
\end{figure}

In \cref{fig:lops_ppzjet_scatter_all} we depict
such events as dots in a scatter plot, where the $x$ coordinate is
the angular distance between the two $b$ quarks in the event, $\Delta
R_{b\bar{b}}$, and the $y$ coordinate is the transverse momentum
fraction, $p_{t,b} / p_{t,b\text{-jet}}$, carried by the $b$ quark in
the $b$ jet as identified with anti-\kt (if there are two $b$
(anti-)quarks, $p_{t,b}$ is the larger of the two \pt's).
Roughly, one would expect a majority of events in the bottom-left
region of the plot to have originated from soft large-angle gluon
splitting $g \to b\bar{b}$, where one $b$ entered the anti-\kt jet,
while the other did not. 
Regions of relatively hard $b$-jet constituents and angular
separations of order one or larger, $\Delta R_{b\bar{b}} \gtrsim 1$, will typically
be populated by genuine $b$ jets produced in the hard collision (a
large  $\Delta R_{b\bar{b}}$ generally indicates that the other
$b$ quark that is outside the jet is collinear to the beam, generated
by an initial-state $g\to b\bar b$ splitting).
We colour events in grey if the leading jet is $b$ labelled by
\textit{both} net-flavour anti-\kt and the flavoured algorithm; in
contrast, if the jet is $b$ labelled by anti-\kt but not by the
flavoured algorithm (because, e.g., the latter neutralised the $b$ flavours), we show these points in red.

\begin{figure}
    \centering
    \includegraphics[width=0.49\linewidth,page=5]{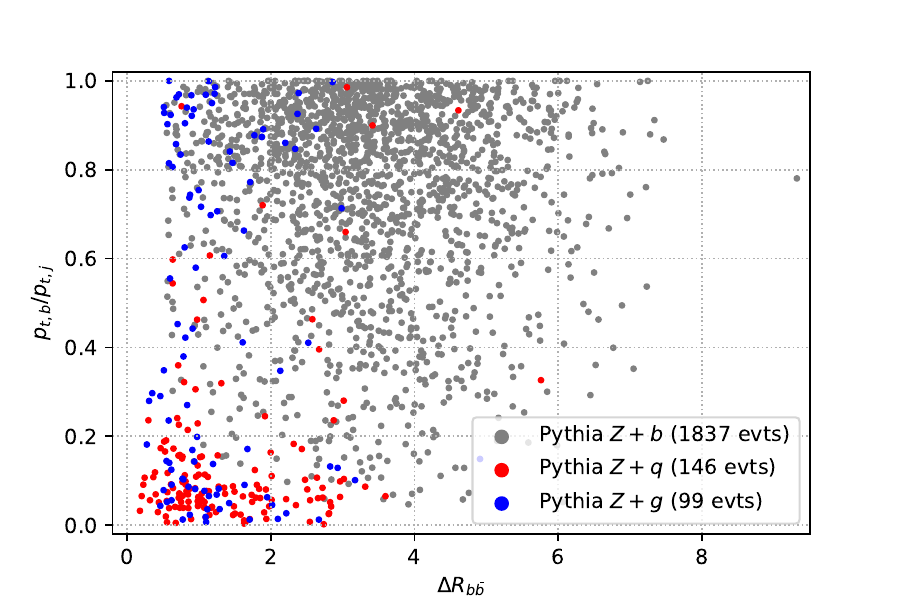}
    \caption{Same as \cref{fig:lops_ppzjet_scatter_all}, with the
      colour of the MC data points determined by the truth-level
      flavour of the hard process generated by \pythia: $Z+b$ (in grey,
      mostly genuine $b$-jets), $Z+q$ (in red, mostly light-quark jets
      with a soft $b$ contaminating them), $Z+g$ (in blue, with some
      soft-$b$ contamination giving ``fake'' $b$ jets and some hard
      $g\to b\bar b$ splitting giving genuine $b$ jets.}
    \label{fig:lops_ppzjet_scatter_pythia}
\end{figure}

As evidenced by the set of plots in
\cref{fig:lops_ppzjet_scatter_all}, the flavoured jet algorithms
remove $b$ flavour to varying degrees, with IFN removing the most.
For comparison, in \cref{fig:lops_ppzjet_scatter_pythia} we show
the same jets ($b$ labelled by net-flavour anti-\kt), with a colouring
that corresponds to the (truth level) flavour generated in the hard
process by \pythia at leading order:
$Z+b$ events (in grey)  are the most common source of $b$ jets, and one
may think of them as producing ``genuine'' $b$ jets;
$Z+q$ events (in red) would typically not lead to hard $b$ jets, but
can occasionally do so, most frequently when an additional soft gluon
from the parton shower splits into a $b$ and a $\bar b$, one of which
enters the leading jet, while the other goes outside
the jet (physically, we would consider this to be a fake $b$ jet);
for $Z+g$ events (in blue), the situation is similar, but additionally
the hard gluon can itself split to $b\bar b$, which is the source of
those blue scatter points that are at large $p_{t,b}$ with a moderate
$\Delta R_{b\bar b}$ (physically, we would consider these to be
genuine $b$ jets).
For the same selection, we also showed the distribution of $\frac{p_{t,b}}{p_{t,b\textrm{-jet}}}\Delta R_{b\bar b}$ in \cref{fig:summary_lops_ppzjet} (right), which highlights the separation of events with a hard $b$ (at high values of the variable) from those with a hard $g$ or a light $q$ (at low values), similarly to the scatter plots of \cref{fig:lops_ppzjet_scatter_all,fig:lops_ppzjet_scatter_pythia}. At the lower end, where standard anti-$k_t$ labels many hard $g/q$-initiated jets as $b$ jets, the jet flavour algorithms differ in how they assign $b$ flavour, with IFN being the closest to the truth labelling from the hard matrix element in \pythia.
One should keep in mind that in \pythia the map from the Born flavour
to the final event depends on the details of the parton shower.
Nevertheless a comparison between
\cref{fig:lops_ppzjet_scatter_pythia,fig:lops_ppzjet_scatter_all} gives an idea of the degree of
similarity between the flavour classification performed by the jet
algorithms in \cref{fig:lops_ppzjet_scatter_all}, and
a jet flavour as determined by a LO matrix-element-based likelihood.

\sectionAuthors{\texorpdfstring{$p p \to \Z + c$}{p p -> Z + c} in \lhcb kinematics: NNLO and NLO+PS}{Alexander Huss and Giovanni Stagnitto}\label{app:ppzcharm_nnlo}

In this appendix, we consider the associated production of a \Z boson with a charmed jet in the forward region, in particular focusing on the \lhcb measurement~\cite{LHCb-PAPER-2021-029} at $\s = \SI{13}{\TeV}$.
Such a measurement is valuable for the determination of intrinsic charm component in the proton~\cite{Brodsky:1980pb,Ball:2022qks,Guzzi:2022rca}.
Precise theory predictions for $\Z + c$-jet production have been provided in Ref.~\cite{Gauld:2023zlv}, both at the fixed-order level (up to NNLO) and with NLO predictions matched to a parton shower.
The purpose of the current study is to extend the analysis of Ref.~\cite{Gauld:2023zlv} to the new generation of flavoured jet algorithms.

The fixed-order theory predictions up to NNLO are obtained with the public \nnlojet code~\cite{NNLOJET:2025rno}, with the additional flavour tracking procedure at parton level developed for the studies in Refs.~\cite{Gauld:2019yng,Gauld:2020deh,Gauld:2023zlv,Gehrmann-DeRidder:2023gdl}.
For the purpose of the present analysis, we have interfaced \nnlojet to \fastjet to access the implementation of the flavoured algorithms.
Both the flavour tracking capabilities in \nnlojet and the interface to \fastjet will be included in a future version of \nnlojet.

We further consider NLO predictions matched with a parton shower (NLO+PS), obtained within the \mgaNLO (v.\ 2.7.3)~\cite{Alwall:2014hca} framework interfaced to \pythia (v.\ 8.243)~\cite{Sjostrand:2014zea} (default $p_{{\rm T}}$-ordered parton shower) or \herwig (v.\ 7.2.2)~\cite{Bahr:2008pv,Bellm:2015jjp,Bellm:2019zci} (default angular-ordered parton shower).
Both the NLO+PS predictions shown here are at the hadron level, with stable heavy hadrons.
It was noticed in Ref.~\cite{Gauld:2023zlv} that Multiple Particle Interaction (MPI) effects are very sizeable in the forward region and for charmed observables. Hence, we will show the impact of MPI effects on NLO+PS predictions, by turning on and off the MPI flag in the \pythia sample (we have checked that the impact of MPI is similar when using \herwig).

For the sake of completeness we report here the numerical setup and the fiducial cuts of Ref.~\cite{Gauld:2023zlv}.
We use the \texttt{PDF4LHC21} Monte Carlo PDF set~\cite{PDF4LHCWorkingGroup:2022cjn}.
We work in the electroweak $G_{\mu}$ scheme, using a complex mass scheme for the unstable internal particles, with the following values for the input parameters: $M_{\mathrm{Z}}^\mathrm{os} = \SI{91.1876}{\GeV}$, $\Gamma_{\mathrm{Z}}^\mathrm{os} =
\SI{2.4952}{\GeV}$, $M_{\mathrm{W}}^\mathrm{os} = \SI{80.379}{\GeV}$, $\Gamma_{\mathrm{W}}^\mathrm{os} = \SI{2.085}{\GeV}$, and $G_\mu = \SI{1.1663787e-5}{\GeV^{-2}}$.

We retain events featuring jets with $\SI{20}{\GeV} < p_{t,j} < \SI{100}{\GeV}$ and $2.2 < \eta_{j} < 4.2$ (note the upper cut on the transverse momentum of jets). Charged leptons are required to have $p_{t,\ell} > \SI{20}{\GeV}$ and $2.0 < y_{\ell} < 4.5$, with an invariant mass $M_{\ell\bar{\ell}} \in \SI[parse-numbers=false]{[60,120]}{\GeV}$. A minimum angular distance between leptons and jets is further imposed, $\Delta R(j,\ell) > 0.5$. 

To define (flavoured) jets, we adopt several IRC-safe algorithms with variation of parameters:
\begin{itemize}
\item GHS algorithm, applied to anti-\kt jets with $R=0.5$, with parameters ($\alpha = 1$, $\omega = 2$);
\item IFN algorithm, with anti-\kt distance and $R=0.5$, with parameters ($\alpha = 1$, $\omega = 2$) or ($\alpha = 2$, $\omega = 1$);
\item CMP algorithm, with $R=0.5$, with parameter $a = 0.1$ or $a = 0.25$;
\item SDF algorithm, applied to anti-\kt jets with $R=0.5$, with parameters ($\beta = 1$, $z_{\mathrm{cut}} = 0.1$), ($\beta = 2$, $z_{\mathrm{cut}} = 0.1$) or ($\beta = 2$, $z_{\mathrm{cut}} = 0.05$).
\end{itemize}
In the case of NLO+PS predictions, we further consider two IRC unsafe variants. We first reconstruct jets with the flavour-agnostic anti-\kt algorithm with $R=0.5$ and in the end we assign flavour in two different ways:
\begin{itemize}
\item ``anti-\kt naive'': we assign flavour to jets according to their flavoured constituents (here we adopt the mod2 recombination).
\item ``cone labelling'' (matching the definition of charm jets used in Ref.~\cite{LHCb-PAPER-2021-029}): we check for the presence of at least one flavoured hadron satisfying both $p_{t,c{~\text{hadron}}} > \SI{5}{\GeV}$ and $\Delta R(j,c{~\text{hadron}}) < 0.5$.
\end{itemize}
In addition, events are only retained if the flavour-labelled $c$ jet is the jet carrying the largest transverse momentum of reconstructed jets passing the selection cuts.

For each set of plots, we will consider two observables: the transverse momentum of the leading-$p_{t}$ charm jet, $p_{t,c}$, and the rapidity of the Z-boson, $y^{\mathrm{Z}}$.
%
%
We will further denote as ``NLO+PY8'' the NLO+PS results with \pythia, and ``NLO+HW7'' the NLO+PS results with \herwig.

\begin{figure}[t]
  \centering
  \includegraphics[width=0.32\linewidth,page=1]{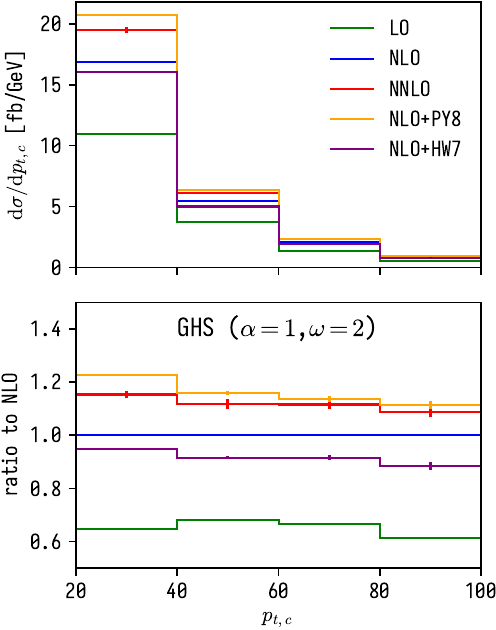}\,
  \includegraphics[width=0.32\linewidth,page=3]{figures/nnlo_ppzc_LHCb/ZcLHCb-crop.pdf}\,
  \includegraphics[width=0.32\linewidth,page=4]{figures/nnlo_ppzc_LHCb/ZcLHCb-crop.pdf}
  \caption{Comparison of predictions at fixed-order and NLO+PS for the transverse momentum of the leading-$p_{t}$ charm jet, $p_{t,c}$.}
  \label{fig:zcA1}
\end{figure}

\begin{figure}[t]
  \centering
  \includegraphics[width=0.32\linewidth,page=19]{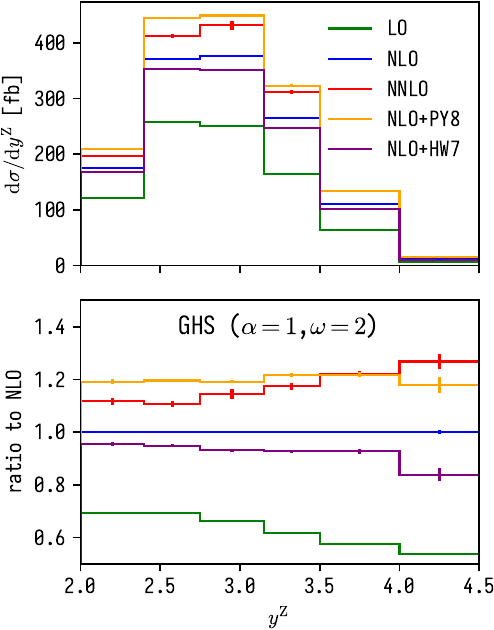}\,
  \includegraphics[width=0.32\linewidth,page=21]{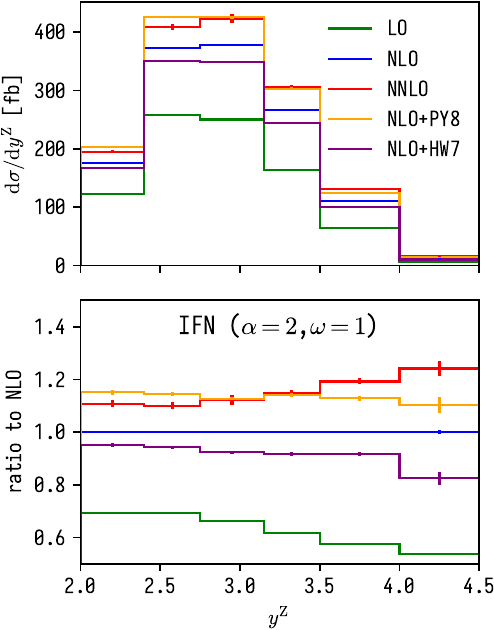}\,
  \includegraphics[width=0.32\linewidth,page=22]{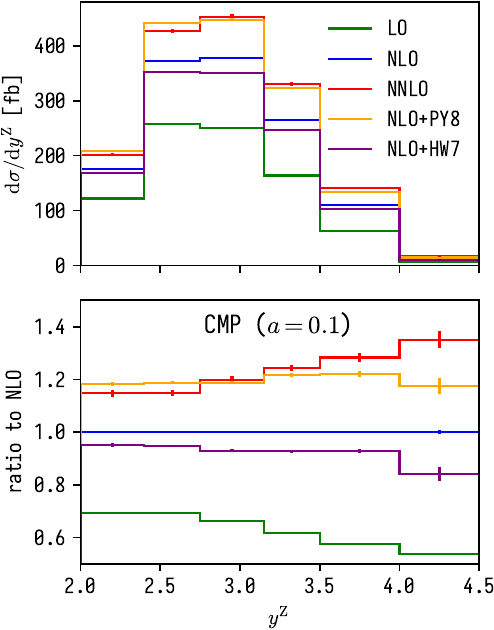}
  \caption{Comparison of predictions at fixed-order and NLO+PS for the rapidity of the Z-boson, $y^{\mathrm{Z}}$.}
  \label{fig:zcA3}
\end{figure}

In \cref{fig:zcA1,fig:zcA3} we study the relative agreement between fixed-order predictions and NLO+PS predictions for each flavoured jet algorithm.
The algorithms or variations of parameters not shown in \cref{fig:zcA1,fig:zcA3} are very similar to the ones considered.
For both observables, we see that NLO+PY8 predictions are very similar to the fixed-order NNLO results, both consistently above NLO by 20\%. NLO+HW7 is instead similar or slightly smaller than the NLO result. 
This may due to the fact that \pt-ordered parton showers are better able to populate the forward region starting from hard events in the central region compared to angular-ordered parton showers.
The same behaviour seems to hold for all of the jet algorithms. 

\begin{figure}[t]
  \centering
  \includegraphics[width=0.49\linewidth,page=28]{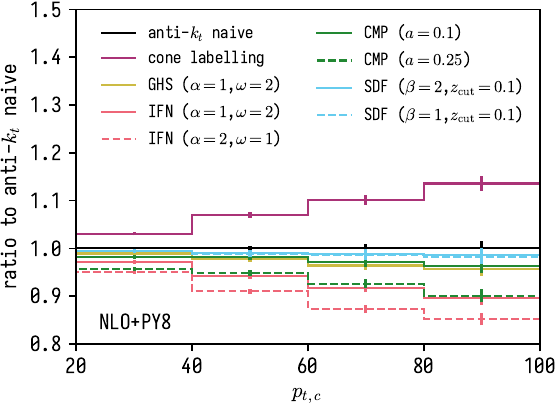}
  \includegraphics[width=0.49\linewidth,page=29]{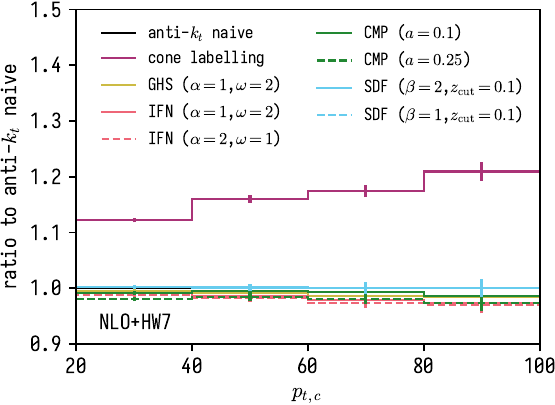} \\
  \includegraphics[width=0.49\linewidth,page=30]{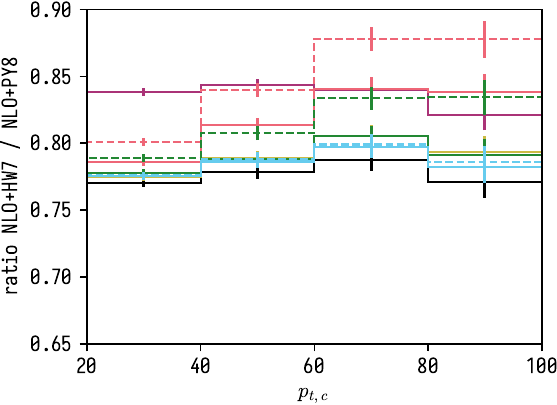}
  \includegraphics[width=0.49\linewidth,page=31]{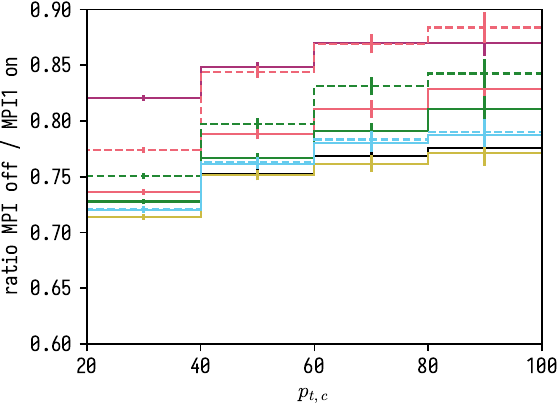}
  \caption{NLO+PS predictions with \pythia (upper right) and \herwig (upper left)
    for the transverse momentum of the leading-$p_{t}$ charm jet, $p_{t,c}$.
    For each algorithm, we further compare \pythia and \herwig (lower left)
    and we study the size of MPI effects within \pythia (lower right).}
  \label{fig:zcB1}
\end{figure}

\begin{figure}[t]
  \centering
  \includegraphics[width=0.49\linewidth,page=36]{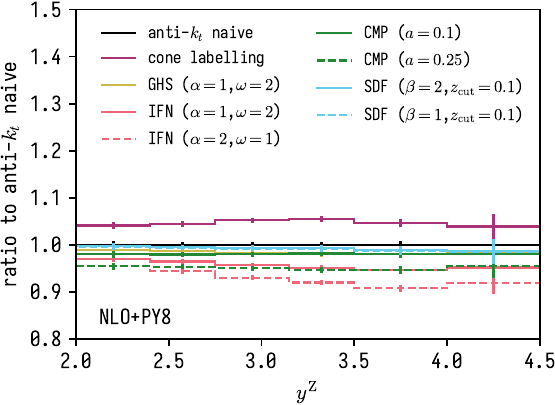}
  \includegraphics[width=0.49\linewidth,page=37]{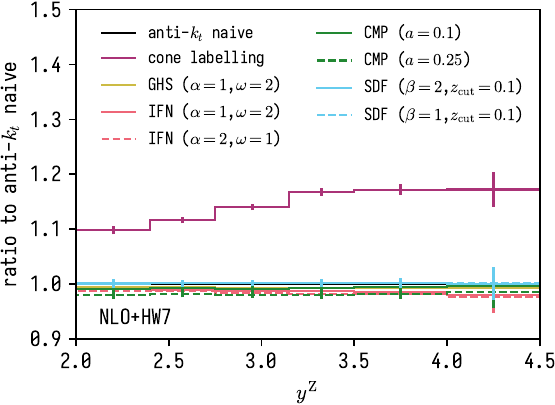} \\
  \includegraphics[width=0.49\linewidth,page=38]{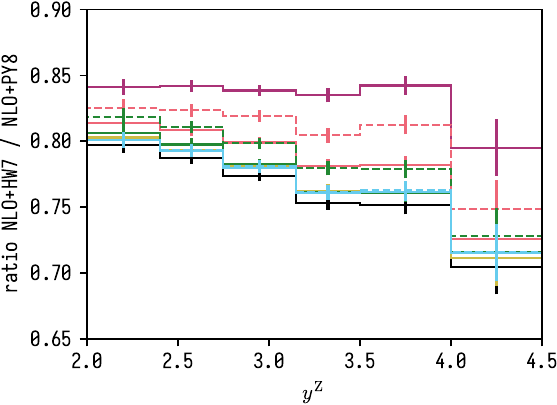}
  \includegraphics[width=0.49\linewidth,page=39]{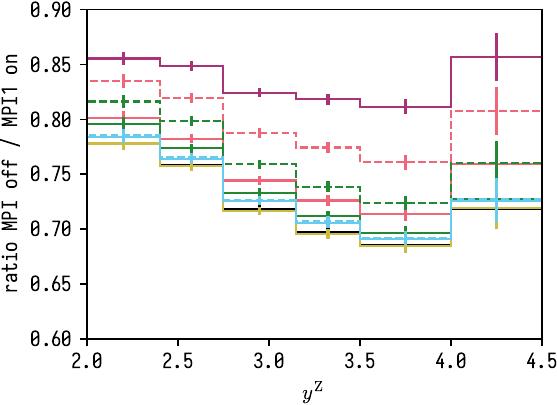}
  \caption{NLO+PS predictions with \pythia (upper right) and \herwig (upper left)
    for the rapidity of the Z-boson, $y^{\mathrm{Z}}$.
    For each algorithm, we further compare \pythia and \herwig (lower left)
    and we study the size of MPI effects within \pythia (lower right).}
  \label{fig:zcB3}
\end{figure}

In \cref{fig:zcB1,fig:zcB3} we compare flavoured jet algorithms, both IRC-safe and IRC-unsafe variants, at the NLO+PS level. The ``cone labelling'' stands as an outlier, both for NLO+PY8 and NLO+HW7: compared to ``anti-$k_t$ naive'' (used as reference in the upper plots of \cref{fig:zcB1,fig:zcB3}), ``cone labelling'' is systematically higher than predictions with other algorithms, ranging from 5\% to 10\% above ``anti-$k_t$ naive'' in NLO+PY8 and up to 20\% above ``anti-$k_t$ naive'' in NLO+HW7. 
This is likely due to the difference between the IRC-unsafe requirement of having at least one flavoured particle inside a jet to declare it as flavoured (as done in ``cone labelling'') and the IRC-safe mod2 prescription for the sum of flavours within a jet (as done in all of the other options).
Focusing on the NLO+PY8 sample, we see that SDF and GHS are similar to ``anti-$k_t$ naive'' prescription, whereas IFN is systematically lower.
This is line with the other studies of this paper: IFN seems to be more aggressive in removing $c\bar{c}$ pairs from the event.
The behaviour in the NLO+HW7 sample is similar, though with reduced differences between flavoured algorithms (at most 3\%).
If we compare the predictions for NLO+PY8 and NLO+HW7 for a fixed algorithm, we observe a ratio close to 0.8, in agreement with the behaviour of the NLO+PS curves in \cref{fig:zcB1,fig:zcB3}.
Finally, we study the effect of MPI in the NLO+PY8 sample, which turns out to be sizeable, from 10\% to 30\% according to the algorithm and the observable considered.
If we compare the different algorithms, we see that the algorithms more aggressive in removing flavour from the event (for which the NLO+PY8 predictions were smaller) are also more effective in reducing the MPI contribution, as already noted in Ref.~\cite{Caola:2023wpj}.
Again, the ``cone labelling'' is an outlier: in this case it seems to be the algorithm most resilient against the effects of MPI. In this case, the minimum $p_{t,c{~\text{hadron}}}$ requirement, imposed only in ``cone labelling'' and absent in the other algorithms because of IRC safety, is effective in suppressing the MPI contribution.

\begin{figure}[t]
  \centering
  \includegraphics[width=0.49\linewidth,page=41]{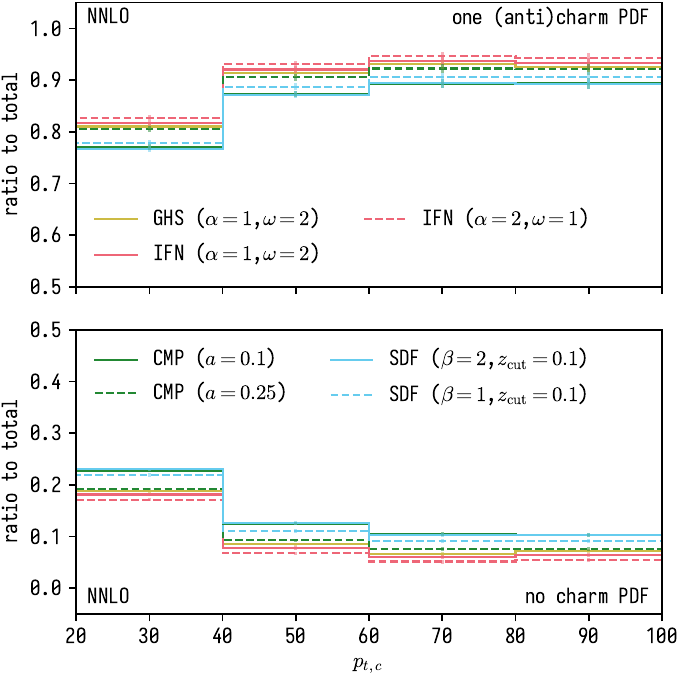}
  \includegraphics[width=0.49\linewidth,page=45]{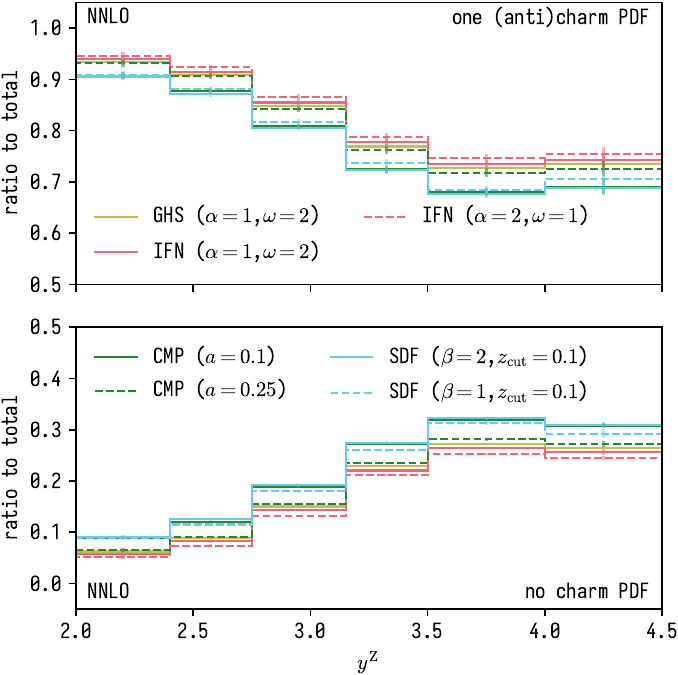}
  \caption{Channel breakdown at NNLO for the transverse momentum of the leading-$p_{t}$ charm jet, $p_{t,c}$ (left) and for the rapidity of the Z-boson, $y^{\mathrm{Z}}$ (right).}
  \label{fig:zcCC}
\end{figure}

In \cref{fig:zcCC} we study the size of individual contributions associated to different partonic luminosities for each of the algorithms. 
Our aim is to determine which algorithm is the best in preserving the dominance of Born-like events (with a single charm PDF) at higher orders.
We then consider contributions with one $c$ or $\bar{c}$ PDF, or contributions with no $c$ or $\bar{c}$ PDF (contributions with two $c$ or $\bar{c}$ PDFs are negligible). 
We see that the differences between algorithms at NNLO are at the order of a few percent, with IFN and GHS providing similar results, followed by CMP with $a=0.25$. The set of SDF variations (and CMP with $a=0.1$) seems to increase the contributions coming from channels with no charm PDF.

\begin{figure}[t]
  \centering
\includegraphics[width=0.49\linewidth,page=46]{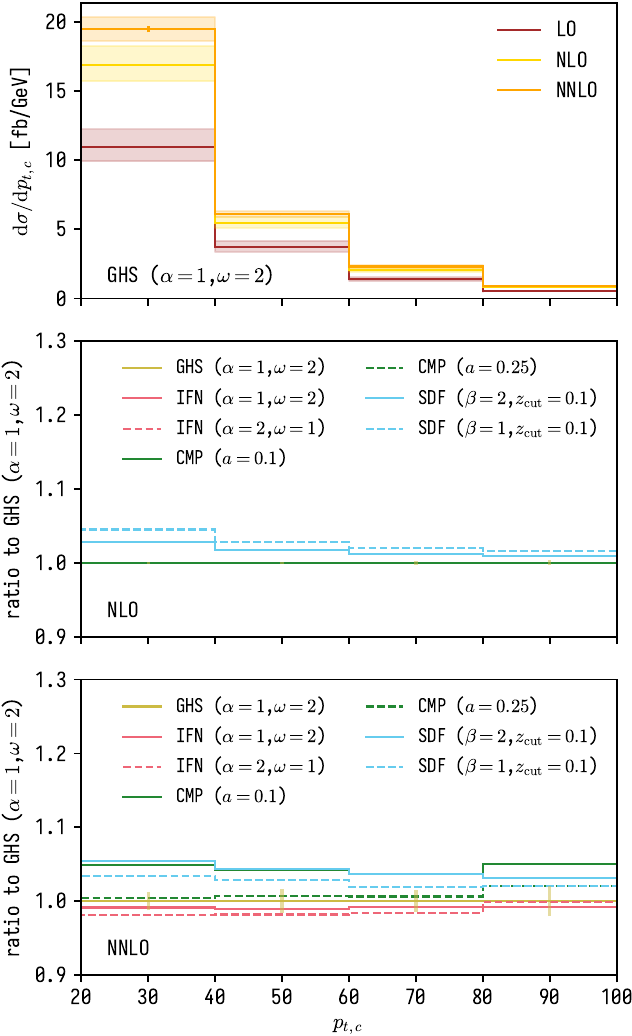}\,  \includegraphics[width=0.49\linewidth,page=48]{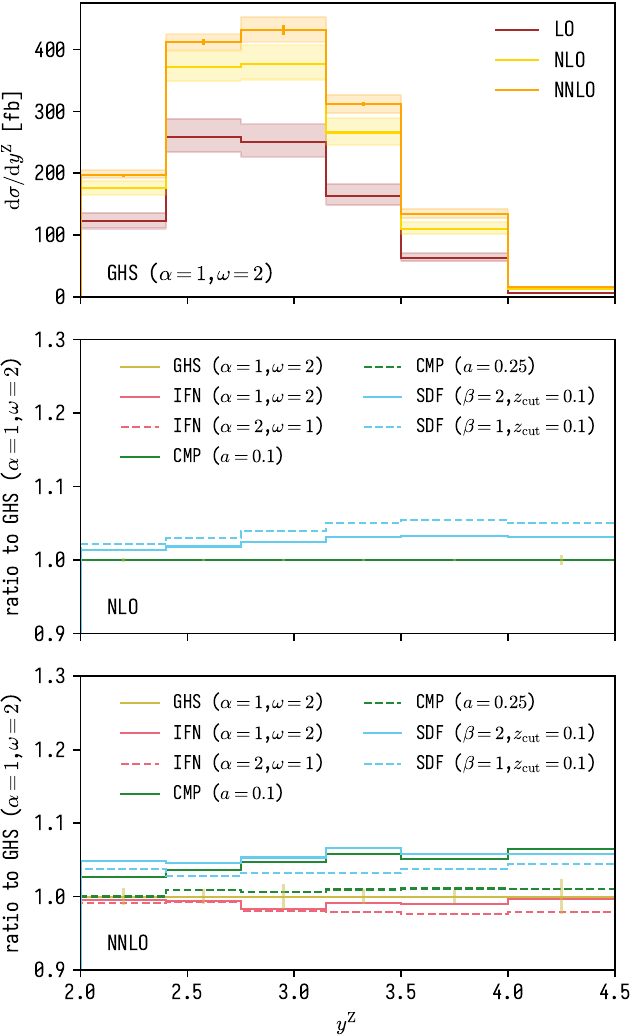}
\caption{Comparison of fixed-order predictions for the transverse momentum of the leading-$p_{t}$ charm jet, $p_{t,c}$, and for the rapidity of the Z-boson, $y^{\mathrm{Z}}$.}
  \label{fig:zcDD}
\end{figure}

Finally, in \cref{fig:zcDD}, we compare the flavoured jet algorithms at the fixed-order level.
For the two observables and for all of the algorithms, the NNLO corrections are sizeable, outside of the NLO uncertainty band. A similar behaviour was observed in Ref.~\cite{Gauld:2023zlv}.
At NLO, we note that SDF differs from the other algorithms: while all the IRC-safe algorithms provide identical results at this order (by construction), there are events that are labelled differently with SDF.
These are configurations with a $c\bar{c}$ pair from a gluon splitting, with one of the two quarks removed by the Soft Drop criterion, resulting in a flavoured jet.
The impact of this effect depends on the specific value for the SDF parameters, that in turn enters the Soft Drop criterion and drives the percentage of discarded events.
At NNLO, we note at most differences of the order of 5\% between algorithms: GHS provides results similar to IFN for both choices of parameters, whereas the value of the cross section with SDF and CMP with $a=0.1$ is a bit larger.

\FloatBarrier\clearpage
\sectionAuthors{LO+PS \texorpdfstring{$p p$}{p p} \texorpdfstring{$2 \to 2$}{2 -> 2} QCD with \texorpdfstring{$c$/$b$}{c/b} in \lhcb kinematics}{Ezra D.~Lesser and Ruide Xu}\label{app:lhcb}

Many recent experimental studies of heavy-flavour-tagged (HF) jet substructure utilise jets which contain a particular heavy-flavour hadron explicitly reconstructed from its individual decay daughters~\cite{ALICE_review_2022, LHCb-PAPER-2025-009, LHCb-PAPER-2025-010}. These `exclusive' HF jet measurements cannot make any discrimination on the perturbative process which initiated the jet formation, and in principle can come from any leading-order process, including from a gluon or light-quark-initiated jet, where the HF is created from a gluon splitting ($g \rightarrow q \bar{q}$). Performing a study of HF jet tagging in a simulated MC sample of events in which all leading-order processes are allowed to contribute therefore offers valuable insight into how these recent HF jet flavour tagging algorithms might perform when used in similar experimental conditions.

In order to study this effect, a MC simulation was generated using \pythiav{8.244}~\cite{Sjostrand:2014zea} configured with the default Monash 2013 tune and \texttt{HardQCD:all = on}, which allows all hard QCD \texorpdfstring{$2 \to 2$}{2 -> 2} processes to contribute. Approximate \lhcb kinematics are chosen for this study. Since the production of HF particles is rare relative to gluons and light flavour in this configuration, optimisations were implemented to reduce the running time. First, the simulation was stopped after the parton shower, and the quark of interest (either $c$ or $b$, depending on the flavour of the final-state hadron of interest) was required to be among the list of particles. The quark is also required to be within one unit of pseudorapidity from the final hadron-level \lhcb kinematic requirements, allowing for some kinematic smearing between parton and hadron levels. If a suitable HF quark was not found in the event after the parton shower, the next event is generated. If a suitable HF quark is found, the event is allowed to hadronise, and the event is used for jet reconstruction.

Simulations were performed using $c$-tagged jets via the \Dzero hadronic channel and $b$-tagged jets via the $\Bpm$ hadronic channel. All $c$ and $b$ hadrons are prevented from decaying, except for the tag hadrons, which are decayed through the $\kam \pip$ (+ c.c.) and $\jpsi \, (\rightarrow \mup \mun)\,  \kapm$ channels, respectively, and then promptly reconstructed before jet reconstruction. Following previous \lhcb jet analyses~\cite{LHCb-PAPER-2021-029, LHCb-PAPER-2016-064, LHCb-PAPER-2019-012, LHCb-PAPER-2022-013, LHCb-PAPER-2024-021}, anti-\kt jets are reconstructed using a moderate resolution parameter ($R=0.5$) with low transverse momenta (between $5 < \pTjet < \SI{100}{\GeV}$).  Results are compared to the WTA, IFN, CMP, and GHS flavour tagging algorithms; SDF was not used for these studies, due to difficulties building the plug-in libraries for use in the Python 3 + \texttt{cppyy} simulation framework~\cite{heppyy}.

\begin{figure}[tb!]
    \centering
    \includegraphics[width=0.48\linewidth]{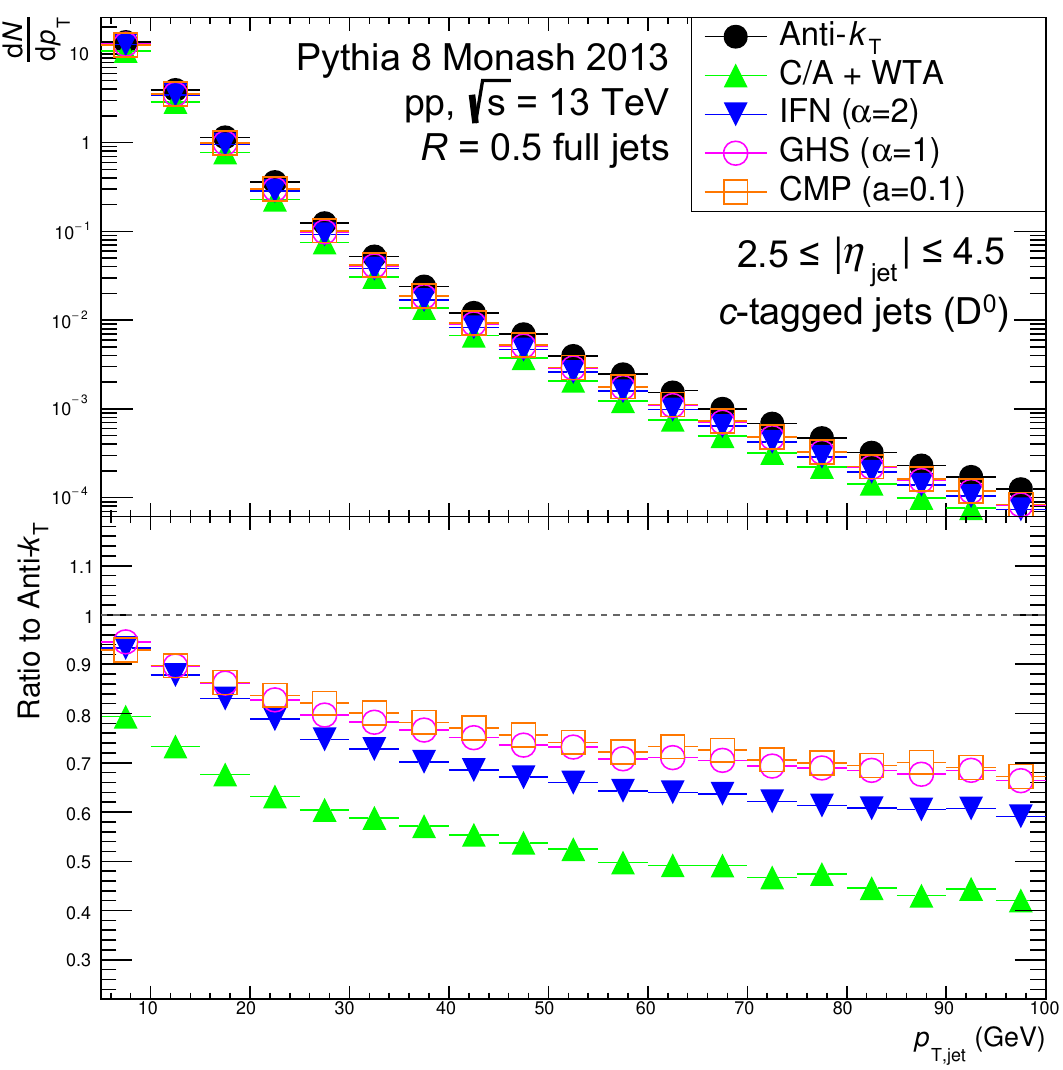}
    \includegraphics[width=0.475\linewidth]{figures/lops_pp_qcdall_LHCb/Pythia8_bjet_jetpt_R0.5.pdf}
    \caption{Differential cross section for $c$-tagged (left) and $b$-tagged (right) jets from $5 < \pTjet < \SI{100}{\GeV}$ using several flavour tagging algorithms. Simulations are produced in \pythiav{8} using all possible LO QCD $2 \rightarrow 2$ processes. The tagging rates are found to be similar for $b$ and $c$ jets, except for C/A + WTA which exhibits a higher tagging rate for $b$ jets.}
    \label{fig:QCDall_pT_xsec}
\end{figure}

The differential \pTjet cross section for $c$- and $b$-tagged jets produced using this framework is shown in \cref{fig:QCDall_pT_xsec}. 
The tagging fraction of the new flavour algorithms relative to traditional anti-\kt tagging is observed to drop rapidly as a function of \pTjet, especially for $c$-tagged jets, where gluonic production is larger (since $m_c < m_b$). Jets reclustered with the Cambridge/Aachen algorithm~\cite{C/A_alg_1, C/A_alg_2} and tagged using the Winner-Take-All axis (C/A + WTA)~\cite{Caletti:2022glq} are observed to drop most rapidly as a function of \pTjet, followed by IFN, GHS, and CMP. At these forward rapidities the tagging fraction is reduced from nearly 100\% at $\pTjet = \SI{5}{\GeV}$ to about 40-70\% ($c$ tagged) and 60-70\% ($b$ tagged) at $\pTjet = \SI{100}{\GeV}$, depending on the choice of algorithm.

\begin{figure}[tb]
    \centering
    \includegraphics[width=0.45\linewidth]{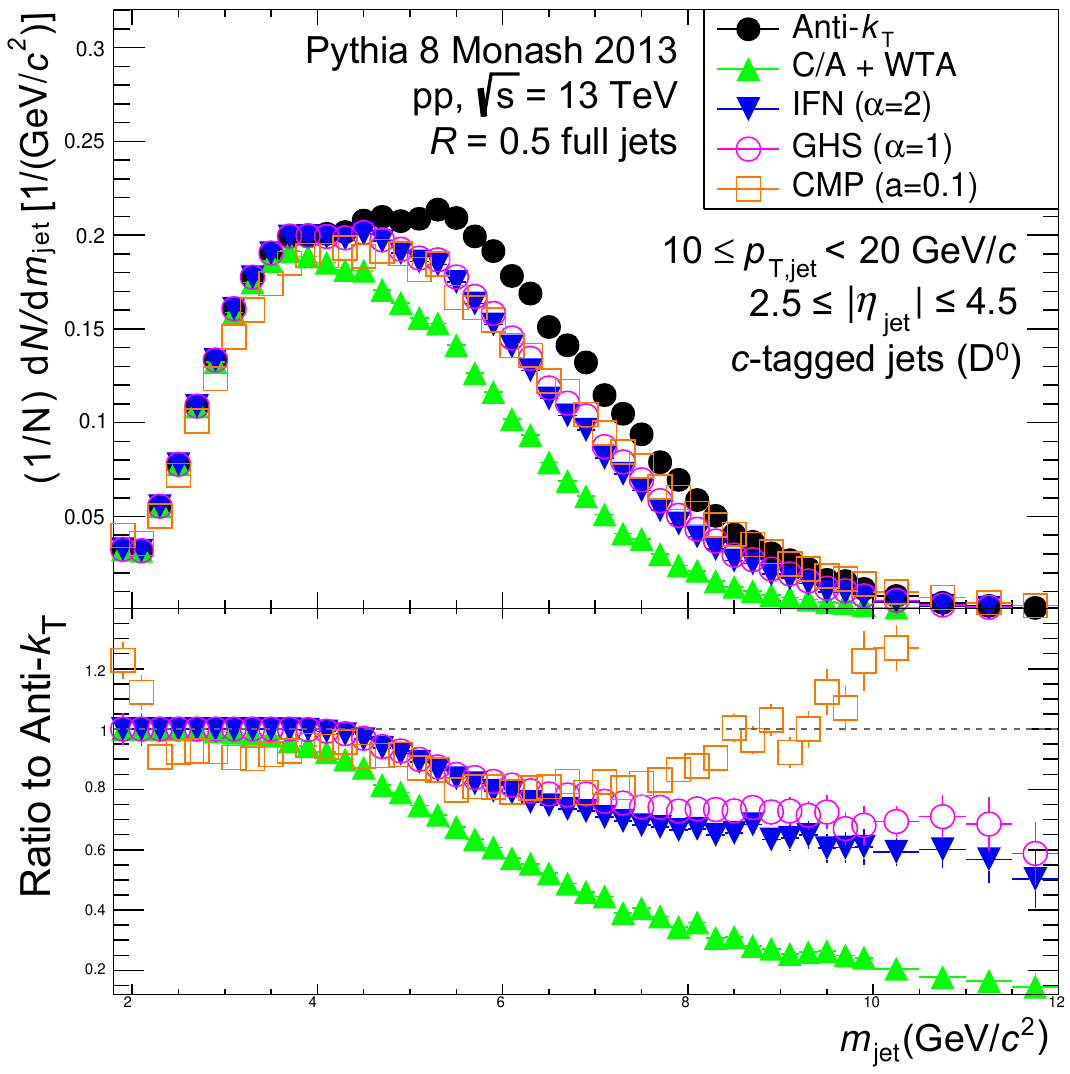}
    \includegraphics[width=0.45\linewidth]{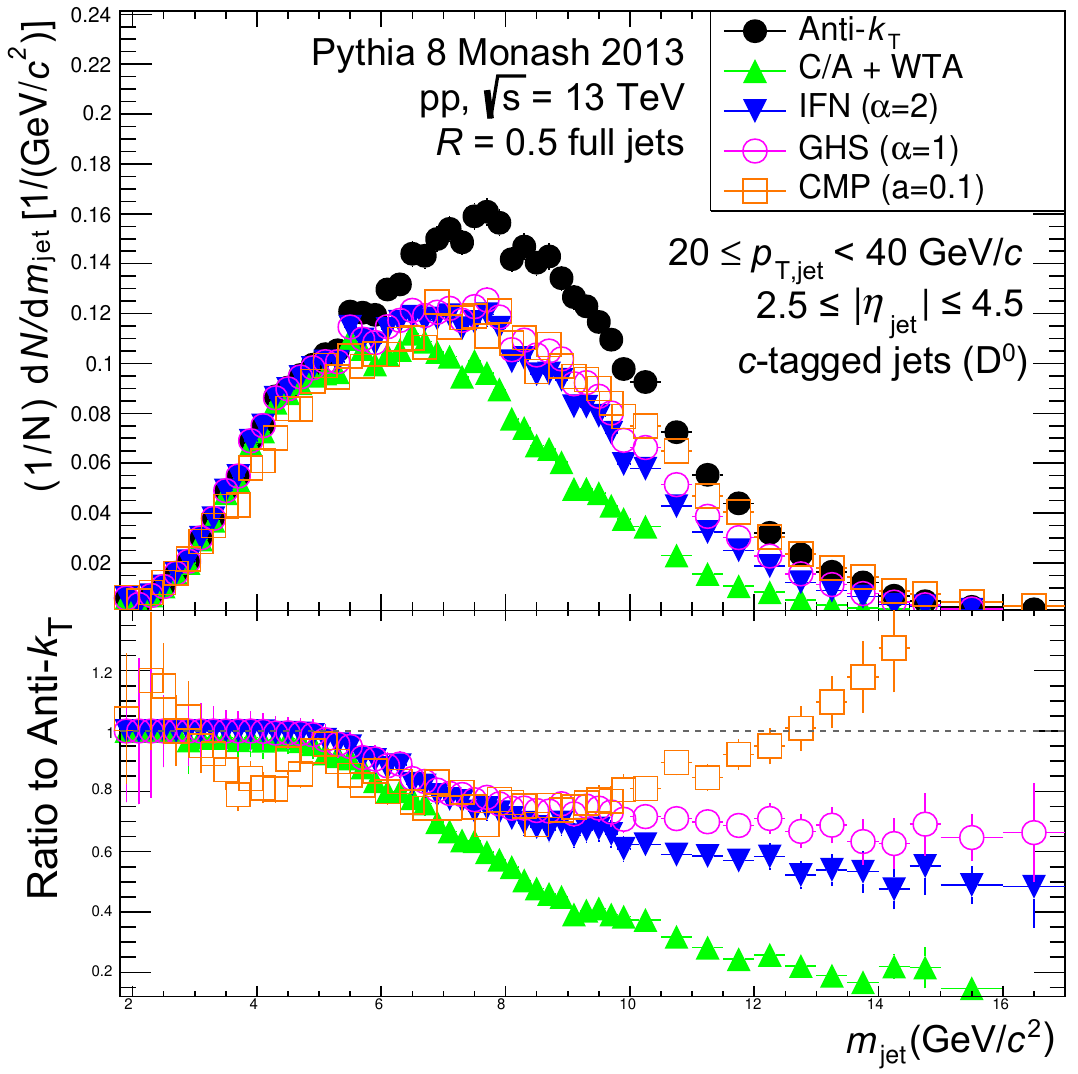}
    \caption{Distributions of the $c$-tagged jet invariant mass for jets containing a $\Dzero$ hadron at low \pTjet. Simulations are produced in \pythiav{8} using all possible LO QCD $2 \rightarrow 2$ processes. A two-peaked structure is observed, with the second \mjet peak observed around $\mjet \sim 2 \cdot m_c \sim \SI{5}{\GeV}$, corresponding to jets containing two $c$ hadrons. These jets are suppressed by the new flavour tagging algorithms (with the exception of CMP).\vspace{2em}}
    \label{fig:QCDall_c_mass}
\end{figure}

\begin{figure}
    \centering
    \includegraphics[width=0.45\linewidth]{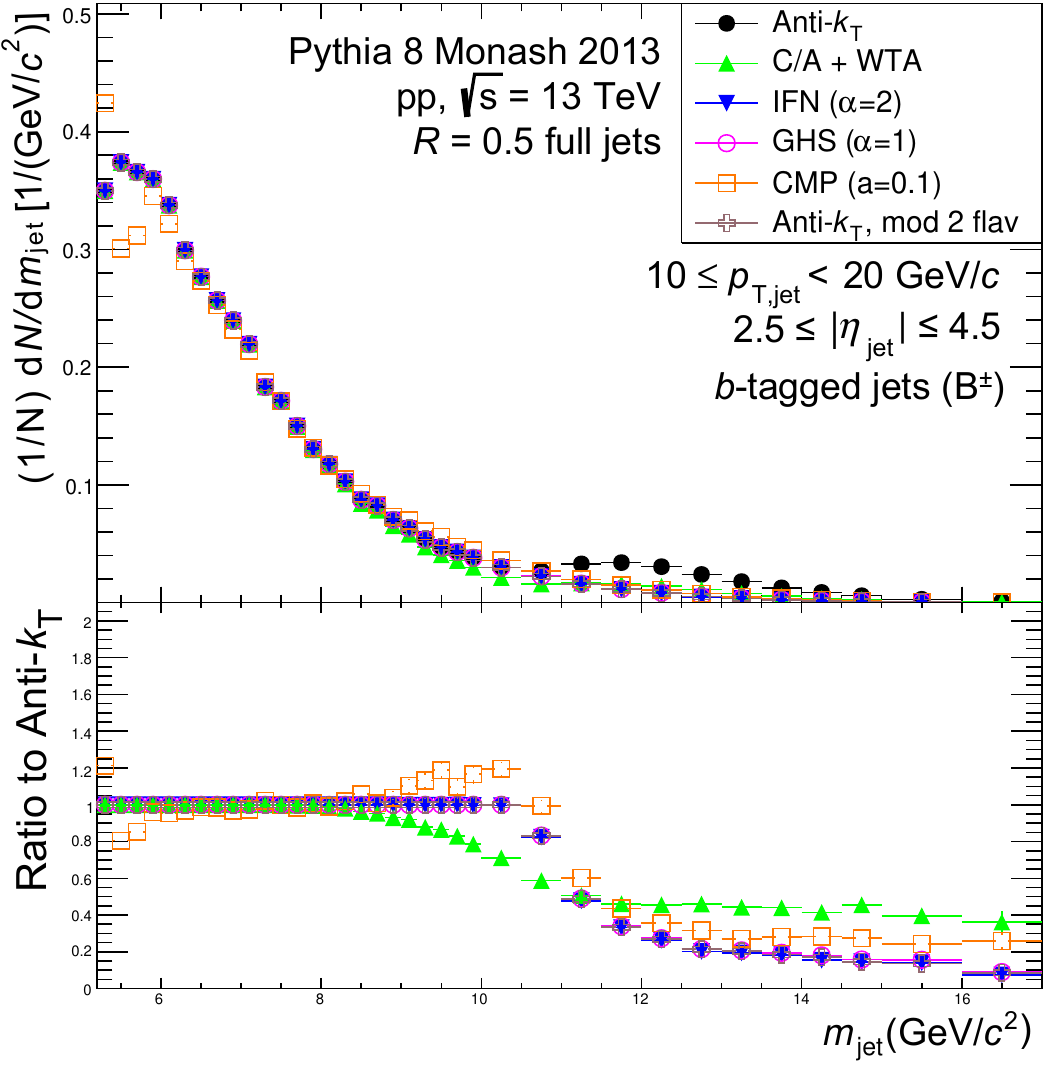}
    \includegraphics[width=0.45\linewidth]{figures/lops_pp_qcdall_LHCb/Pythia8_bjet_pt20-40_R0.5_mass.pdf}
    \caption{Distributions of the $b$-tagged jet invariant mass for jets containing a $\Bpm$ hadron at low \pTjet. Simulations are produced in \pythiav{8} using all possible LO QCD $2 \rightarrow 2$ processes. A two-peaked structure is observed, with the second \mjet peak observed around $\mjet \sim 2 \cdot m_b \sim \SI{10}{\GeV}$, corresponding to jets containing two $b$ hadrons. These jets are suppressed by the new flavour tagging algorithms.}
    \label{fig:QCDall_b_mass}
\end{figure}

In order to study the effect of this tagging fraction on the jet substructure, this sample of HF-tagged jets was used to calculate the jet invariant mass, $\mjet = \sqrt{E^2_\mathrm{jet} + p^2_\mathrm{jet}}$, where $E_\mathrm{jet}$ is the jet energy and $p_\mathrm{jet}$ is its total momentum. The \mjet distributions at low \pTjet, where the tagging fractions are largest, are shown in \cref{fig:QCDall_c_mass} for $c$-tagged jets and \cref{fig:QCDall_b_mass} for $b$-tagged jets. In both cases, a two-peaked structure is observed, with the second \mjet peak observed around $\mjet \sim 2 \cdot m_{c/b}$, corresponding to jets containing two HF hadrons; these jets can be thought of as originating from a gluon splitting process. This second peak of HF jets associated with gluon splitting are suppressed by the new flavour tagging algorithms in comparison to anti-\kt tagging, resulting in a substantial dip in the tagging fraction in the tail of the \mjet spectrum. The primary mass peak, where $m_{c/b} \lesssim \mjet \lesssim 2 \cdot m_{c/b}$ and the presence of two HF hadrons is kinematically forbidden, is relatively unaffected by the choice of flavour tagging algorithm.

\afterpage{\clearpage}

Although the overall tagging fractions depicted in \cref{fig:QCDall_pT_xsec} are relatively large at low \pTjet, the tagging fractions for the new algorithms in the \mjet tail are much lower. This is due to the fact that \mjet is more strongly biased by the jet production process than the \pTjet cross section. Since the relative number of counts in the tails of the \mjet distributions is small relative to the primary mass peak, a significant loss of counts in that region only accounts for an overall small change in the total tagging fraction. This effect is aggravated by going to higher \pTjet:
the second mass peak grows, as the phase space for gluon splitting increases, which corresponds to the overall greater reduction in the tagging fraction seen in \cref{fig:QCDall_pT_xsec}.

It is worthwhile to study whether these differences are particular to a bias in the jet invariant mass or whether the shape differences between algorithms can be reduced by a careful redefinition of the observable. For example, an imaginary substructure observable which is completely insensitive to whether or not an HF-tagged jet contains 1 or 2 HF hadrons will have an equal fraction of entries discarded across the entire spectrum by the elimination of gluon splitting HF jets. This would mean that, after renormalising the distributions to the number of flavour-tagged jets, there would be no difference in the shapes of the spectrum. Observables which are insensitive to these tagging effects could be useful for future experimental measurements, where reconstruction of 2 or more HF hadrons suffers from low combinatorial efficiencies, which can only be corrected by MC simulations.

\begin{figure}[tb!]
    \centering
    \includegraphics[width=0.48\linewidth]{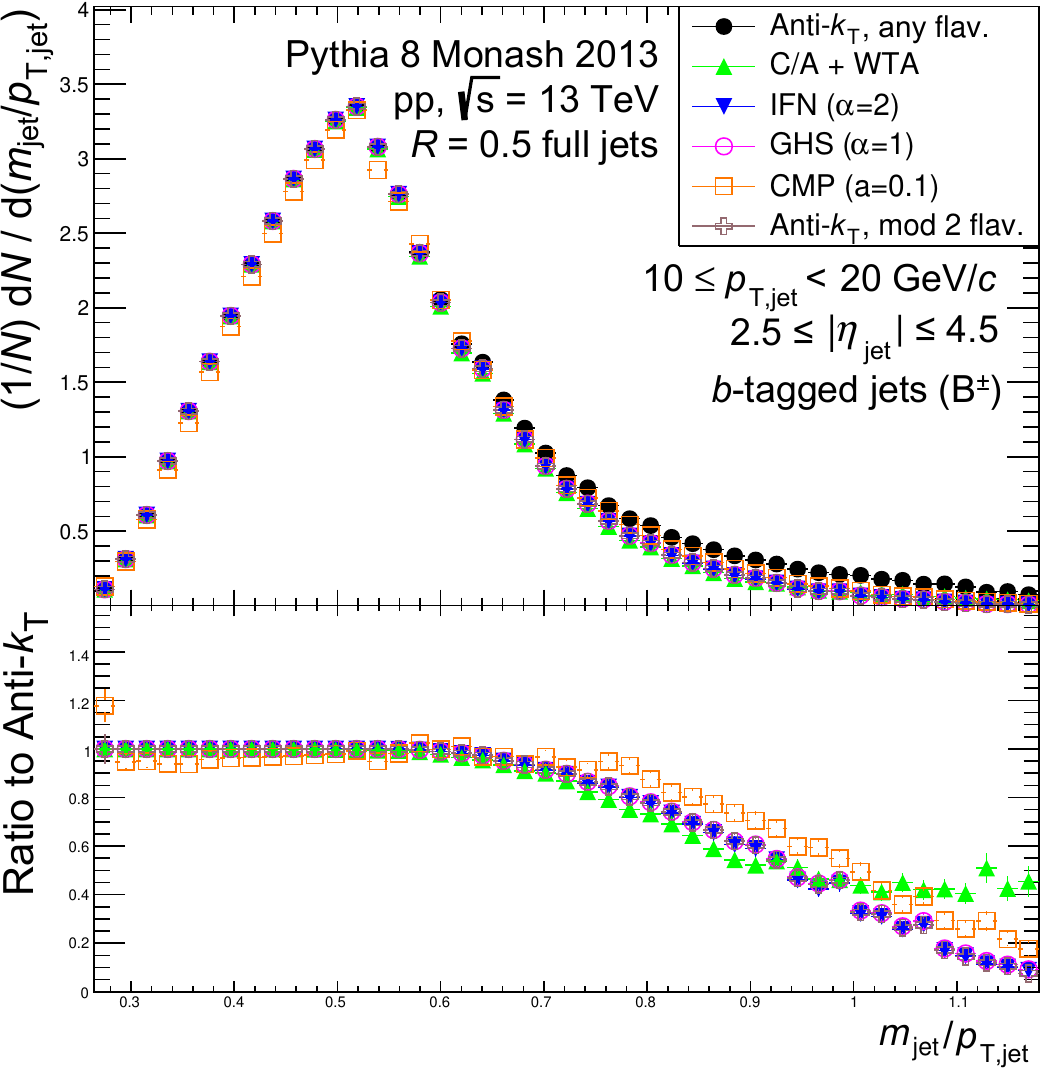}
    \includegraphics[width=0.48\linewidth]{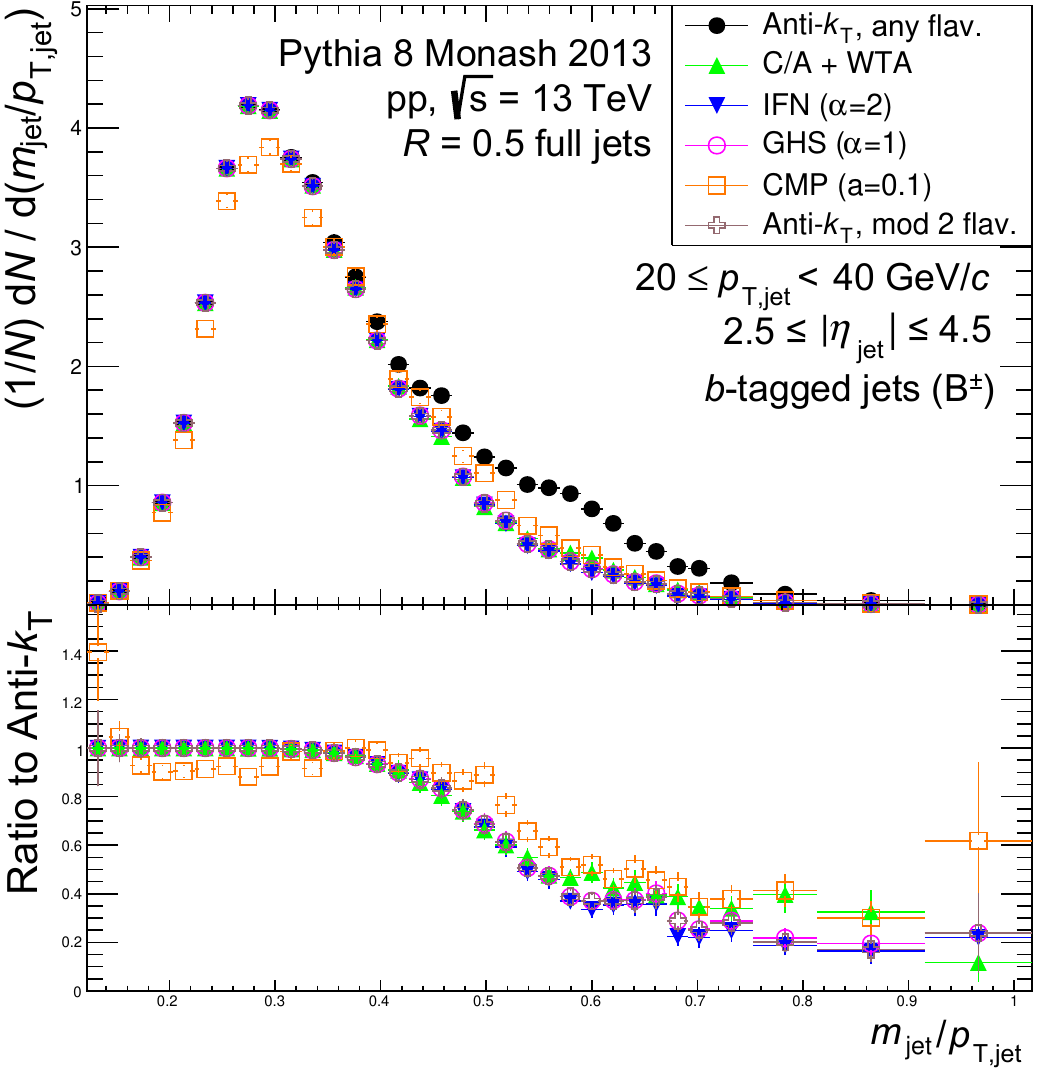}
    \caption{Distributions of the $b$-tagged \mjet/\pTjet for jets containing a $\Bpm$ hadron at low \pTjet. Simulations are produced in \pythiav{8} using all possible LO QCD $2 \rightarrow 2$ processes.}
    \label{fig:QCDall_b_mass_by_pTjet}
\end{figure}

One such observable of possible interest is the jet invariant mass relative to the jet transverse momentum, \mjet/\pTjet. While there is some correlation between \mjet and \pTjet, inducing a degree of smearing in the higher, di-HF \mjet peak such that entries appear across the spectrum could improve the overall shape resiliency. Example distributions of the $b$-tagged \mjet/\pTjet spectrum are shown in \cref{fig:QCDall_b_mass_by_pTjet}, generated using the same sample of jets. The \pTjet scaling marginally affects the shape of the distributions, but the tagging fractions are largely unaffected; however, the shape of the CMP tagging algorithm is observed to most closely approximate anti-\kt, with excesses and dips in the unscaled \mjet distribution being smeared out by the \pTjet scaling.

As observed in central rapidity, the predominant difference between standard ``any flavour'' anti-\kt and the new algorithms is the modified flavour recombination scheme, followed by the choice of algorithm. This is explicitly confirmed in the $b$-jet case, where comparisons to anti-\kt with ``mod2 flavour'' are included. The main text of this article establishes this difference as originating primarily from the exclusion of gluon splitting. However, this has not been explicitly verified for WTA flavour, which exhibits a similar tagging fraction for $b$ jets but appears more selective for $c$ jets.

The origin of jet suppression for WTA flavour is studied by producing another simulation using the same parameters as above, but allowing all heavy-flavour hadrons to decay except the one reconstructed experimentally (i.e., the \Bpm hadron). This more closely matches the ALICE and LHCb experimental methods, where heavy-flavour candidates are typically only reconstructed in a single decay channel. Any events containing more than one of the same heavy-flavour hadron are rejected due to limited experimental efficiencies. The identities of the LO ``parent'' partons, labelled 5 and 6 in the \pythia scheme, are saved. Jets are then reconstructed at hadron level, and the jet containing the \Bpm hadron is evaluated using the WTA flavour tagging algorithm.

\begin{figure}
    \centering
    \includegraphics[width=0.6\linewidth]{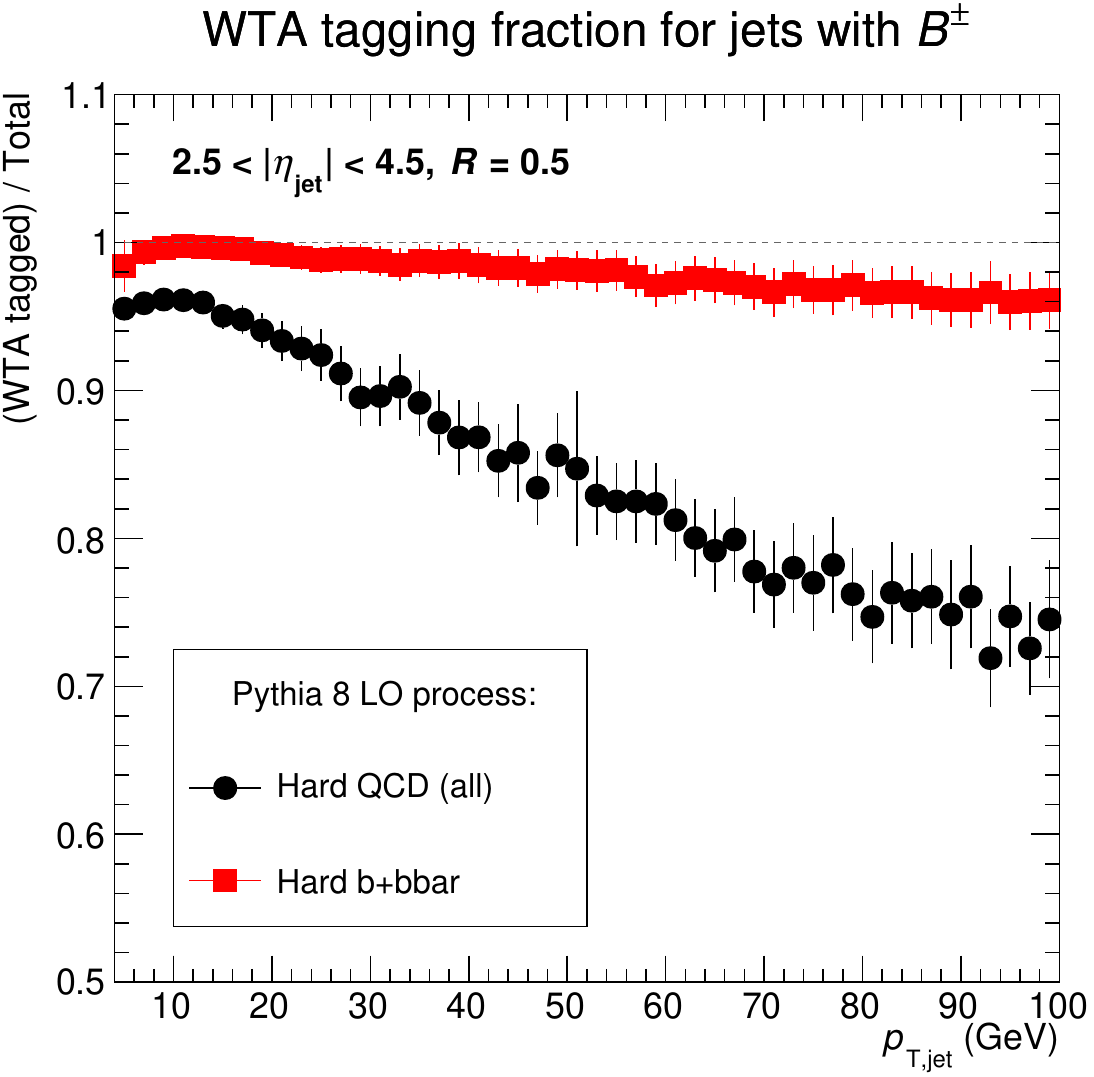}
    \caption{WTA tagging fraction for \Bpm jets in events using two different LO production mechanisms: all hard $2 \rightarrow 2$ QCD channels (black), or hard $b \bar{b}$ only (red). }
    \label{fig:WTAdist_jetpt}
\end{figure}

The fraction of jets tagged with WTA flavour as a function of \pTjet is shown in \cref{fig:WTAdist_jetpt} for  events generated with all hard $2 \rightarrow 2$ QCD processes as well as events generated purely with a hard $b\bar{b}$ pair generated at LO. At $\pTjet = \SI{100}{\GeV}$, the tagging fraction is observed to be roughly 95\% for the latter $b\bar{b}$ sample while only 70-75\% for the former inclusive sample, showing that the inclusion of $b$ jets generated via gluon splitting naturally decreases the fraction of jets positively tagged as $b$ flavoured using WTA.

\begin{figure}
    \centering
    \includegraphics[width=0.49\linewidth]{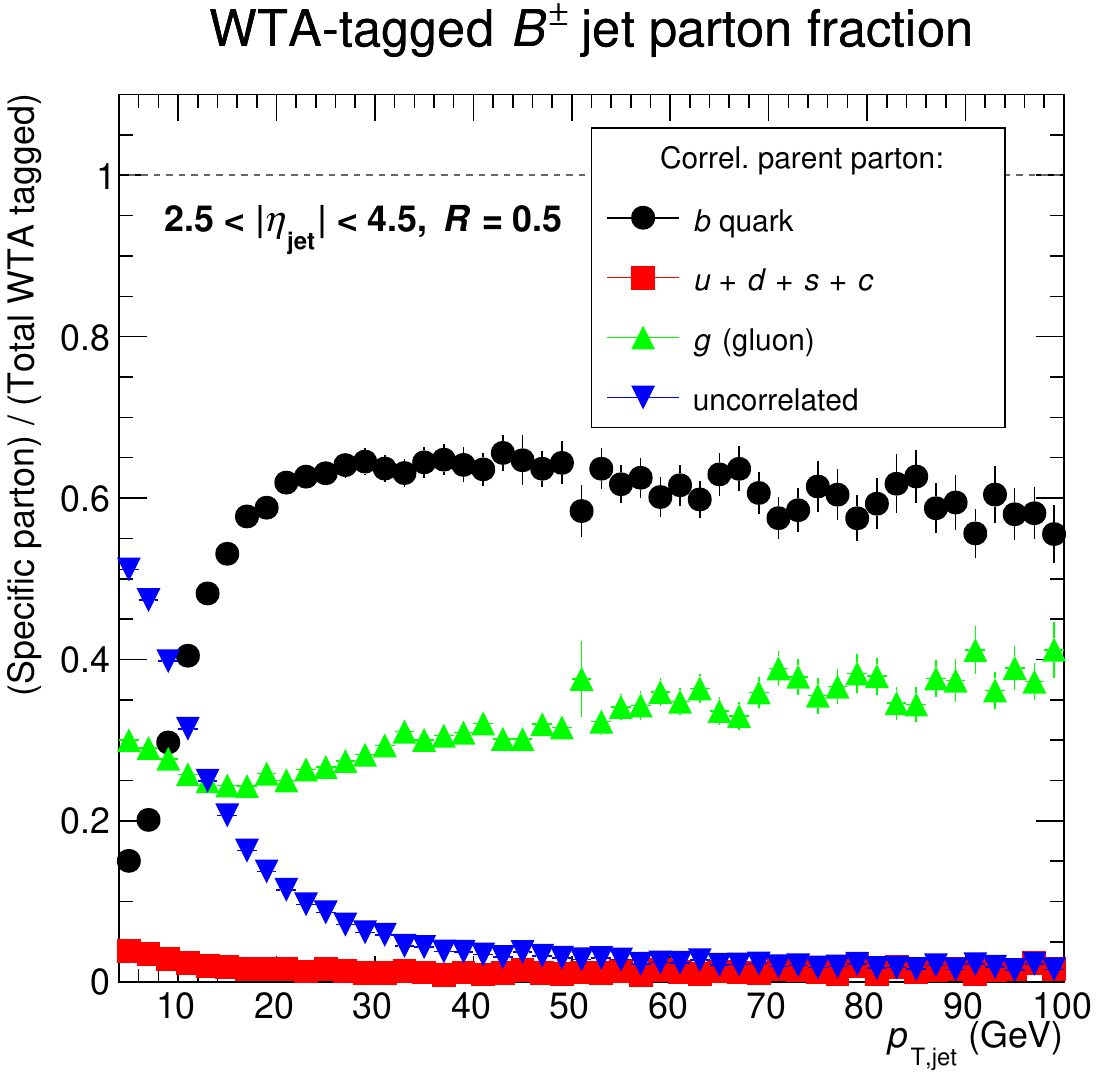}
    \includegraphics[width=0.49\linewidth]{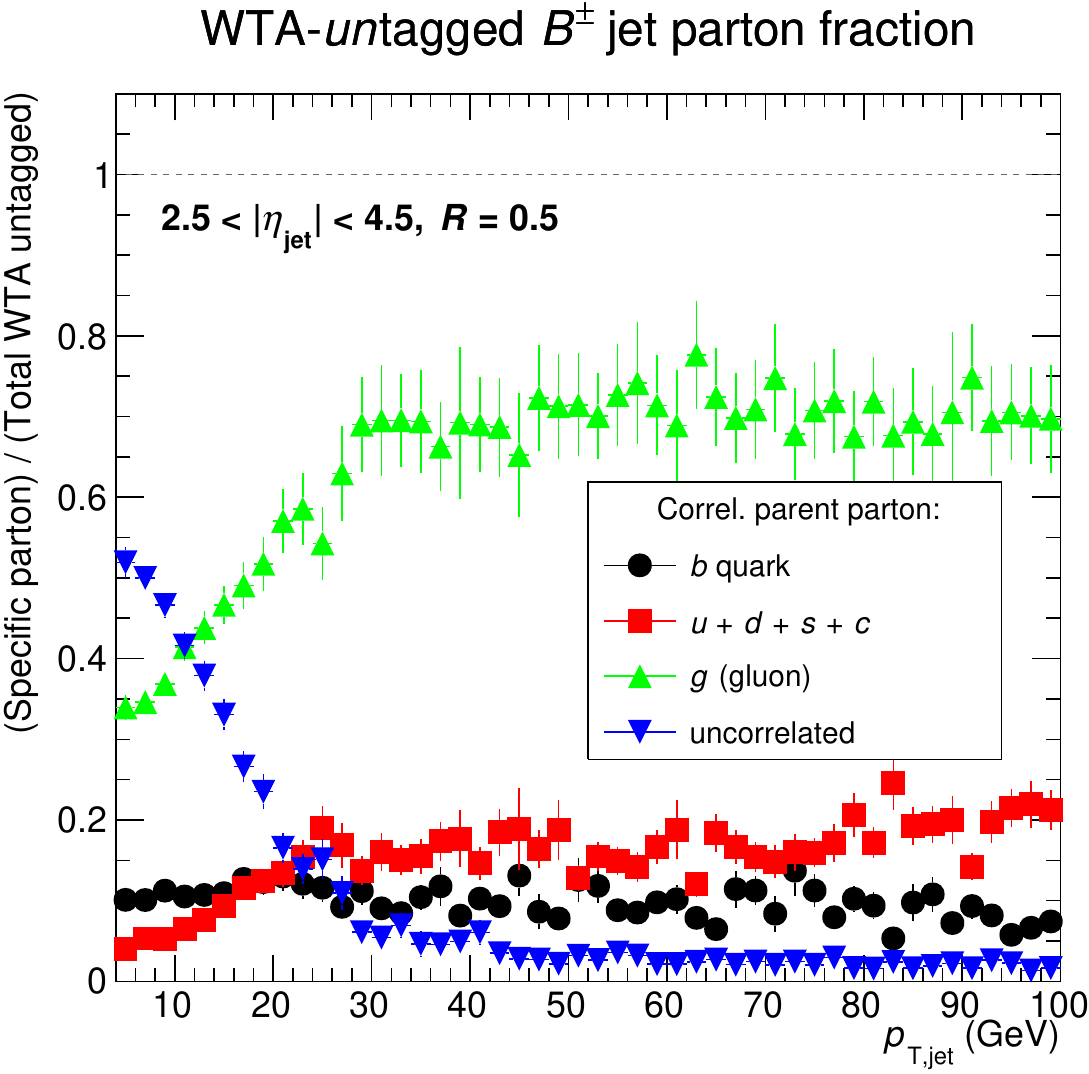}
    \caption{Correlations between LO parent partons and WTA tagged (left) or rejected (right) \Bpm jets, as evaluated in \pythiav{8}. A jet's axis is required to be within $\Delta R = 2$ of either parent parton's four-momentum to be correlated; otherwise, it is labelled as `uncorrelated'.}
    \label{fig:WTAdist_tag_untag}
\end{figure}

To understand this further, jets are matched to either parent parton by checking whether the standard jet axis (using $E$-scheme recombination) is within $\Delta R = 2$ of the momentum vector of either parent parton. If there is a match to one or both partons, the entry is stored as correlated to that parton's PID in a histogram. Otherwise, if the jet is not correlated with either parent parton within the geometrical limit, then it is saved as ``uncorrelated''. The resulting correlation fractions are shown in \cref{fig:WTAdist_tag_untag}. For jets tagged with WTA flavour, the most likely parent parton is observed to be a $b$ quark ($\sim 60$\% of the $\Bpm$-tagged jet sample), though a substantial fraction ($\sim$30-40\%) of jets also correlate to parent gluons, with this gluon-initiated fraction slowly increasing as a function of \pTjet. 

For jets that are rejected by WTA flavour tagging, the fraction of jets originating from these partons is substantially different. Most of the jets rejected by WTA flavour tagging ($\sim 70$\%) are correlated to parent gluons, whereas only a small fraction ($\sim 10$\%) are correlated with a hard $b$ quark. This means that, as with the other flavour tagging algorithms, WTA tagging primarily rejects gluon splitting and therefore increases the purity of HF quark-initiated jets in the LO+PS picture.

\FloatBarrier
\sectionAuthors{ATLAS \texorpdfstring{$Z+\text{jets}$}{Z+jets} studies} {Francesco Giuli, Rados\l aw Grabarczyk, Alberto Rescia, and Federico Sforza}\label{app:atlas}

Here we investigate the effect of the new flavoured jet algorithms in an experimental context. We aim to understand how these algorithms can affect current particle-level definitions of jet flavour, if at all, and seek to gain insight on how flavour labelling can be improved in the LHC collaborations.

Our study focuses specifically on comparing the algorithms to ATLAS and CMS labelling. The former uses cone labelling while the latter utilises ghost labelling to define $b$ jets, i.e., the requirement is that an ultra-soft $b$ hadron is clustered together with the other jet constituents.

As a caveat, it is important to note that this study is \emph{not an apples-to-apples} comparison. Such a study is impossible, as the input particles for the new algorithms and the experimental particle-level jet definitions differ in a significant way: while the algorithms require undecayed $b$ hadrons to meaningfully ascribe flavour to a jet, the experimental jets are defined with an eye towards realism. Indeed, in these cases jets are clustered using the visible decay products of the hadron. This difference inevitably leads to \emph{ab initio} kinematic discrepancies.

The study focuses on the $\Z + b$-jet or $\Z + c$-jet final states, where the \Z boson decays leptonically to either electrons or muons. Leptons are required to have a $\pt > \SI{27}{\GeV}$ due to trigger thresholds and a pseudorapidity $|\eta| < 2.5$. The dilepton pair must have a mass compatible with that of the \Z boson, within the range \SIrange{76}{106}{\GeV}.

Jets are clustered according to the ATLAS prescription for \texttt{AntiKt4TruthJets}, with final state particles within $|\eta| < 4.5$ and with a radius $R = 0.4$. The particles used as input differ as above. Jets found to have a $\pt > \SI{20}{\GeV}$ and $|\eta| < 2.5$ are labelled by the appropriate means, and final states containing at least one $b$-jet are selected. When more $b$ jets are found, the two $b$ jets with the highest \pt are selected.

Samples are simulated with \mgaNLO (v3.5.8)~\cite{Alwall:2014hca} interfaced with \pythia (v8.313)~\cite{Sjostrand:2014zea} for the parton shower. Matrix elements are generated at LO, requiring final states containing a \Z boson along with a $b\bar{b}$ pair, or a $c\bar{c}$ pair.

\begin{figure}
    \centering
    \includegraphics[width=0.49\linewidth]{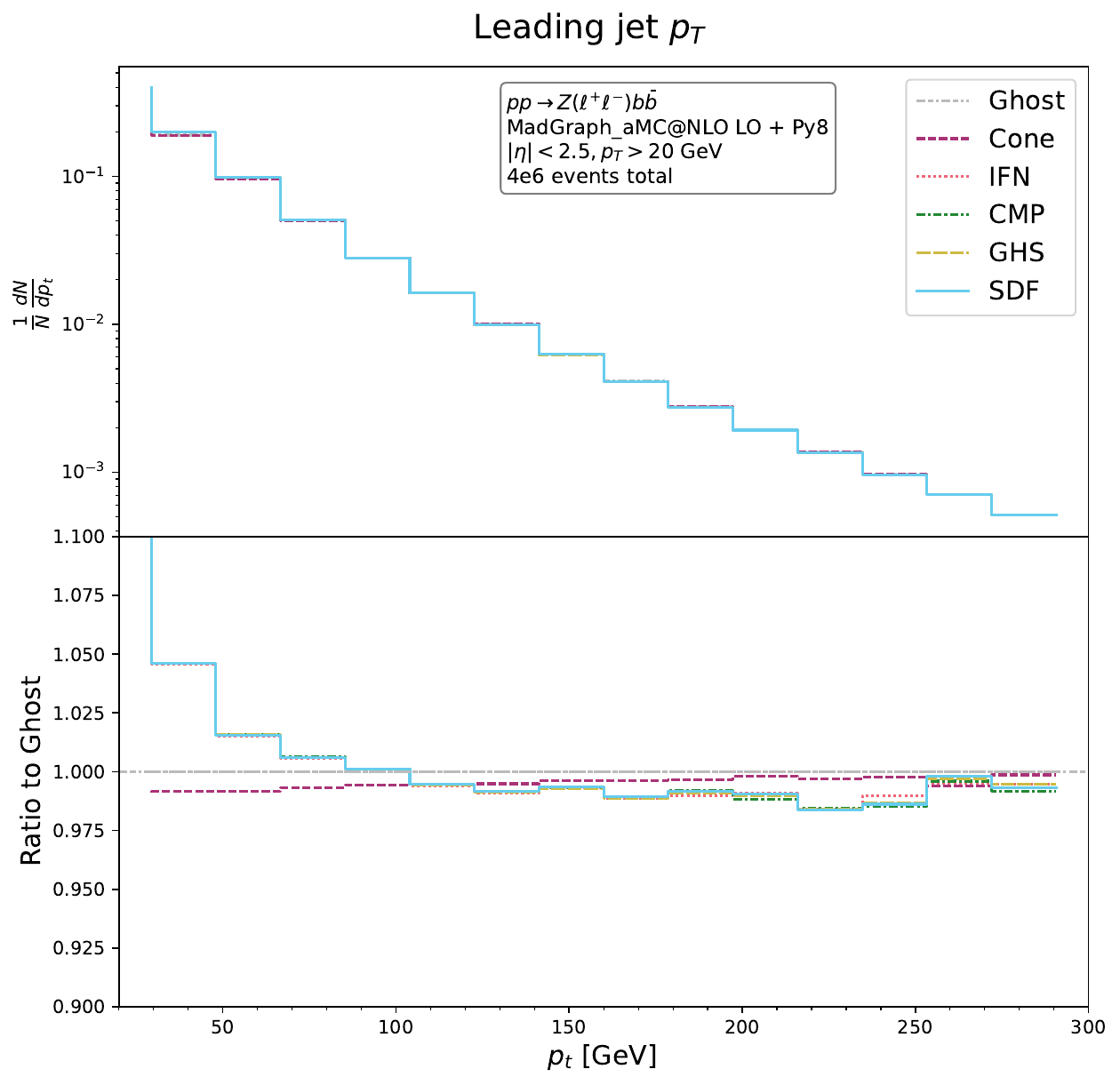}
    \includegraphics[width=0.49\linewidth]{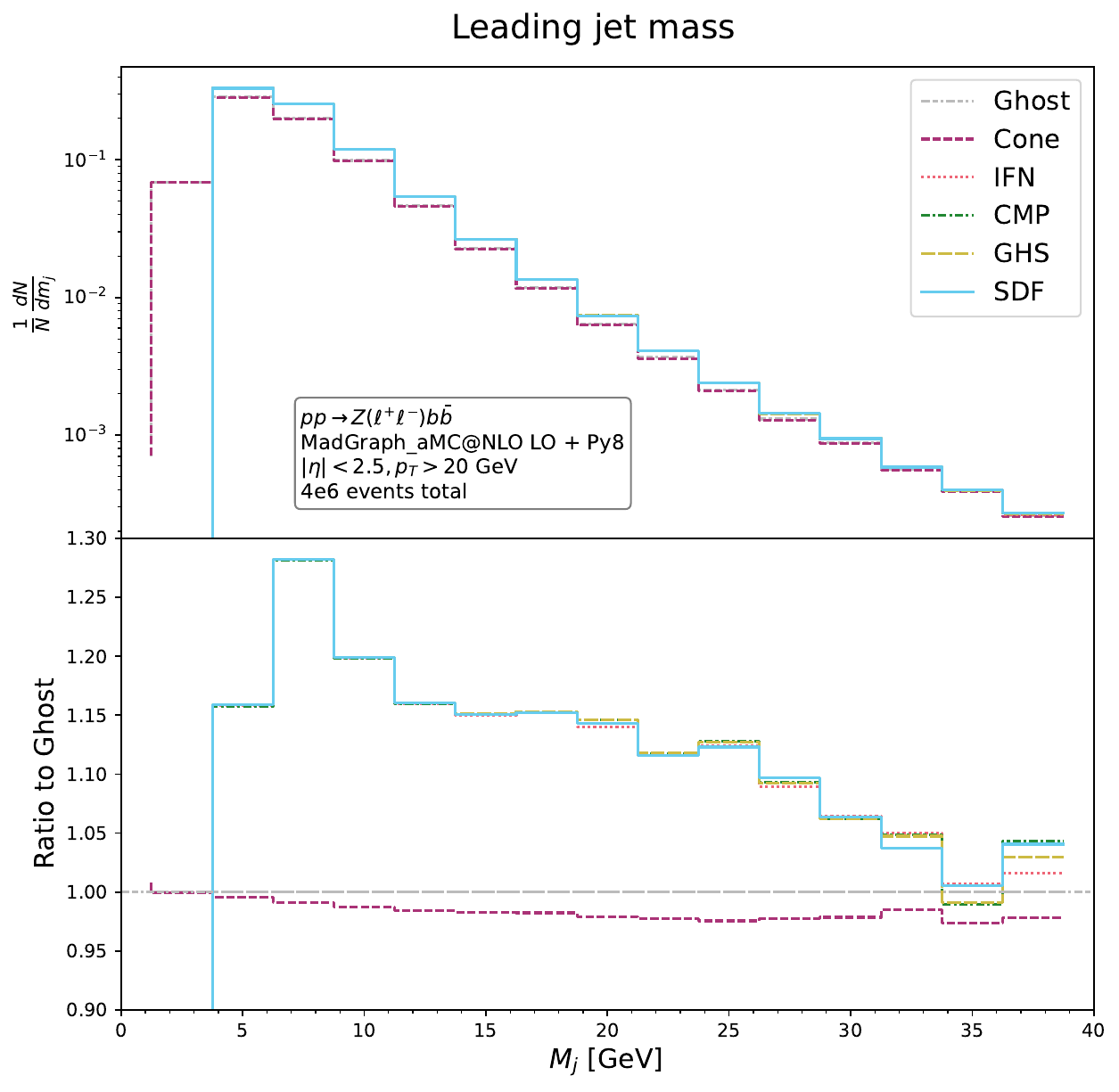} \\
    \includegraphics[width=0.49\linewidth]{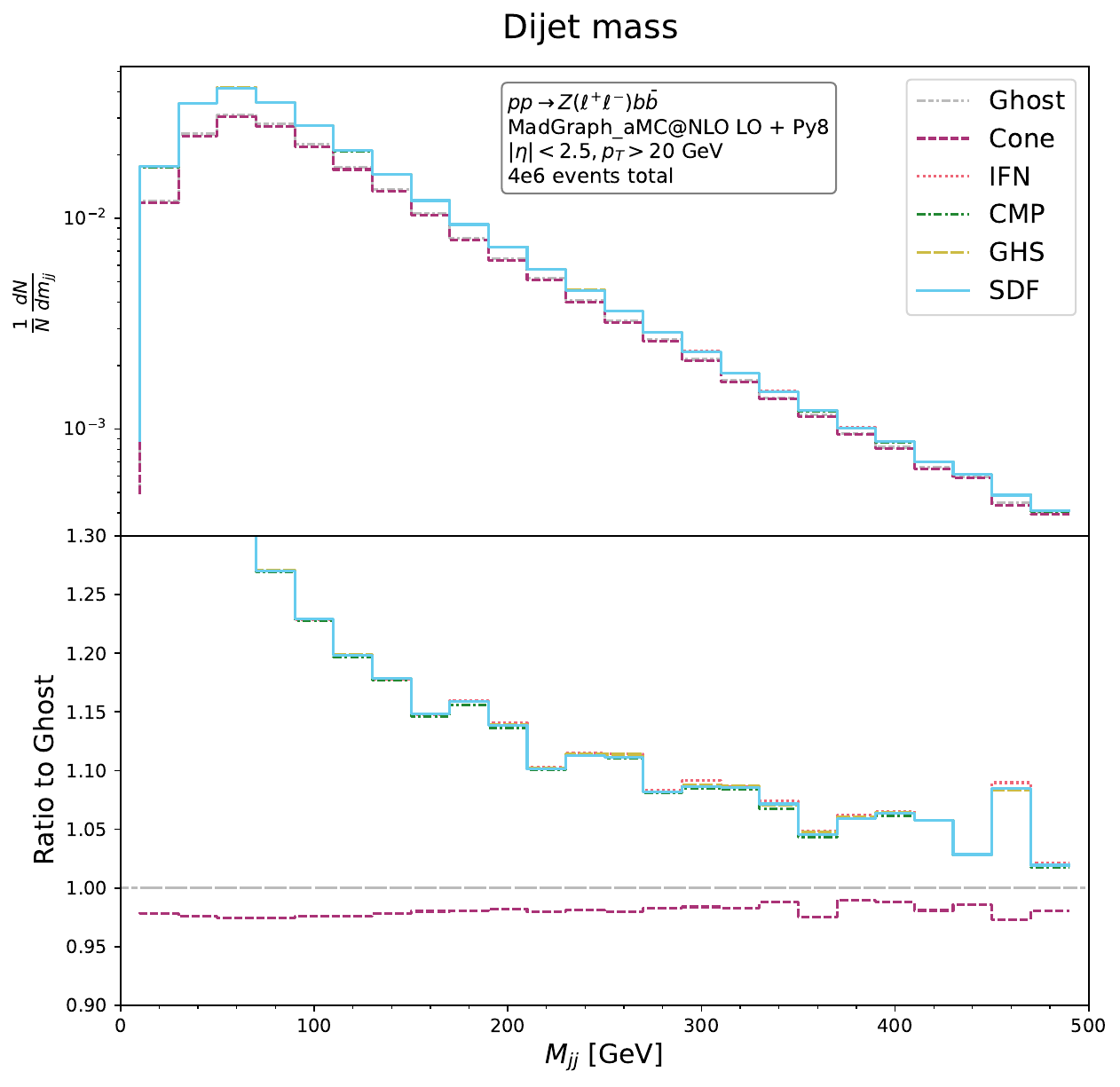}
    \caption{The \pt distribution of leading $b$ jets identified with the cone, ghost, and flavour algorithms (top left), the leading jet mass distribution (top right), and the dijet mass distribution of the two highest-\pt $b$ jets identified in the same manner (bottom). Ratios are compared to the distribution for jets clustered with ghost labelling.}
    \label{ATLAS b-jet kinematics}
\end{figure}

As can be seen in \cref{ATLAS b-jet kinematics}, the various algorithms form two distinct populations. The first is composed of cone and ghost labelled jets, which, although they differ slightly in the way they ascribe flavour labels to jets, use jets clustered including the decayed $b$ hadrons as mentioned previously. Another population includes jets clustered with the flavour algorithms. As can be seen in the ratio plots, the \pt and mass distributions obtained in this population nearly overlap, although in some regions, such as at high \pt, the differences between the flavour algorithms become apparent. 

\begin{figure}
    \centering
    \includegraphics[width=0.49\linewidth]{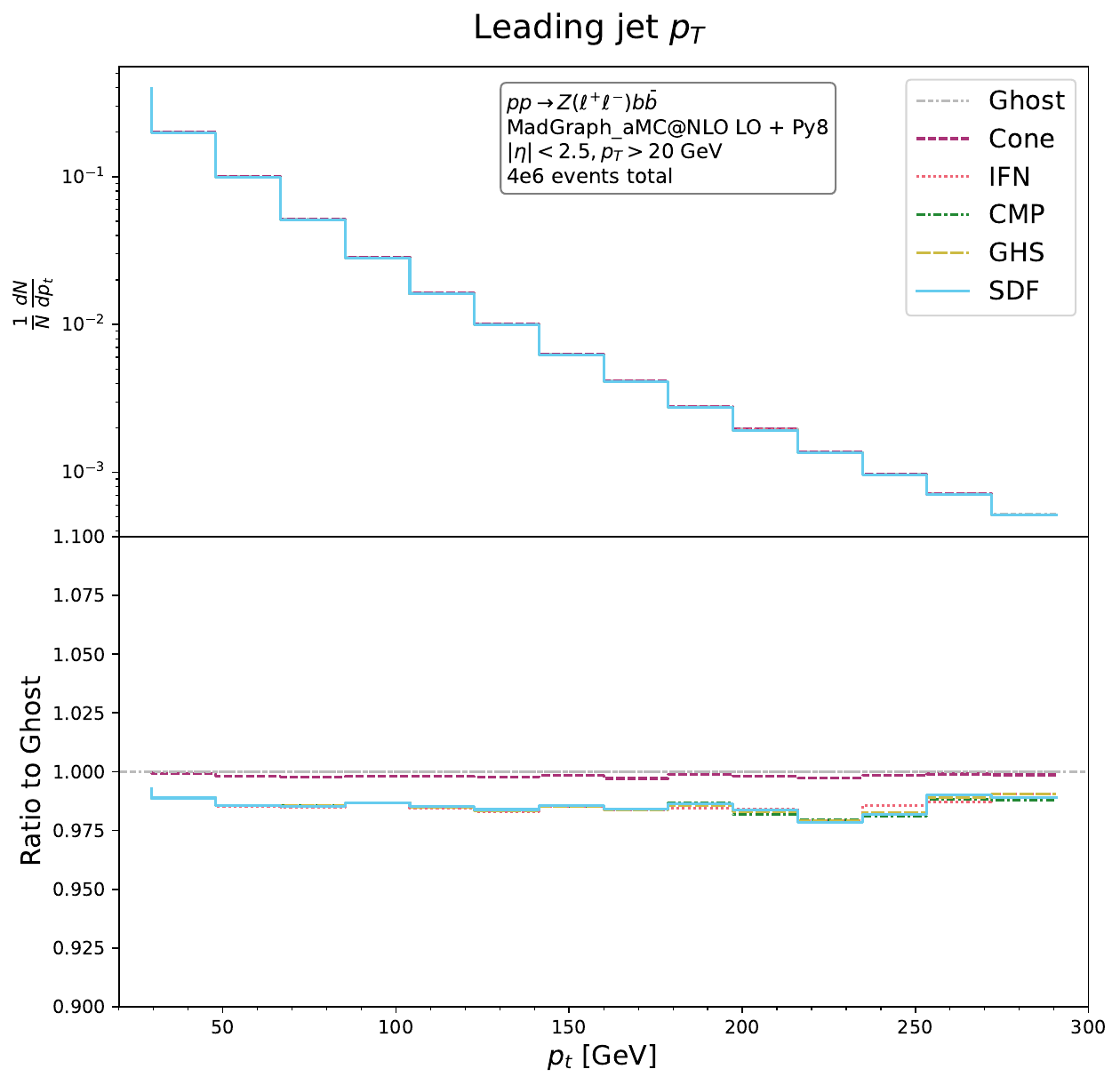}
    \includegraphics[width=0.49\linewidth]{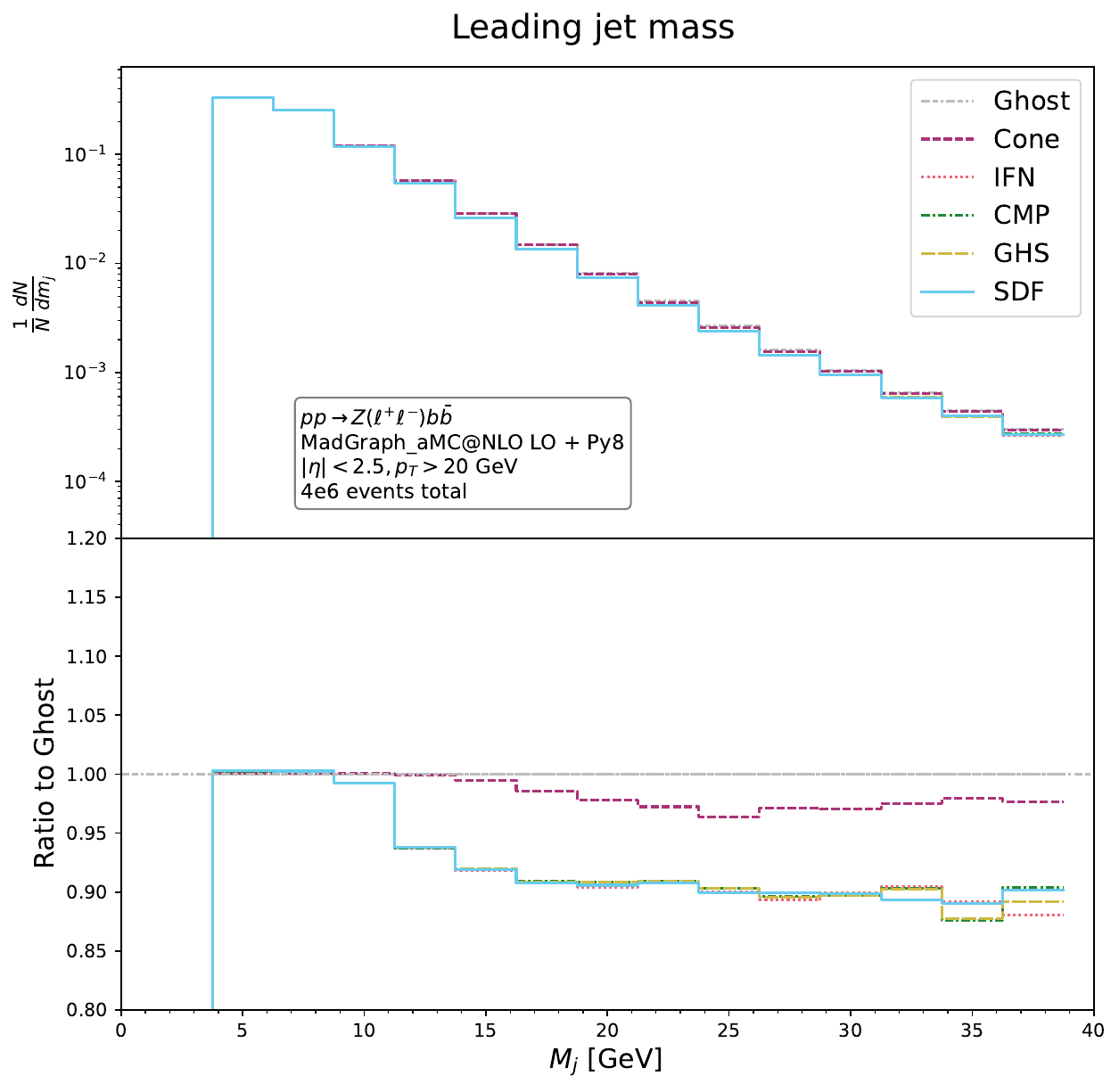} 
    \includegraphics[width=0.49\linewidth]{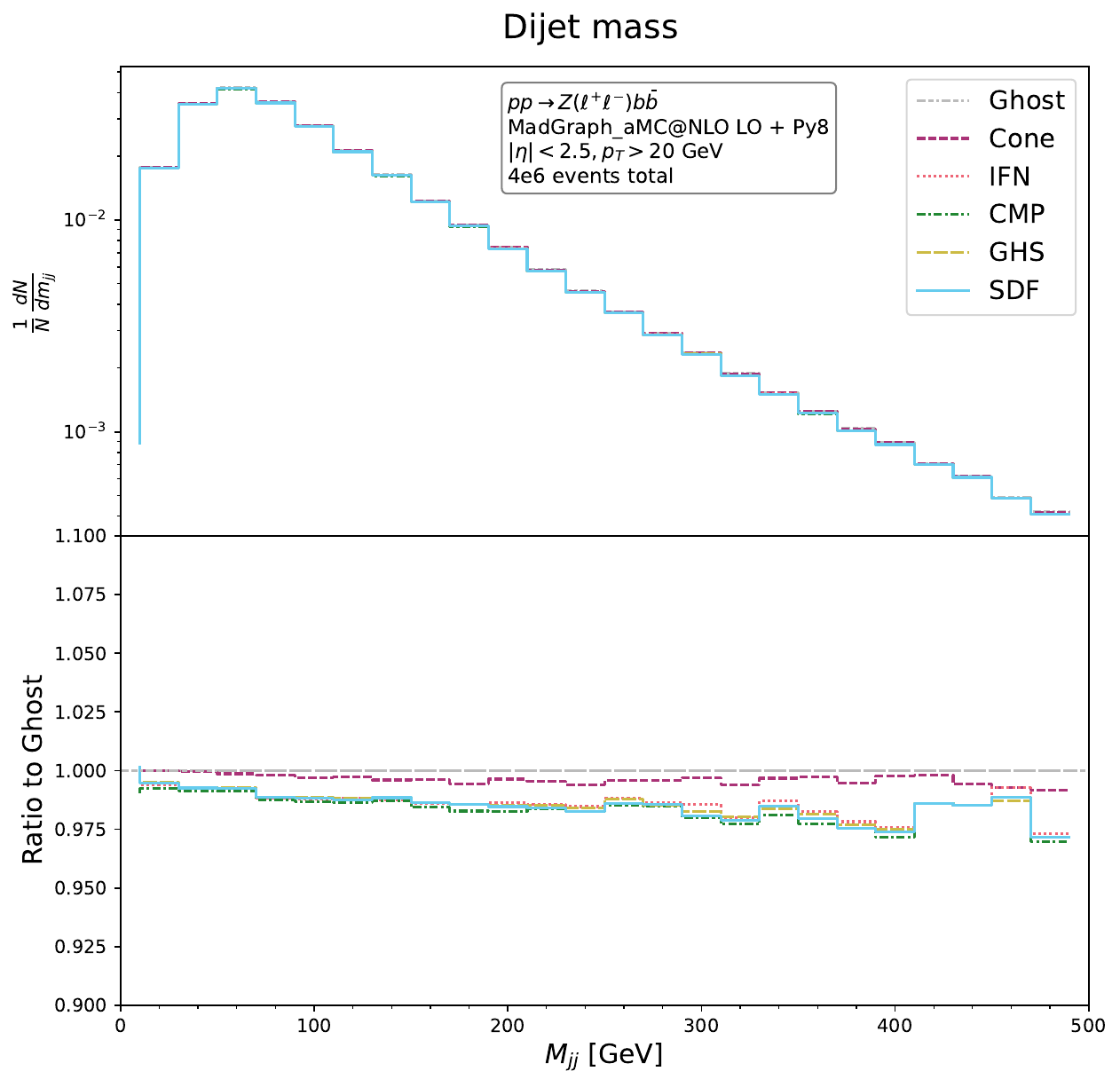}
    \caption{The \pt distribution of leading $b$ jets identified with the cone, ghost, and flavour algorithms with undecayed hadrons (top left), the leading $b$-jet mass (top right), and the dijet mass distribution of the two highest-\pt $b$ jets identified in the same manner (bottom). Ratios are compared to the distribution for jets clustered with ghost labelling.}
    \label{ATLAS b-jet kinematics undecayed}
\end{figure}

Surprisingly, \cref{ATLAS b-jet kinematics} shows that the flavour algorithms ascribe flavour to more jets than the experimental algorithms. Naively, this is unexpected as the flavour algorithms should be \emph{cancelling} flavour pairs originating from gluon splitting, which the experimental algorithms would be unable to identify. This difference can be attributed to the difference in constituents. This can be especially seen in the leading jet mass plot in \cref{ATLAS b-jet kinematics}, as it shows that the ghost and cone algorithms find a significant proportion of jets with a mass below that of a $b$ hadron. When considering undecayed hadrons and ascribing flavour with all algorithms, results match more closely, as can be seen in \cref{ATLAS b-jet kinematics undecayed}.

We have also investigated some aspects of jet substructure beyond just the jet mass. In particular, we have looked at any differences in the Lund Jet Plane~\cite{Dreyer:2018nbf} which may arise due to the flavour labelling strategy. For jet substructure measurements, it is often best to exploit the higher resolution of tracking detectors compared to calorimeters. For this reason, measurements focus on the charged component of the jet to elucidate more information. However, in this study we choose to study the substructure of the jet as a whole, thus imposing no cut on either the \pt of the constituents or on their charge. 

Another important point to mention is the fact that some substructure effects, such as the dead cone effect~\cite{Dokshitzer:1991fd} (see e.g., Refs.~\cite{Ghira:2023bxr,Caletti:2023spr,Dhani:2024gtx} for recent developments on heavy-flavour jet substructure), rely on the identification of the heavy-flavour hadron and accounting for its four-momentum. For jets obtained with the algorithms, this is straightforward as the hadron is already included in the jets' constituents. On the other hand, experimentally it is necessary to either identify the heavy hadron, as the ALICE collaboration has done~\cite{ALICE:2021aqk}, or reconstruct it as the CMS collaboration has done~\cite{CMS:2024gds}. In this study, we have opted to only consider undecayed heavy hadrons.

\begin{figure}
    \centering
    \includegraphics[scale=0.65]{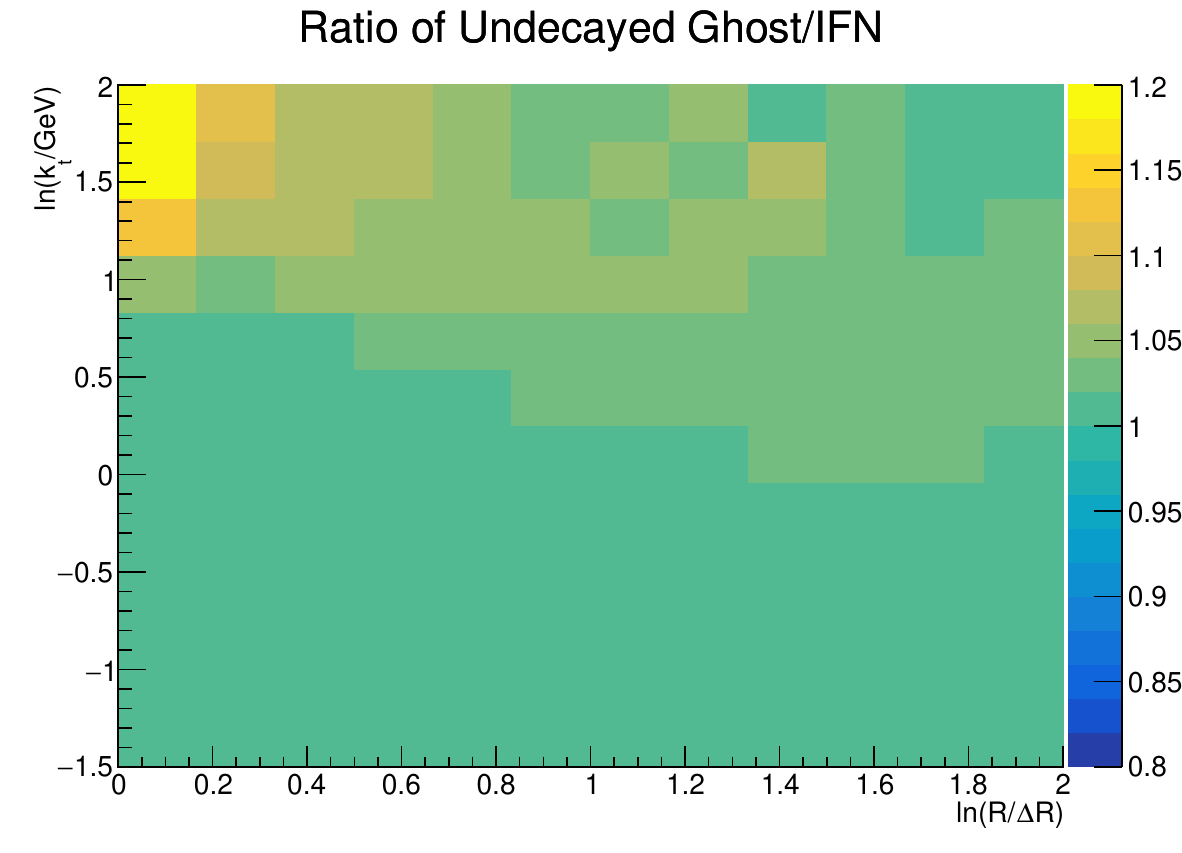} 
    \caption{The ratio of the Lund Jet Plane for leading $b$ jets identified with ghost labelling containing undecayed $b$ hadrons and  those identified with IFN.}
    \label{GHOST-IFN b}
\end{figure}

To compare the Lund Plane reconstructed using one of the flavour algorithms and one of the standard experimental labelling algorithms, in \cref{GHOST-IFN b} we show the ratio of the two-dimensional Lund Plane for the leading $b$ jets found with ghost and IFN labelling. It is clear that in the central part of the Lund Plane the results are very similar, while a mild excess is visible in the high-$k_t$ and small angle regions of the Lund Plane. We understand the excess at high-$k_t$ and wide angle to be due to contributions from initial state radiation.

The motivation for this work is to ensure that jets labelled as heavy-flavour jets are indeed such, and not misidentified due to higher-order effects. Experimentally, it is of course extremely difficult, if not infeasible to disentangle the true origin of a $b$ hadron, though for labelling purposes, a simple way to correct the experimental labelling procedure currently implemented is by imposing a 2-$b$-tag veto. By doing so, we require jets to contain (in the ghost case) or be matched to (in the cone case) \emph{only one} $b$ hadron. This explicitly avoids some problematic configurations, such as soft gluons splitting within jets.

When accounting for this, we can try to see if the distributions obtained with experimental flavour labelling approach those for the flavour algorithms. 

\begin{figure}
    \centering
    \includegraphics[width=0.49\linewidth]{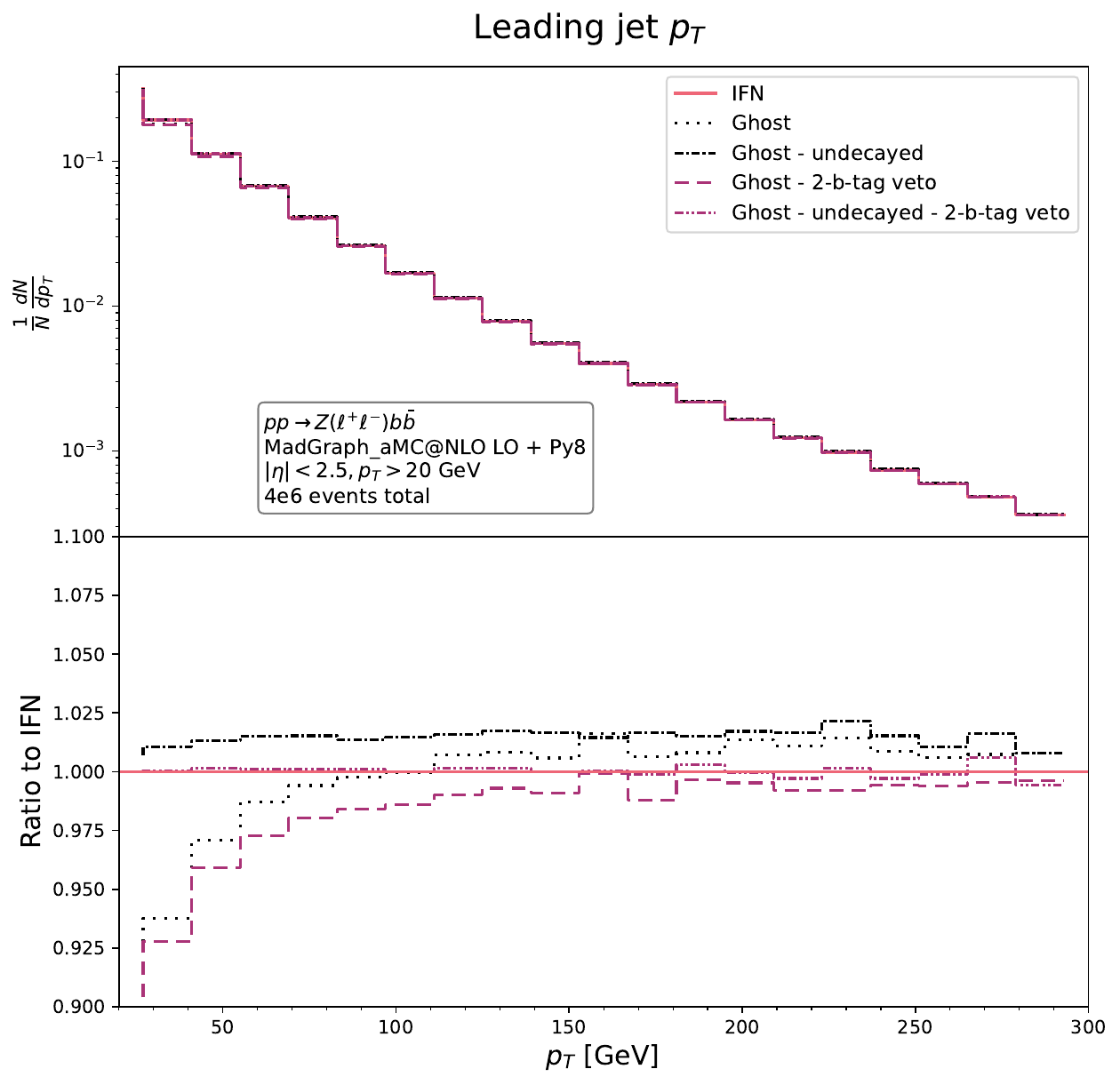} 
    \includegraphics[width=0.49\linewidth]{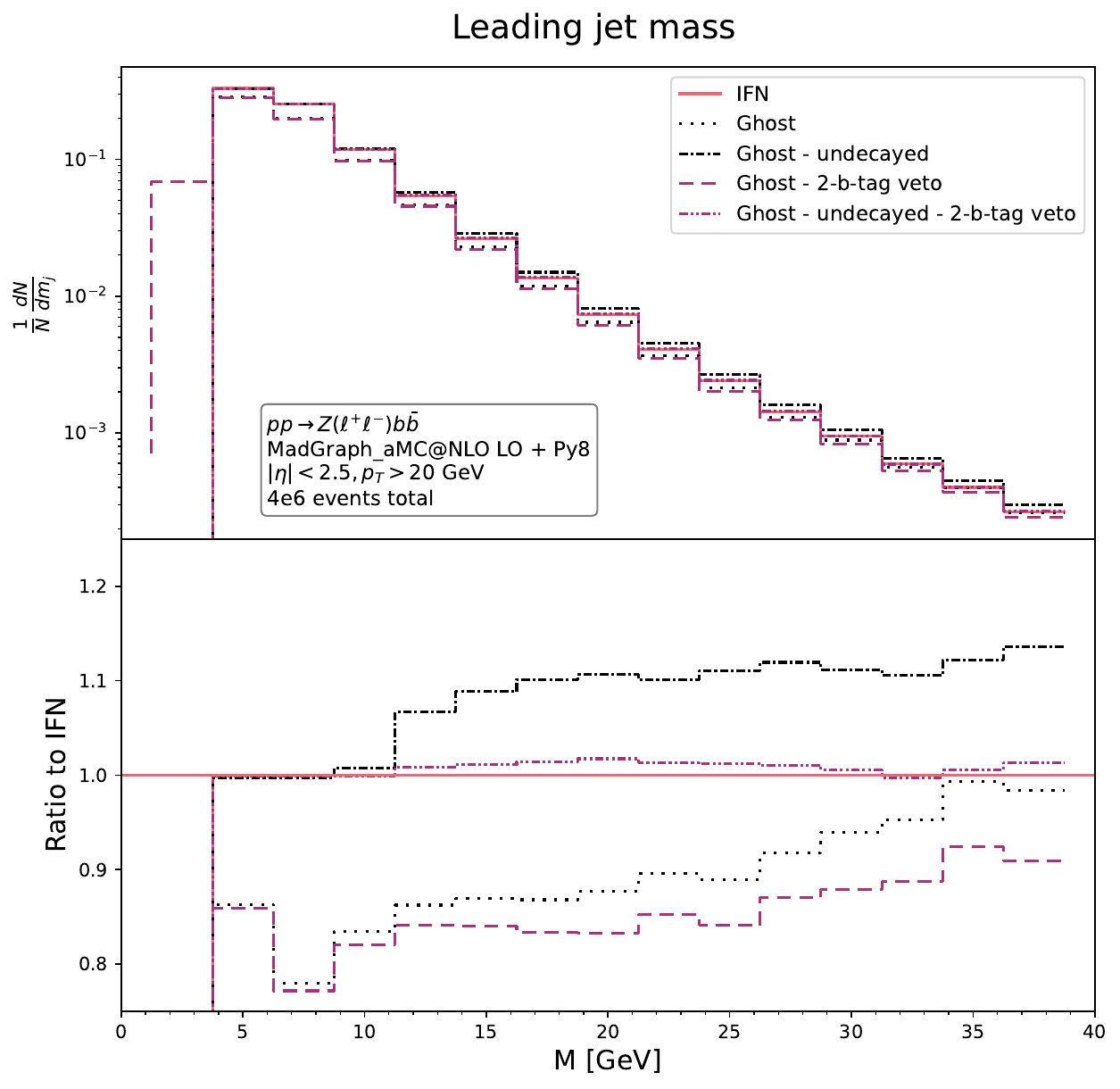}\\
    \includegraphics[width=0.49\linewidth]{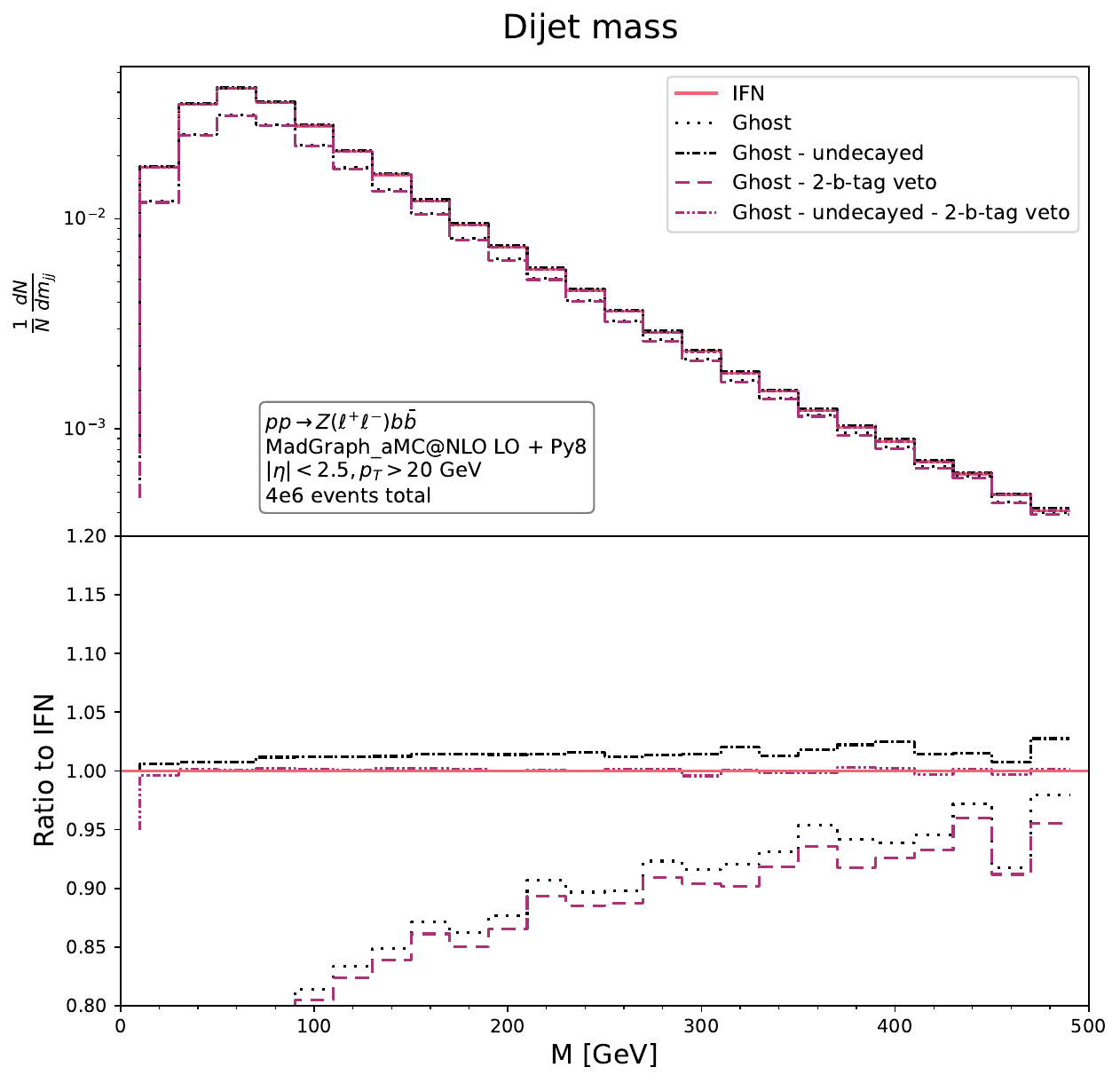}
    \caption{The \pt distribution of the leading $b$ jet (top left), the leading jet mass (top right), and the dijet mass (bottom), comparing $b$ jets clustered with IFN (continuous line) to those whose flavour was identified with the ghost prescriptions. The dotted and the dashed lines correspond to the standard ghost definitions (decayed hadrons) without or with 2-$b$-tag veto, respectively, while the dotted-dashed lines refer to the distributions with undecayed hadrons.}
    \label{ATLAS 2-b-tag veto cmp}
\end{figure}

As can be seen in \cref{ATLAS 2-b-tag veto cmp}, applying a 2-$b$-tag veto has a small effect on the result of the corresponding distribution for all observables considered for the LO samples considered, of the order of a few percent, both in the case of decayed and undecayed heavy hadrons. This effect, however, is negligible with respect to the kinematic differences between jets with different constituent definitions, as previously stated.

\begin{figure}
    \centering
    \includegraphics[width=0.49\linewidth]{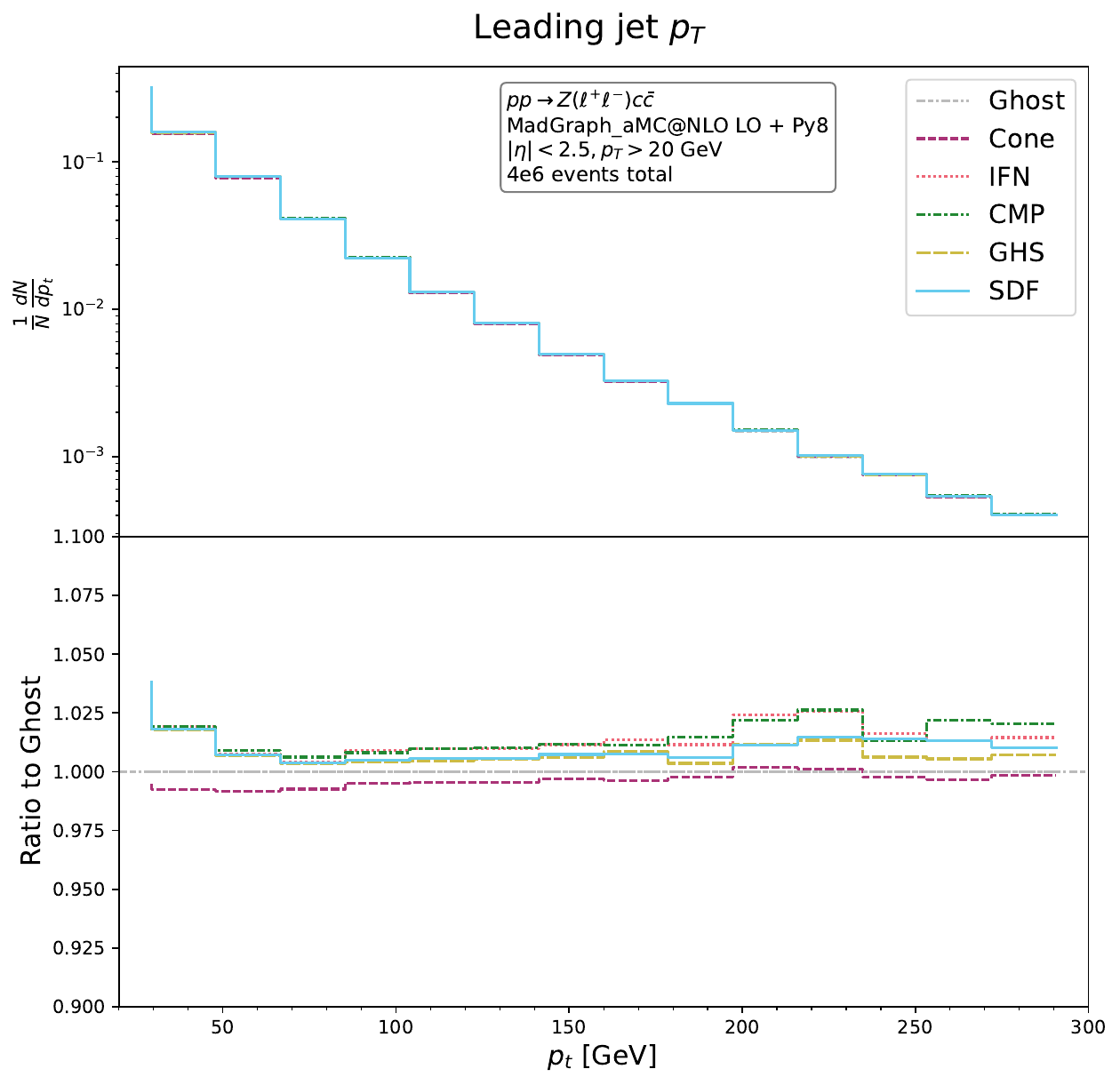} 
    \includegraphics[width=0.49\linewidth]{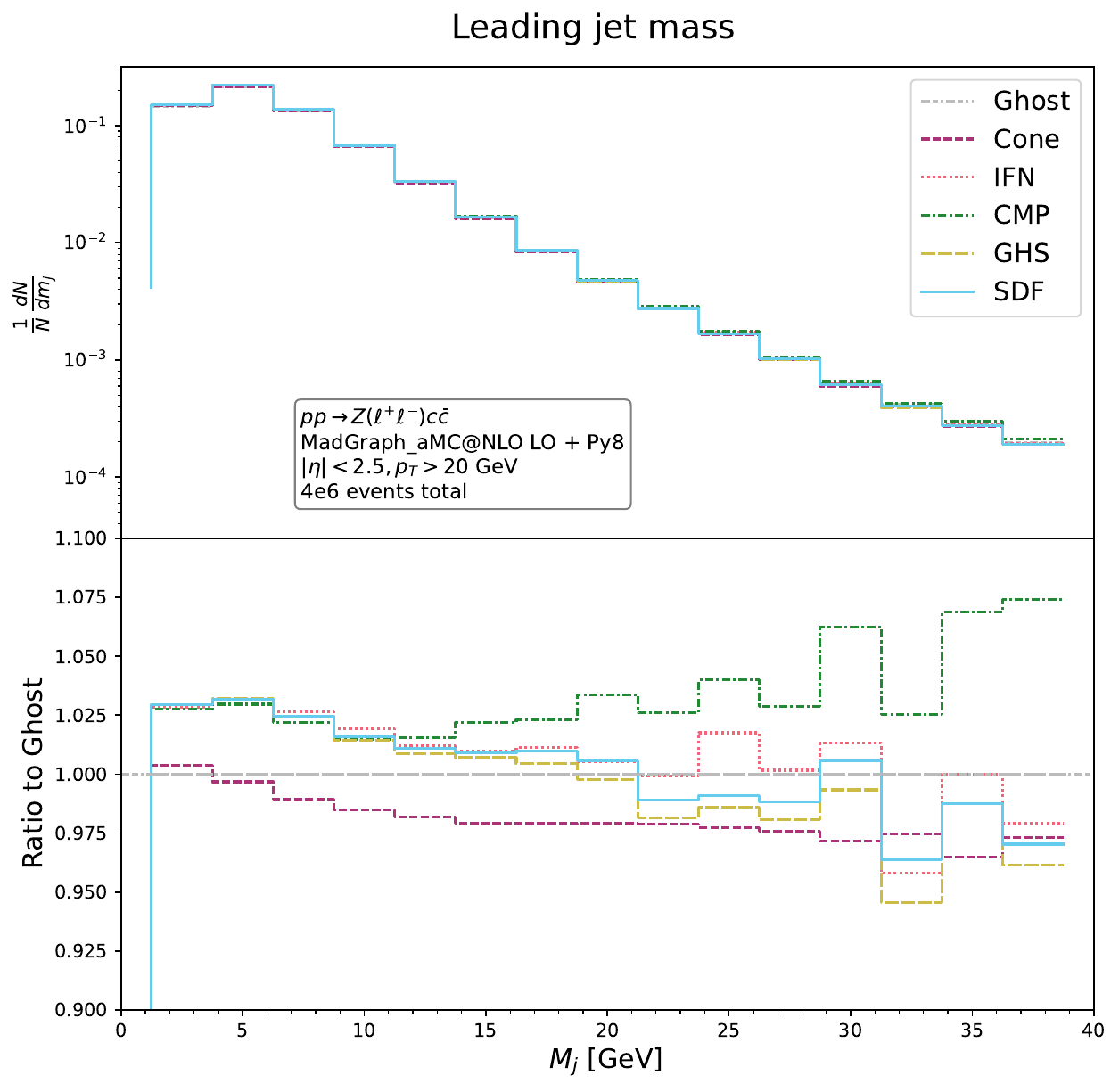}\\
    \includegraphics[width=0.49\linewidth]{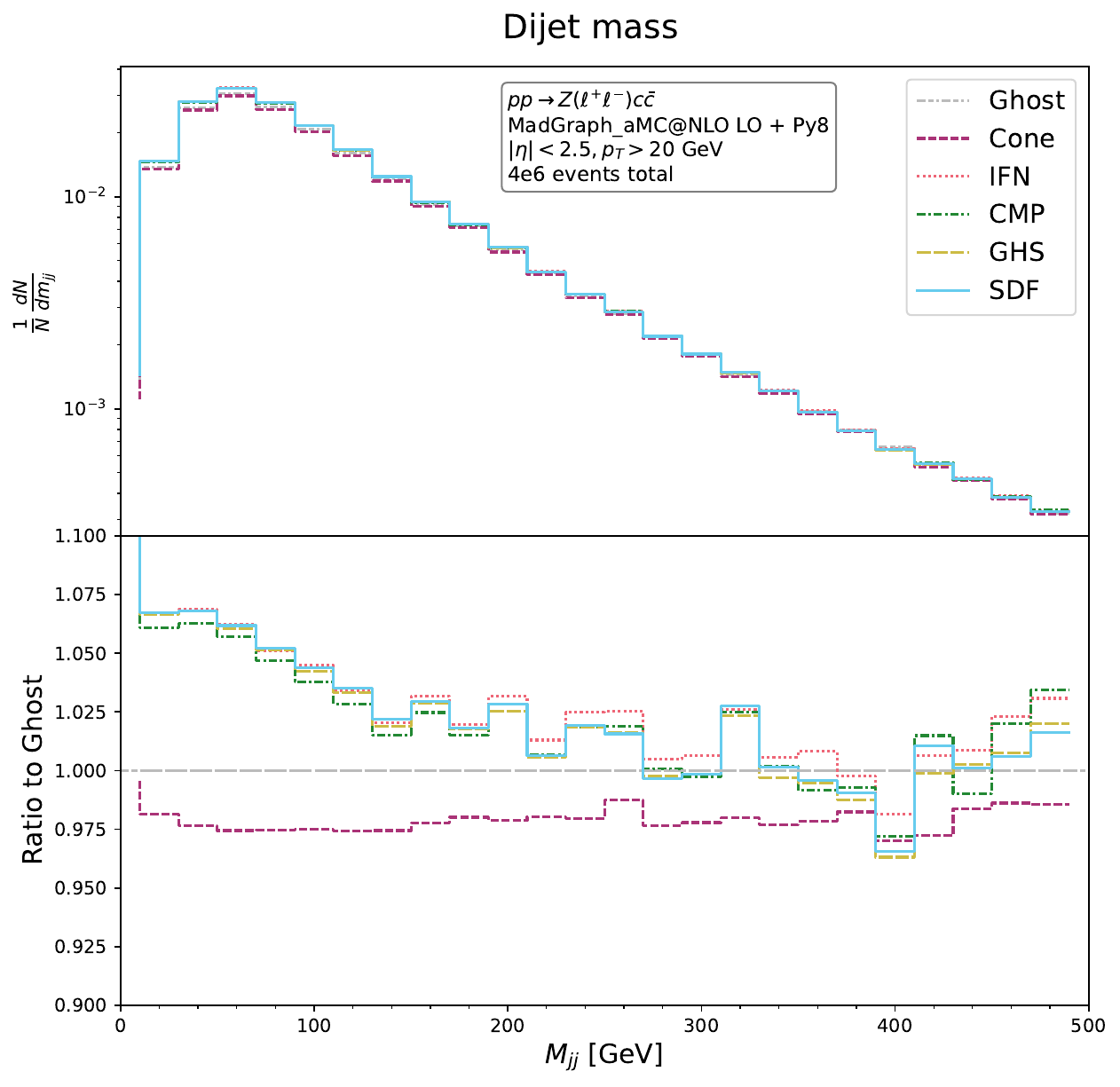}
    \caption{The \pt distribution of leading $c$ jets identified with the cone, ghost, and flavour algorithms (top left), the leading jet mass (top right), and the dijet mass distribution of the two highest-\pt $c$ jets identified in the same manner (bottom). Ratios are compared to the distribution for jets clustered with ghost labelling.}
    \label{ATLAS c-jet kinematics}
\end{figure}

\begin{figure}
    \centering
    \includegraphics[width=0.49\linewidth]{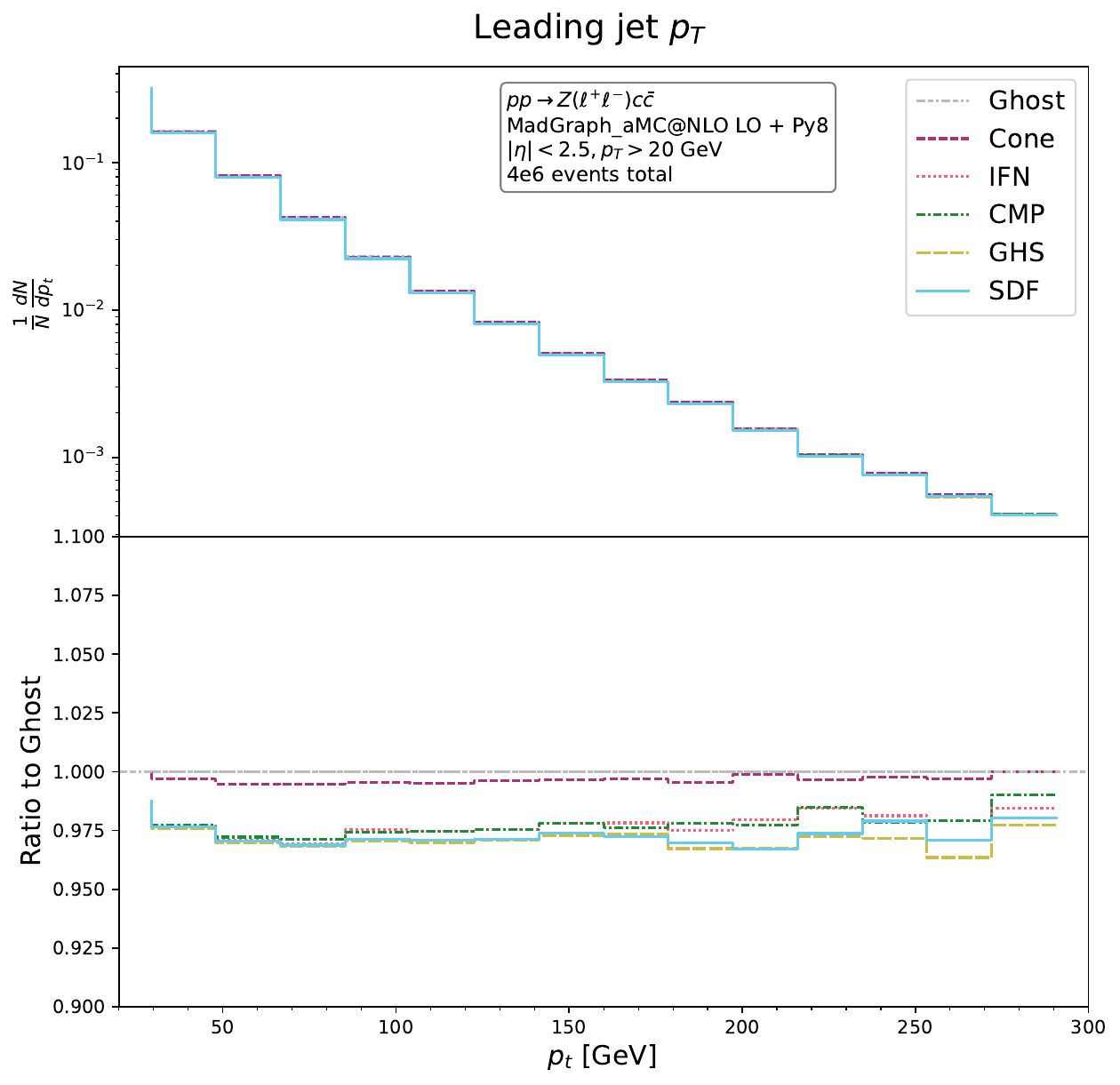} 
    \includegraphics[width=0.49\linewidth]{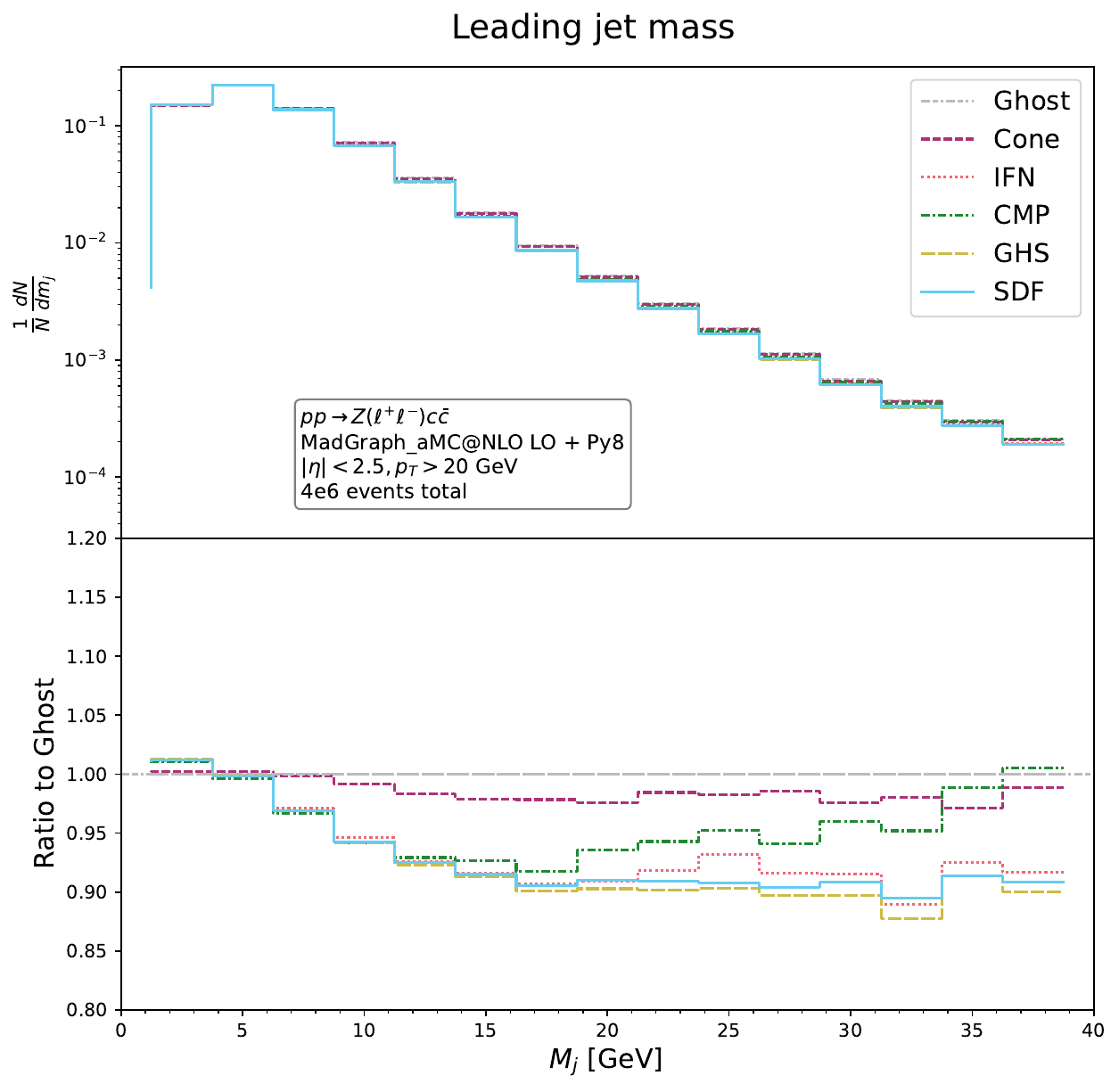} 
    \includegraphics[width=0.49\linewidth]{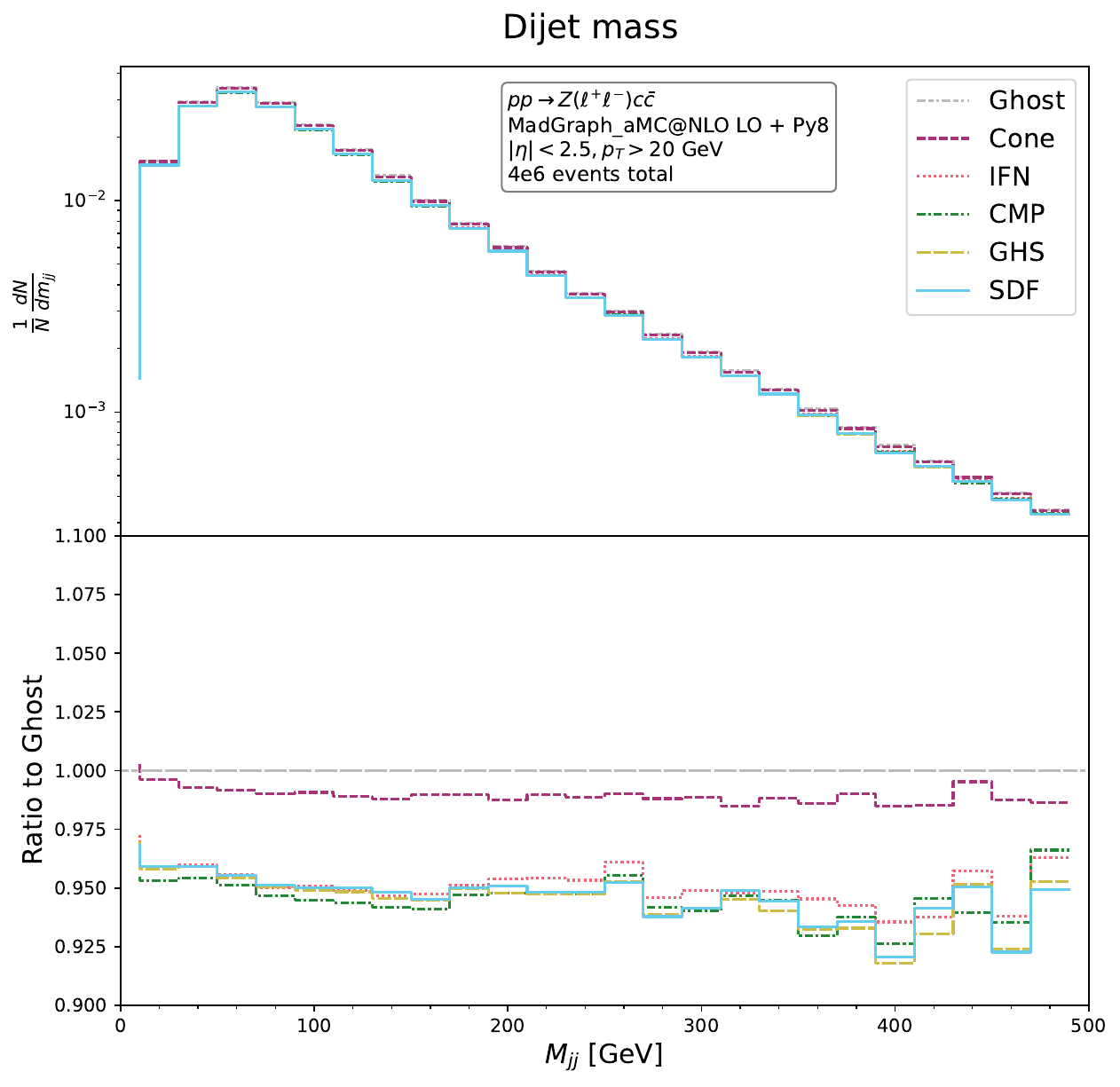}
    \caption{The \pt distribution of leading $c$ jets identified with the cone, ghost, and flavour algorithms with undecayed hadrons (top left), the leading jet mass distribution (top right), and the dijet mass distribution (bottom) of the two highest-\pt $c$ jets identified in the same manner. Ratios are compared to the distribution for jets clustered with ghost labelling.}
    \label{ATLAS c-jet kinematics undecayed}
\end{figure}

\begin{figure}
    \centering
    \includegraphics[scale=0.65]{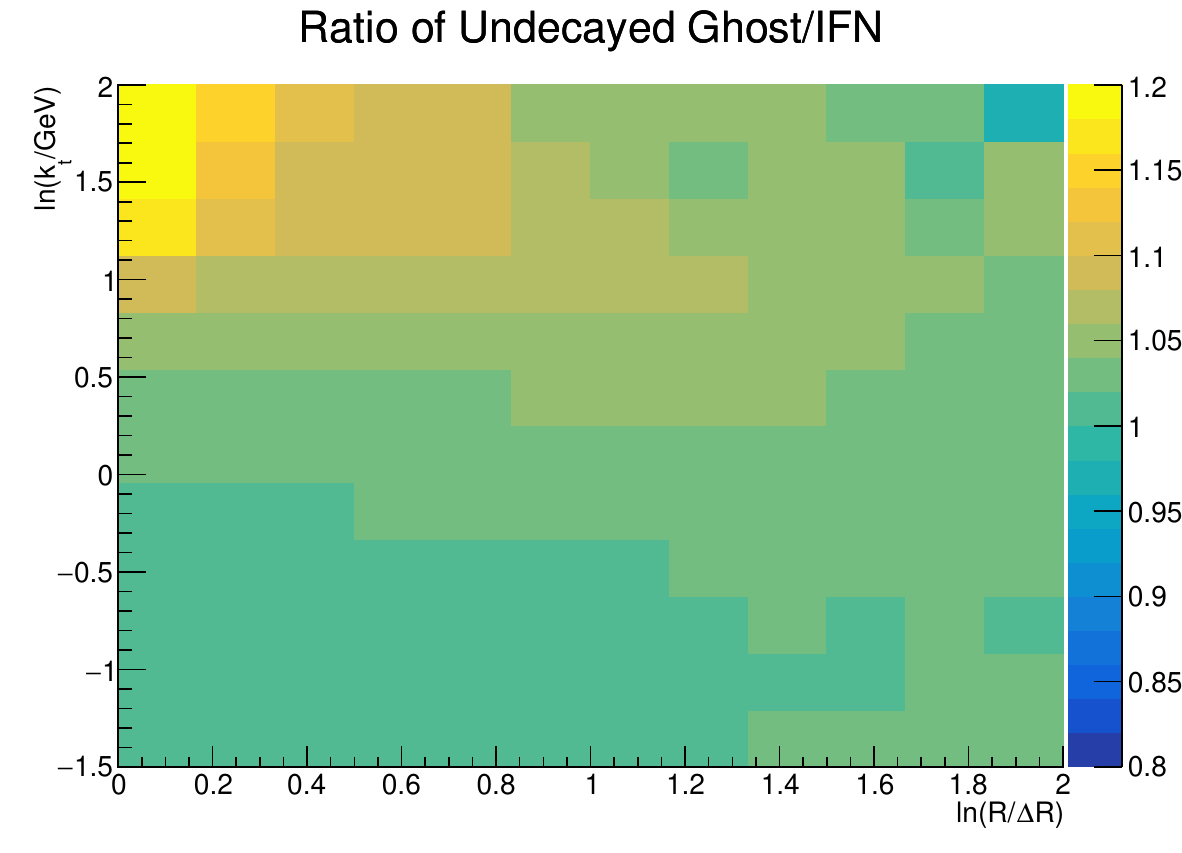} 
    \caption{The ratio of the Lund Jet Plane for leading $c$ jets identified with ghost labelling containing undecayed $c$ hadrons and  those identified with IFN.}
    \label{GHOST-IFN c}
\end{figure}

\begin{figure}
    \centering
    \includegraphics[width=0.49\linewidth]{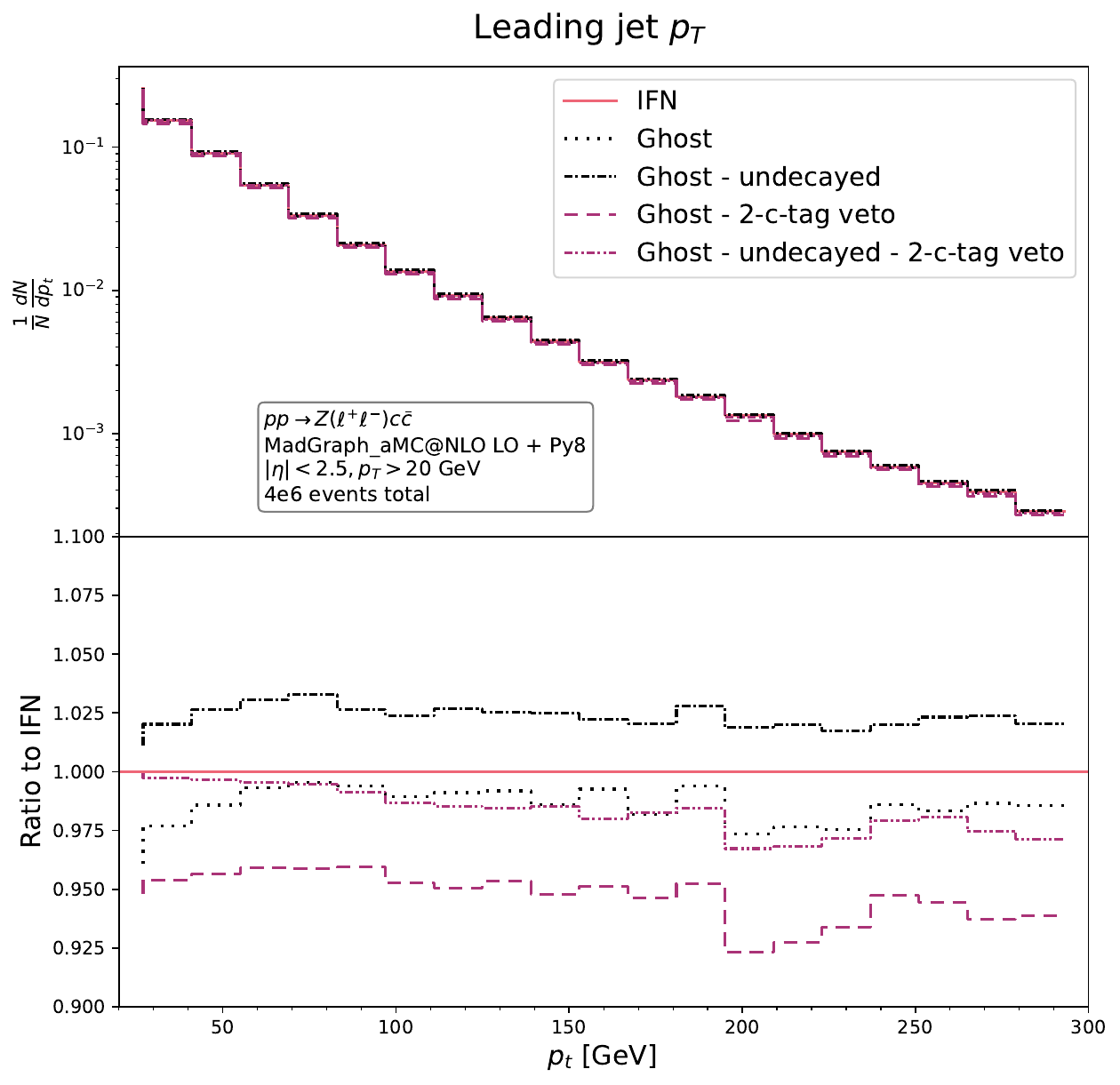}
    \includegraphics[width=0.49\linewidth]{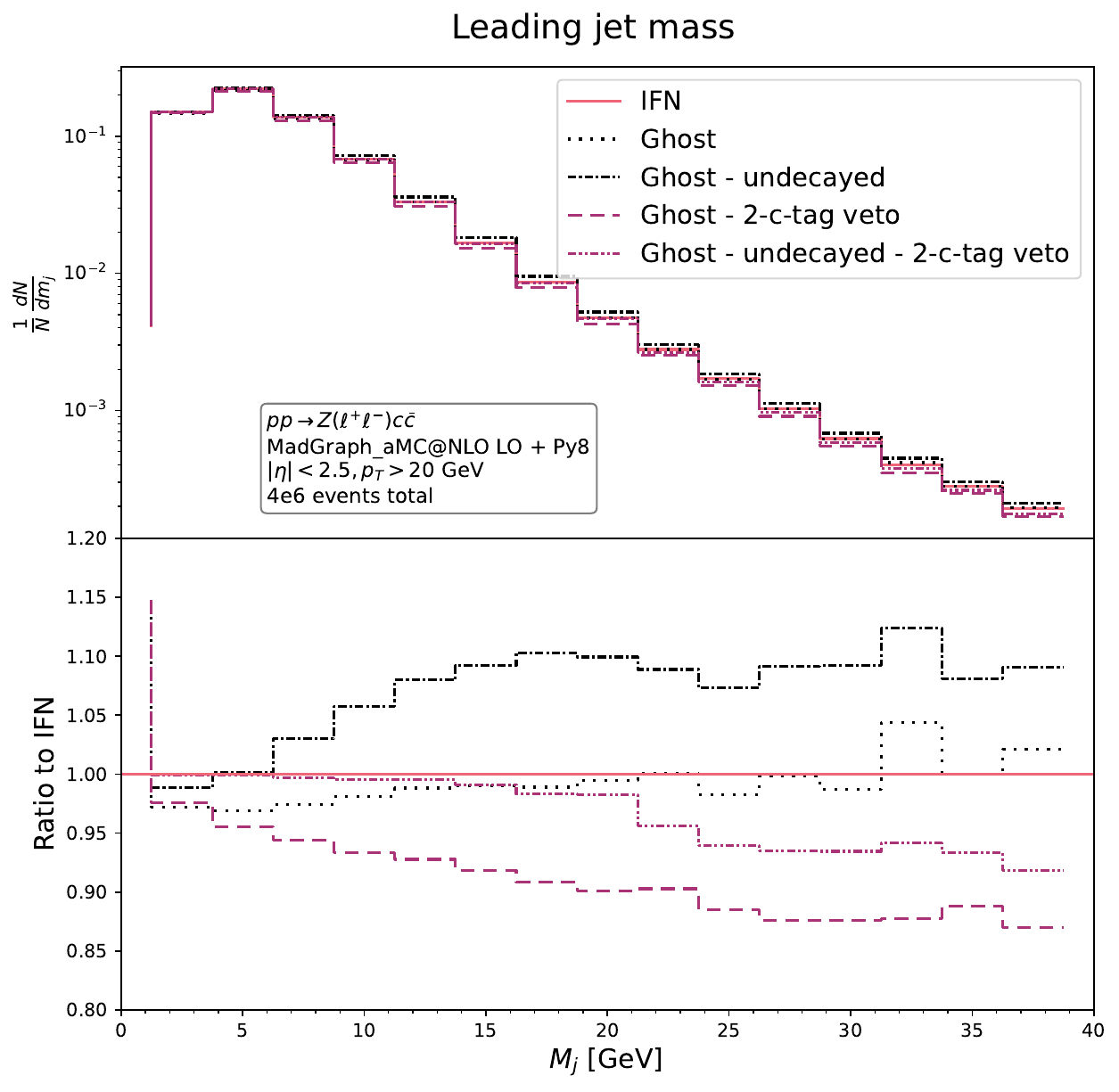}\\
    \includegraphics[width=0.49\linewidth]{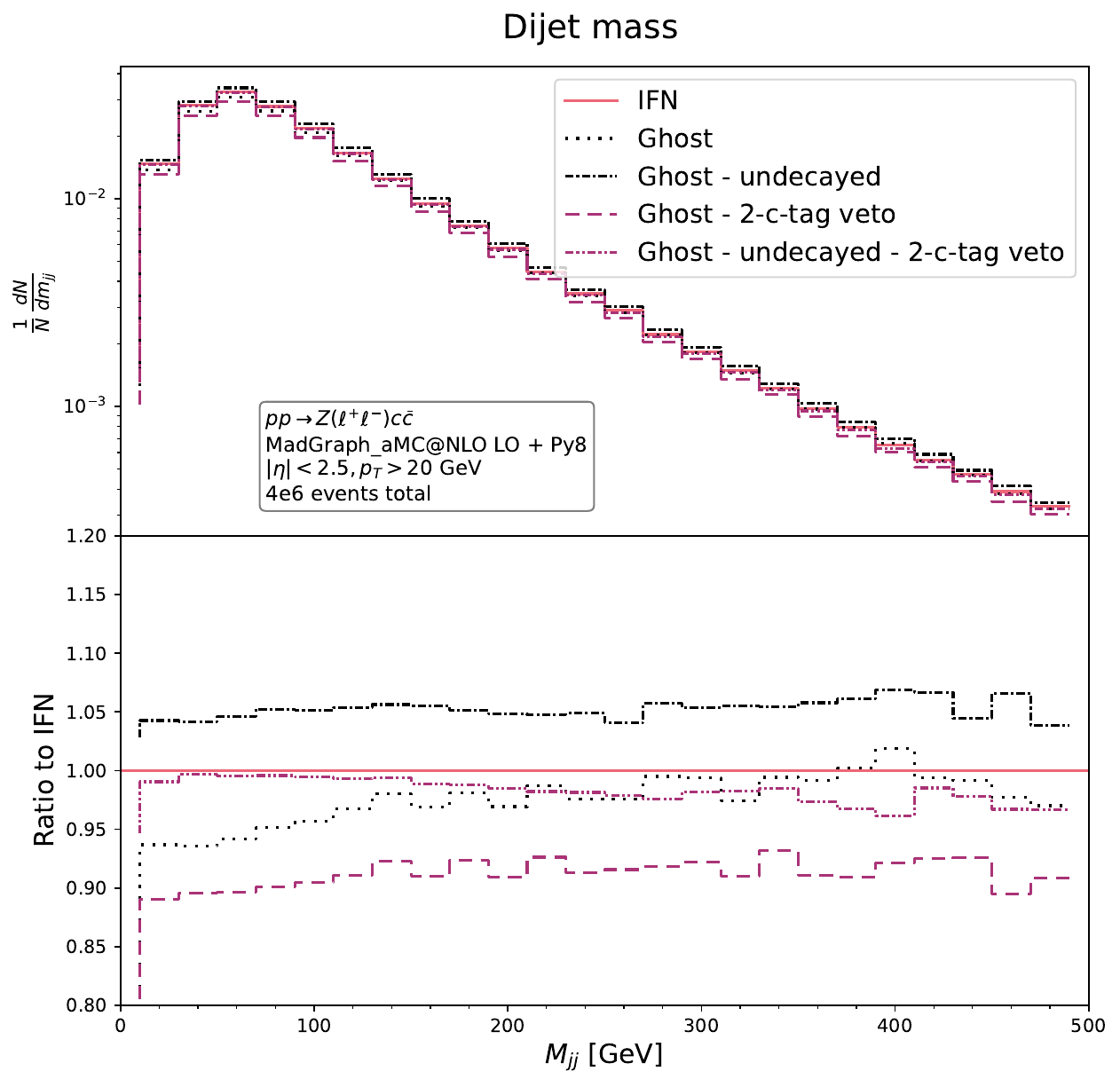}
    \caption{The \pt distribution of the leading $c$ jet (top left), the leading jet mass distribution (top right), and the dijet mass (bottom) comparing $c$ jets clustered with IFN (continuous line) to those whose flavour was identified with the ghost prescriptions. The dotted and the dashed lines correspond to the standard ghost definitions (decayed hadrons) without or with 2-$c$-tag veto, respectively, while the dotted-dashed lines refer to the distributions with undecayed hadrons.}
    \label{ATLAS 2-c-tag veto cmp}
\end{figure}

The study was repeated for $Z+c\bar{c}$ events, as shown in \crefrange{ATLAS c-jet kinematics}{ATLAS 2-c-tag veto cmp}. Some effects, such as the kinematic differences in the jets due to the presence of the heavy-hadron decay products, are not as strong as in the $b$-jet case due to the lower particle multiplicity of the decays compared to those for $b$ hadrons, as can be seen in \cref{ATLAS c-jet kinematics}. On the other hand, differences between the algorithms are more prominent due to the higher likelihood of $g\rightarrow c\bar{c}$ splittings. Overall, however, the same trends and conclusions hold. 

\FloatBarrier

\sectionAuthors{Impact on jet flavour tagging in ATLAS} {Francesco Giuli, Rados\l aw Grabarczyk, Alberto Rescia, and Federico Sforza}\label{app:atlas_ftag}

Jet flavour tagging in ATLAS is done by training neural network classifiers on Monte Carlo simulated data. In the case of high-\pt jets, for which the differences between different jet flavour algorithms are enhanced, the training sample is obtained by simulating a heavy resonance $Z'$. This resonance is set to decay into $b\bar{b}$ 1/3rd of the time, $c \bar{c}$ 1/3rd of the time, and light flavours with equal branching ratios otherwise. The width of $Z'$ is tuned to ensure a large spread of different transverse momenta of the jets. The event generation is done entirely with \pythia. Following simulation of the ATLAS detector response and reconstruction, all jets that satisfy $|\eta| < 2.5$ and $\pt > \SI{250}{\GeV}$ enter the training, which means that they can often originate from the parton shower instead of the hard event. 

\begin{figure}
    \centering
    \includegraphics[width=0.49\linewidth]{figures/ATLAS_JSS/NETno-grey-lineBLACKANTIKT-MAINCIFN-CMP-ATLAS-compare-pt-with-ratio.pdf} 
    \includegraphics[width=0.49\linewidth]{figures/ATLAS_JSS/NETno-grey-line-BLACKANTIKT-MAINCBFN-CMP-ATLAS-compare-pt-with-ratio}\\
    \includegraphics[width=0.49\linewidth]{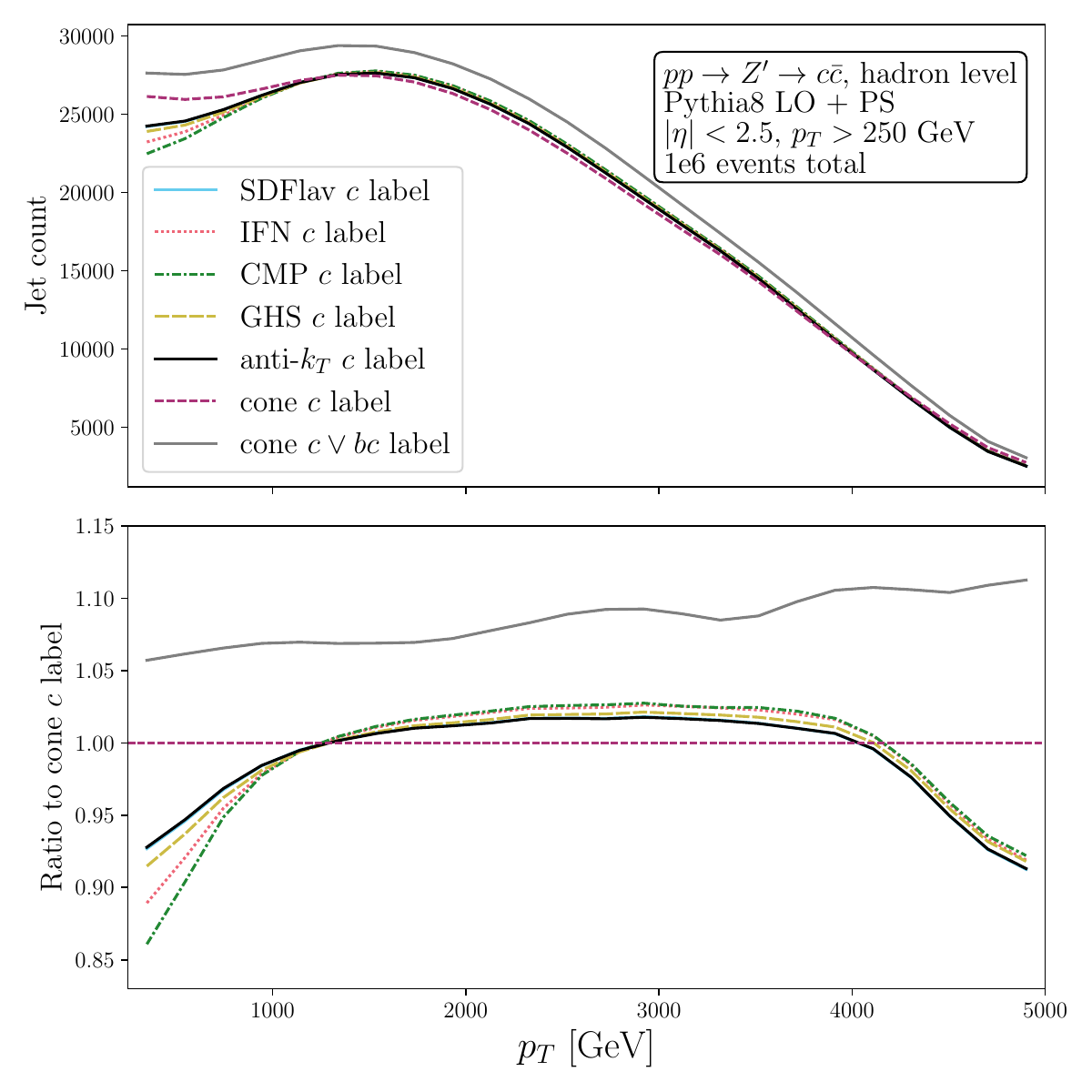}
    
    \caption{The \pt spectra of $c$ jets (left) and $b$ jets (right) found in $pp \rightarrow Z' \rightarrow c \bar{c}$ events, simulated at LO and showered with \pythiav{8}. All plots use labelling schemes with ``net-flavour'' recombination scheme except cone, for which a difference is also shown if $c$-jet definition is prioritised over the presence of a $b$ hadron or not.}
    \label{fig:ATLAS-FTAG-comparison}
\end{figure}

To investigate the effect of using jet flavour algorithms on these types of events, we simulate the same $Z'$ resonance with \pythiav{8}, with the modification that it decays into $c \bar{c}$ pairs 100\% of the time, which means that any $b$ jets found in this sample originate from the parton shower. After hadronisation and decay of unstable hadrons, we identify all decay products of $b$ and $c$ hadrons, replace them with their parent hadrons and remove them. Next, we cluster the event with different jet flavour algorithms using the mod2 recombination scheme. The results are shown in \cref{fig:ATLAS-FTAG-comparison}. In the case of $c$ jets, we find that the differences between cone labelling and other algorithms are largest at the lower and higher ends of the \pt spectrum, where we expect more gluon splitting into heavy flavour generated by the parton shower. 

Interestingly, in the middle of the \pt spectrum, IRC safe jet flavour algorithms yield more $c$-labelled jets than anti-\kt or cone labelling. This is due to a disagreement between labelling schemes regarding jets having \textit{both} $b$ and $c$ labels, most often arising in this sample from a hard $c$ quark and a soft, large-angle $b\bar{b}$ pair. The most ``aggressive" IRC-safe labelling schemes will tend to only assign a $c$ label in this case. Other algorithms may label it as $bc$, which, due to $b$ label always taking priority, results in a final $b$ label of the jet. To show that this effect is significant, we provide the result of using a ``cone $c \vee bc$" labelling scheme, where the $c$ label takes priority instead.

In the case of $b$ labelling, due to the nature of our sample, we are guaranteed to see effects of gluon splittings. On the lower end of the \pt spectrum, similarly to the $c$-jet case, CMP is the most aggressive algorithm. On the higher end, IFN becomes the most aggressive one. The effect of collinear gluon splittings into $b\bar{b}$, signified by the difference between the cone-labelled $b$ jets and anti-\kt-labelled ones slowly increases with \pt, but is likely to be significant for all transverse momenta in the range considered.

We conclude that collinear gluon splittings into heavy flavour can have a sizeable effect on the ATLAS flavour tagging training sample throughout the whole \pt spectrum, while differences between jet flavour algorithms may become relevant especially at transverse momenta just above \SI{250}{\GeV}, where a significant amount of jets that enter the training do not originate from the hard event.

\clearpage
\bibliography{lh23,ATLAS,ATLAS-PubNotes,CMS,LHCb-PAPER,LHCb-DP}

\end{document}